\PassOptionsToPackage{unicode}{hyperref}
\PassOptionsToPackage{hyphens}{url}
\PassOptionsToPackage{dvipsnames,svgnames,x11names}{xcolor}
\documentclass[
12pt]{article}

\usepackage{amsmath,amssymb}
\usepackage{iftex}
\usepackage{comment}
\usepackage{setspace}
\usepackage{mathtools}
\usepackage{algorithm,algpseudocode,float}
\usepackage{bm}
\usepackage{amsthm}
\usepackage{natbib}
\ifPDFTeX
\usepackage[T1]{fontenc}
\usepackage[utf8]{inputenc}
\usepackage{textcomp} 
\else 
\usepackage{unicode-math}
\defaultfontfeatures{Scale=MatchLowercase}
\defaultfontfeatures[\rmfamily]{Ligatures=TeX,Scale=1}
\fi
\usepackage{lmodern}
\ifPDFTeX\else  
\fi
\IfFileExists{upquote.sty}{\usepackage{upquote}}{}
\IfFileExists{microtype.sty}{
	\usepackage[]{microtype}
	\UseMicrotypeSet[protrusion]{basicmath} 
}{}
\makeatletter
\@ifundefined{KOMAClassName}{
	\IfFileExists{parskip.sty}{%
		\usepackage{parskip}
	}{
		\setlength{\parindent}{0pt}
		\setlength{\parskip}{6pt plus 2pt minus 1pt}}
}{
	\KOMAoptions{parskip=half}}
\makeatother
\usepackage{xcolor}
\makeatletter
\ifx\paragraph\undefined\else
\let\oldparagraph\paragraph
\renewcommand{\paragraph}{
	\@ifstar
	\xxxParagraphStar
	\xxxParagraphNoStar
}
\newcommand{\xxxParagraphStar}[1]{\oldparagraph*{#1}\mbox{}}
\newcommand{\xxxParagraphNoStar}[1]{\oldparagraph{#1}\mbox{}}
\fi
\ifx\subparagraph\undefined\else
\let\oldsubparagraph\subparagraph
\renewcommand{\subparagraph}{
	\@ifstar
	\xxxSubParagraphStar
	\xxxSubParagraphNoStar
}
\newcommand{\xxxSubParagraphStar}[1]{\oldsubparagraph*{#1}\mbox{}}
\newcommand{\xxxSubParagraphNoStar}[1]{\oldsubparagraph{#1}\mbox{}}
\fi
\makeatother

\providecommand{\tightlist}{%
	\setlength{\itemsep}{0pt}\setlength{\parskip}{0pt}}\usepackage{longtable,booktabs,array}
\usepackage{calc} 
\usepackage{etoolbox}
\makeatletter
\patchcmd\longtable{\par}{\if@noskipsec\mbox{}\fi\par}{}{}
\makeatother
\IfFileExists{footnotehyper.sty}{\usepackage{footnotehyper}}{\usepackage{footnote}}
\makesavenoteenv{longtable}
\usepackage{graphicx}
\makeatletter
\def\maxwidth{\ifdim\Gin@nat@width>\linewidth\linewidth\else\Gin@nat@width\fi}
\def\maxheight{\ifdim\Gin@nat@height>\textheight\textheight\else\Gin@nat@height\fi}
\makeatother
\setkeys{Gin}{width=\maxwidth,height=\maxheight,keepaspectratio}
\makeatletter
\def\fps@figure{htbp}
\makeatother

\makeatletter
\@ifpackageloaded{caption}{}{\usepackage{caption}}
\AtBeginDocument{%
	\ifdefined\contentsname
	\renewcommand*\contentsname{Table of contents}
	\else
	\newcommand\contentsname{Table of contents}
	\fi
	\ifdefined\listfigurename
	\renewcommand*\listfigurename{List of Figures}
	\else
	\newcommand\listfigurename{List of Figures}
	\fi
	\ifdefined\listtablename
	\renewcommand*\listtablename{List of Tables}
	\else
	\newcommand\listtablename{List of Tables}
	\fi
	\ifdefined\figurename
	\renewcommand*\figurename{Figure}
	\else
	\newcommand\figurename{Figure}
	\fi
	\ifdefined\tablename
	\renewcommand*\tablename{Table}
	\else
	\newcommand\tablename{Table}
	\fi
}
\@ifpackageloaded{float}{}{\usepackage{float}}
\floatstyle{ruled}
\@ifundefined{c@chapter}{\newfloat{codelisting}{h}{lop}}{\newfloat{codelisting}{h}{lop}[chapter]}
\floatname{codelisting}{Listing}

\makeatother
\makeatletter
\@ifpackageloaded{caption}{}{\usepackage{caption}}
\@ifpackageloaded{subcaption}{}{\usepackage{subcaption}}
\makeatother

\ifLuaTeX
\usepackage{selnolig}  
\fi
\usepackage[]{natbib}
\usepackage{bookmark}

\IfFileExists{xurl.sty}{\usepackage{xurl}}{} 
\hypersetup{
	pdftitle={Safe, Always-Valid Alpha-Investing Rules For Doubly Sequential Online Inference},
	pdfauthor={Zeyu Yao; Wenguang Sun; Bowen Gang},
	pdfkeywords={},
	colorlinks=true,
	linkcolor={blue},
	filecolor={Maroon},
	citecolor={Blue},
	urlcolor={Blue},
	pdfcreator={LaTeX via pandoc}}



\newtheorem{definition}{Definition}
\newtheorem{theorem}{Theorem}
\newtheorem{lemma}{Lemma}
\newtheorem{remark}{Remark}
\newtheorem{proposition}{Proposition}

\newtheorem{condition}{Condition}

\newcommand{\anon}{1}

\let\oldbibliography\thebibliography
\renewcommand{\thebibliography}[1]{%
	\oldbibliography{#1}%
	\setlength{\itemsep}{0pt}
	\setlength{\parsep}{0pt}
	\setlength{\topsep}{0pt}
	\setlength{\partopsep}{0pt}
	\small                          
}

\begin{document}

	\def\spacingset#1{\renewcommand{\baselinestretch}%
		{#1}\small\normalsize} \spacingset{1}


	\if1\anon
	{
		\title{\bf Safe, Always-Valid Alpha-Investing Rules For Doubly Sequential Online Inference}
		\date{ }
		\author{Zeyu Yao
			\hspace{.2cm}\\
			School of Mathematical Sciences and Center for Data Science, Zhejiang University\\
		Wenguang Sun \\
	Center for Data Science and School of Management, Zhejiang University\\
			and\\
			Bowen Gang \\
	        Department of Statistics and Data Science, Fudan University\\
		}
		\maketitle
	} \fi
	
	\if0\anon
	{
		\bigskip
		\bigskip
		\bigskip
		\begin{center}
			{\LARGE\bf Safe, Always-Valid Alpha-Investing Rules For Doubly Sequential Online Inference}
		\end{center}
		\medskip
	} \fi
	
	\bigskip
	\begin{abstract}
		
		Dynamic decision-making in rapidly evolving research domains, including marketing, finance, and pharmaceutical development, presents a significant challenge.  Researchers frequently confront the need for real-time action within a doubly sequential framework characterized by the continuous influx of high-volume data streams and the intermittent arrival of novel tasks. 
		This calls for the development and implementation of new online inference protocols capable of handling both the continuous processing of incoming information and the efficient allocation of resources to address emerging priorities. We introduce a novel class of Safe and Always-Valid Alpha-investing (SAVA) rules that leverages powerful tools including always valid p-values, e-processes, and online false discovery rate methods. The SAVA algorithm effectively integrates information across all tasks, mitigates the alpha-death problem, and controls the false selection rate (FSR) at all decision points. We validate the efficacy of the SAVA framework through rigorous theoretical analysis and extensive numerical experiments. Our results demonstrate that SAVA not only offers effective control of the FSR but also significantly improves statistical power compared to traditional online testing approaches.
	\end{abstract}
	
	\noindent%
	\vfill
	
	\newpage
	\spacingset{1.9} 

	\section{Introduction}\label{sec:Intro}
	\subsection{Doubly sequential experiments: examples and illustrations}

In various data-intensive application domains, researchers are tasked with monitoring extensive data streams and making real-time decisions adaptively, while accounting for incoming data, experimental conditions, and task-specific contexts. To motivate our study and highlight the associated challenges, we present a few specific application scenarios.
	
	\begin{itemize}
		\tightlist
		\item \emph{Online experimentation:} A/B testing has become a fundamental technique for evaluating the effectiveness of various web features and marketing strategies \citep{Feit19, Berman22}. This tool is particularly valuable in website optimization and mobile app development for more informed decision-making and improved user experiences. To identify the most effective user interfaces, page layouts, and innovative functionalities, organizations frequently conduct multiple experiments simultaneously over extended periods. This continuous influx of new features and designs necessitates the development of valid and efficient online inference protocols to provide timely insights and facilitate iterative development, thereby minimizing risk and reducing costs before large-scale implementation. 
		
		\item \emph{Candidate customer screening:} In the finance sector, a crucial task is to identify high-value potential customers, particularly in dynamically changing markets \citep{onlinetrust04,Pot-Cust23}. Agents continuously gather data from emerging user profiles and swiftly identify target users to optimize time and budget expenditures. A critical challenge is to avoid excessive false selections, as a high false selection rate can incur significant financial costs. The development of principled screening protocols not only enhances the efficiency of user acquisition but also ensures that financial resources are allocated judiciously.

		\item \emph{Drug discovery:} High throughput screening involves the screening of a large library of compounds against a specific biological target or assay system \citep{Dueas23}. This screening is typically conducted using automated robotic platforms, which allow for the rapid testing of thousands to millions of compounds. The discovery of reliable candidates is a dynamic decision-making process, where various statistical methods, such as dose-response analysis and hit selection algorithms, have been employed \citep{Parham09, Tansey21}. When a compound shows promise during the initial screening, it is promptly subjected to further experiments to evaluate its therapeutic effectiveness in more detail. In this fast-paced environment, a key objective is to design sequential experiments conducted in parallel, along with screening strategies, to manage resources efficiently and expedite the discovery process. 
		
	\end{itemize}

\begin{figure}
		
		\centering{
			
			\includegraphics[width=4.5in]{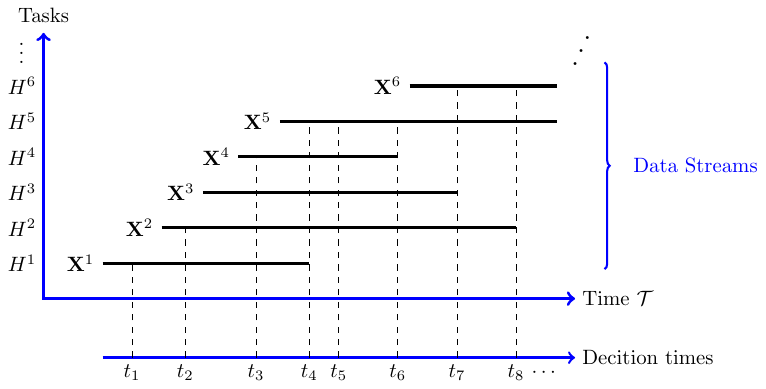}
			
		}
		
		\caption{\label{fig:illus}\small An illustration of the doubly sequential setup. A stream of tasks, $H^1, H^2, \cdots$, each of which collects data streams ($\mathbf X^1, \mathbf X^2, \cdots$) in a sequential manner, arrive sequentially. The process is asynchronous: tasks can start at arbitrary times, and multiple data streams can overlap in time.}
		
	\end{figure}

	The commonality between the above examples lies in the occurrence of sequentiality at both the task and data levels, a configuration known as the doubly sequential setup \citep{onlinereview23,xu2024online}. The situation is depicted in Figure~\ref{fig:illus}. Each task is represented by a horizontal line, symbolizing a data stream. As time progresses, a team of researchers receives a series of tasks $H^1, H^2, \cdots$ intermittently. Upon receiving a task, the researchers engage in sequential data collection while simultaneously evaluating the collected data in real time. This evaluation helps determine whether further data collection is required or if a conclusion can be made.
	
	The dynamic process of doubly sequential experiments is \emph{asynchronous} \citep{Zrnic21}, enabling data sampling to commence and conclude at arbitrary times, while allowing multiple experiments to overlap during specific study periods. This flexibility is particularly valuable in scenarios where researchers operate in a decentralized manner and are assigned diverse responsibilities for facilitating rapid iterations.

\subsection{Existing Methods}\label{subsec:existing}

The analysis of doubly sequential experiments primarily intersects with two significant lines of research: sequential testing and online false discovery rate (FDR; \citealp{BenHoc95}) analysis, which are marked by the horizontal direction (data stream $\rightarrow$) and the vertical direction (task stream $\uparrow$) in Figure~\ref{fig:illus}.

Sequential testing \citep{siegmund1985sequential} is a well-established discipline focused on making decisions based on data collected sequentially over time. This area has gained renewed attention due to the development of  always-valid inference \citep{Johetal22, russac2021b, ramdas2023game, Gruetal24, casgrain2022anytime}, which is critical for contemporary large-scale A/B testing applications. Recent advancements in error rate control within sequential experiments \citep{bartroff2020sequential, jamieson2018bandit, malek2017sequential, SMART} provide useful strategies for addressing the uncertainties inherent in sequential decisions. However, existing research  typically operates within fixed budgets for overall efforts, either in data collection or task management, and focuses on isolated tasks. This single-stream approach falls short in the context of doubly sequential experiments, where multiple data streams are collected and monitored simultaneously for an unknown number of ongoing and asynchronous tasks.

Online FDR analysis provides a principled approach for controlling error rates in real-time decision-making involving a stream of hypotheses. The concept of alpha-investing \citep{foster2008alpha} forms a foundational mechanism for controlling the FDR while allowing adaptive allocation of resources as new hypotheses arrive. This framework enables practitioners to manage their alpha-wealth efficiently, investing in hypotheses that show promising evidence. Several generalizations have been proposed to enhance both the applicability and efficiency of alpha-investing, including LORD \citep{javanmard2018online}, SAFFRON \citep{ramdas2018saffron}, ADDIS \citep{tian2019addis}, SAST \citep{gang2023structure}, e-LOND \citep{xu2024online}, and e-GAI \citep{zhang2025gai}. Recent work has also considered asynchronous settings, in which hypotheses arrive and depart at arbitrary times \citep{Zrnic21}.

Despite these developments, existing sequential or online FDR methods typically address sequentiality along a single dimension: online FDR methods handle streams of tasks with fixed data, whereas sequential testing methods focus on streams of data for a single, isolated task. In the doubly sequential setting, investigators actively determine decision times by dynamically allocating sampling efforts and error budgets across concurrent experiments. This distinct and highly interactive environment introduces fundamentally new challenges, with direct practical implications for large-scale A/B testing and adaptive clinical trials.


{
\subsection{The ``active investigator'' and dynamic decision space}

In the existing online alpha-investing literature, the investigator tacitly acts as a \emph{passive observer} of an exogenous sequence of test statistics. She has no control over how much data to collect for a given hypothesis or when to terminate monitoring of a specific task. This passive observer assumption is violated in important applications such as large-scale A/B testing and adaptive clinical experiments, where the \emph{active investigator} decides during the monitoring process whether to make a definitive decision or to continue collecting data. The interactive relationship between decision-making and data collection fundamentally distinguishes doubly sequential experiments from classical online testing. 

In classical frameworks, a non-rejection at time $t$ carries no permanent commitment: it simply records that evidence was insufficient at that moment, and---as the ARC procedures of \cite{fisher2022online} and \cite{fischer2024online} make explicit---the decision may be overturned at a later stage when additional hypotheses have been tested and the rejection threshold changes. A closer examination reveals that the non-rejection option  conflates two conceptually distinct states: (a) the investigator genuinely believes the null hypothesis is true, or (b) she lacks sufficient evidence to distinguish the null from the alternative. These two situations are indistinguishable within the passive observer framework, where she receives whatever evidence the exogenous stream delivers and records a non-rejection whenever that evidence falls short of the threshold.  

In the doubly sequential framework, the investigator is not required to make a decision while evidence remains ambiguous. We formalize this framework in Section~\ref{subsec:design}, where the active investigator performs adaptive sampling and inference with optional stopping. This generalized, bilateral decision framework provides a strictly richer set of actions, in which neither arm A nor arm B occupies a privileged ``null'' status; the investigator seeks to determine which arm is superior. Furthermore, we incorporate the \textit{Continue} and \textit{Drop} actions to actively manage uncertainty, enabling the investigator to defer a definitive decision to collect additional information or disengage from an unpromising task. 
}

\subsection{A preview of our proposal and contributions}

We introduce a class of Safe, Always-Valid Alpha-investing (SAVA) rules for selective inference in doubly sequential experiments. The SAVA framework, which is built upon existing research on always-valid p-values \citep{Johetal22, ramdas2022admissible}, confidence sequences constructed by e-processes \citep{darling1967confidence, robbins1970statistical, howard2020time, waudby2020estimating}, and alpha-investing rules in online FDR analysis \citep{foster2008alpha, aharoni2014generalized}, presents several methodological and theoretical advancements that address the limitations of existing methods.

Firstly, {SAVA can handle sequentiality occurring in both the task and data dimensions.} We adopt a generic selective inference framework that enhances the flexibility of sequential testing by incorporating directional decisions as well as an \emph{abstention} option. This abstention option substantially broadens the applicability of sequential designs, enabling practitioners to adapt their strategies to evolving conditions across multiple concurrent experiments.

Secondly, we employ safe testing strategies to develop an always-valid selection rule that integrates evidence from two directional p-values, without requiring a designated control arm or prior knowledge of a null distribution. This new rule prevents overlapping selections while ensuring anytime validity in the continuous monitoring of uncertain and fast-paced environments. {Unlike existing methods that treat decision times as exogenous, our framework allows the investigator to actively determine when to continue or stop sampling.}

Thirdly, we devise a novel class of alpha-investing rules tailored for doubly sequential experiments that effectively addresses the ``alpha-death'' issue \citep{ramdas2017online}, enabling the SAVA algorithm to operate continuously and extend to any future time point. {We employ an interactive, expanded decision space, allowing alpha-wealth to regenerate as directional selections are made in response to evolving evidence from the data streams.}

Finally, we develop finite-sample theory to establish the validity of SAVA for FSR control. Our analysis extends the classical leave-one-out technique \citep{javanmard2018online} to a ``leave-one-sequence-out'' framework. This new theory: {(a) addresses the dependencies inherent in doubly sequential experiments, where alpha-wealth allocation depends on evolving trajectories of test statistics rather than static p-values, and (b) accommodates the expanded decision space that existing passive-observer frameworks cannot handle.}

These innovations collectively provide practitioners with a reliable framework for active decision-making that adapts to evolving evidence in a timely and cost-effective manner. Through experiments on both synthetic and real datasets, we demonstrate that SAVA ensures anytime validity and outperforms existing methods in power. 

The rest of the article is structured as follows. Section \ref{sec:formulation} presents the foundational aspects of our online inference protocol for the design and analysis of doubly sequential experiments. In Section \ref{sec:methodology}, we develop the SAVA algorithm and establish its theoretical properties. The empirical performance of SAVA is investigated using real data in Sections \ref{sec:realdata}. A summary of notations, details regarding SAVA algorithms under alternative setups, along with technical proofs and supplementary numerical results, are provided in the supplementary material.

	\section{Basics of the Online Inference Protocol}\label{sec:formulation}
	
	In resource-intensive contexts such as A/B testing and drug discovery, operating under constraints of limited resources and efforts presents two fundamental challenges. First, the design must be adaptive to incoming data, allowing the research team to quickly adopt promising ideas while discarding less viable ones. Second, the control of decision errors becomes critical, given the resource-intensive and time-consuming nature of subsequent studies. The two issues are respectively addressed in Sections \ref{subsec:design} and \ref{subsec:SIframework}.

	\subsection{Adaptive sampling and inference with optional stopping}\label{subsec:design}
	
\begin{figure}
		
		\centering{
			
			\includegraphics[width=6in,height=\textheight]{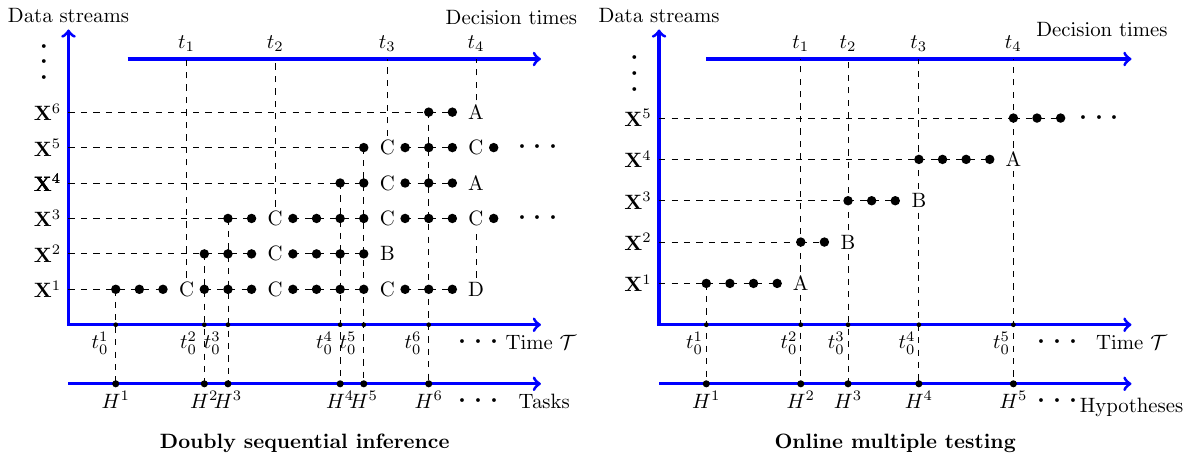}
			
		}
		
		\caption{\label{fig:comparison}\small The contrast of doubly sequential inference and online multiple testing. Black points on each line represent data collected during the active period, producing a data stream for the corresponding task. In online multiple testing, only one hypothesis is considered at each decision time. By contrast, in doubly sequential setups, it is permissible to make decisions for multiple tasks simultaneously, e.g. ${\delta}_{t_2} = (\textrm{C,C,C})$ and ${\delta}_{t_3} = (\textrm{C,B,C,C,C})$.}
		
	\end{figure}

	We present a novel framework for adaptive sampling and dynamic decision-making, specifically designed for large-scale A/B testing arising from doubly sequential experiments. This adaptive framework, visually represented in the left panel of Figure~\ref{fig:comparison}, facilitates the addition of new tasks or the removal of existing tasks in response to the evolving information.  
	
	Let $\mathcal{T}$ denote the temporal domain within which all potential tasks operate and \(\mathbb{T} = \{t_1, t_2, \ldots\}\subset \mathcal{T}\) a countable set of decision times. Let $\mathbb{J} = \{1, 2, \ldots\}$ represent the index set for all tasks. For each task $j \in \mathbb{J}$, denoted as $H^j$, data streams are collected during the time interval $\mathcal{T}^j \subset \mathcal{T}$. Let $t_0^j \coloneqq \min \mathcal{T}^j$ denote the initiation time of task $j$ and \(\mathbf X^j = (X_t^j)_{t\in\mathcal T^j}\) represent the data collected for task $j$ during $\mathcal T^j$.  Without loss of generality, initiation times $\{t_0^j\}_{j\in\mathbb J}$ of different tasks are assumed to be distinct.

	The decision for task \(j\) at time \(t \in \mathbb{T}\) is represented as \(\delta^j_t \in \mathcal{A} = \{\mathrm{A}, \mathrm{B}, \mathrm{C}, \mathrm{D}\}\), where \(\delta^j_t = \mathrm{A}\) or \(\delta^j_t = \mathrm{B}\) signifies the assertion that either arm A or arm B is superior. In contrast, values C and D, representing ``Continue'' and ``Drop'', respectively, signify states of \emph{abstention} when there is insufficient information to reach a definitive conclusion \citep{HerWeg06, Lei14, Sun2015, SMART}. Specifically, \(\delta^j_t = \mathrm{C}\) indicates that data collection for task \(j\) continues at time \(t\), allowing for reassessments as new data  arrives; \(\delta^j_t = \mathrm{D}\) indicates that we drop the arm and discontinue data collection. 
	
	We compare the doubly sequential setup (left panel of Figure~\ref{fig:comparison}) with the online multiple testing framework (right panel of Figure~\ref{fig:comparison}) to highlight two distinctive features of our approach. Firstly, the action space has been expanded to \(\delta_t^j \in \{\textrm{A, B, C, D}\}\). The left panel illustrates that, in addition to choosing either arm A (\(\delta_t^j = \textrm{A}\)) or arm B (\(\delta_t^j = \textrm{B}\)), the team also has the option to refrain from making a definitive decision. Specifically, if the team believes that one arm shows promise but requires more data for a confident decision, they can continue data collection (\(\delta_t^j = \textrm{C}\)). Alternatively, if the team perceives no practically meaningful difference between the two arms, or if the experiment duration  has exceeded a pre-specified tolerance level, they can opt to drop the task (\(\delta_t^j = \textrm{D}\)). This new mechanism offers more flexibility compared to online multiple testing, where the team only has two options (A or B). In particular, the abstention states C and D eliminate the need to rush into making a decision: the team can swiftly abandon the task at hand or gather more evidence before selecting a specific arm.

	Secondly, for temporal structure, traditional online multiple testing assumes a synchronous process, where the current experiment must conclude before the next one begins. While recent asynchronous methods \citep{Zrnic21, xu2024online} allow for task overlap, they generally decouple the sequential monitoring of tasks from global alpha-wealth management. Specifically, in these frameworks, the error budget allocated to a specific hypothesis is typically determined based on the history of past decisions and remains static throughout the task's duration. In contrast, our framework operates on a sequence of decision times, $\mathcal{T}$, at which decisions can be made simultaneously for all actively monitored tasks. By incorporating an explicit abstention option (actions C and D) and an expanded action space $\{\mathrm{A, B, C, D}\}$, our method allows for the continuous re-evaluation of active tasks. This enables the dynamic allocation of alpha-wealth: if a discovery is made in one data stream, the generated wealth can be immediately redistributed to other currently active tasks, a feature not present in existing online testing approaches.

	\subsection{False selection rate with directional errors}\label{subsec:SIframework}
	
	Suppose we are interested in making inference regarding an unknown parameter \(\mu^j\) associated with task \(j \in \mathbb{J}\), which frequently represents the contrast between two mean effects in practice. For instance, in drug discovery, \(\mu^j\) quantifies the incremental impact of a new compound in comparison to a control. In A/B testing, \(\mu^j := \mu^j_B - \mu^j_A\), where \(\mu^j_B\) and \(\mu^j_A\) denote the mean effects of two treatment arms. The true state of nature corresponding to \(\mu^j\) is denoted as \(\theta^j\). Depending on the specific context, \(\theta^j\) may be defined in various ways. This paper focuses on a generic selective inference setup tailored for A/B testing; the classical hypothesis testing setup is briefly addressed in Section \ref{subsec:classical-setup} of the Supplement.
	
	In doubly sequential experiments, the agent is tasked with selecting better arms across multiple experiments over time. Denote \(\theta^j\in\{\mathrm{A}, \mathrm{B}\}\) the specific arm that demonstrates superior performance:
	\vskip-1cm
	\begin{equation}\label{SI}
		\theta^j = \left\{
		\begin{array}{ll}
			\text{A}, & \text{if } \mu^j = \mu^j_B - \mu^j_A < 0 \\
			\text{B}, & \text{if } \mu^j = \mu^j_B - \mu^j_A > 0
		\end{array}
		\right..
	\end{equation}
	\vskip-0.5cm
	Consider a family of decision rules \(\pmb \delta = \{\pmb \delta_t\}_{t \in \mathbb{T}}\), 
	where \(\pmb \delta_t = \{\delta_t^j: j \in \mathbb{J}, t_0^j\leq t\}\) 
	represents the concurrent decisions for experiments that have commenced prior to time \(t\), with \(\delta_{t}^j \in \{\mathrm{A}, \mathrm{B}, \mathrm{C}, \mathrm{D}\}\) 
	denoting the decision for task \(j\) at time \(t \in \mathbb{T}\). We require that once an arm is selected or a task is dropped at time \(t\), data collection is halted and $\delta^j_t$ remains unchanged for the rest of the study period:
	\vskip-1cm
	\begin{equation*}
		\text{If } \delta^j_t \in \{\mathrm{A}, \mathrm{B}, \mathrm{D}\}, \text{ then } \delta^j_s = \delta^j_t \text{ for all } s > t, \, s \in \mathbb{T}.
	\end{equation*}
	\vskip-0.5cm
In contrast, when \(\delta^j_t = \mathrm{C}\), data collection continues, allowing for \(\delta^j_t\) to be revised based on newly acquired information.
	
	Our selective inference framework interprets abstentions ($\delta_{t}^j =\mbox{C or D}$) as missed opportunities rather than decision errors, given that they do not incur significant costs for follow-up studies. Hence, when assessing decision errors at \(t \in \mathbb{T}\), we focus exclusively on the set of selected candidates up to that time:  
	\vskip-1cm
	\begin{equation}\label{eq:St}
		\mathcal{S}_t = \{j \in \mathbb{J} : t_0^j \leq t, \delta_{t}^j = \mathrm{A} \text{ or } \mathrm{B}\}.
	\end{equation}
	\vskip-0.5cm
	To aggregate the decision errors in two possible directions: (i) \(\delta_t^j = \mathrm{A}\) while \(\theta^j = \mathrm{B}\), and (ii) \(\delta_t^j = \mathrm{B}\) while \(\theta^j = \mathrm{A}\), we define the false selection rate (FSR) as the expected proportion of the false selection proportion (FSP):  
	$\mathrm{FSR}_t = \mathbb{E}\{\mathrm{FSP}_t(\pmb{\delta}_t)\}$, where
	\vskip-0.5cm
	\begin{equation}\label{eq:FSP}
		\mathrm{FSP}_t(\pmb{\delta}_t) = \dfrac{\sum_{j:t_0^j \leq t} \mathbb{I}\{\delta^j_t \neq \theta^j, j \in \mathcal{S}_t\}}{|\mathcal{S}_t| \vee 1}, \quad t\in\mathbb T,
	\end{equation}
	\vskip-0.5cm
	\(\mathbb{E}\) is taken over the data set \(\{X_s^j : s \leq t, t_0^j \leq t, s \in \mathcal{T}^j\}\), \(|\mathcal{S}_t|\) denotes the cardinality of the set \(\mathcal{S}_t\), and \(x \vee y=\max\{x,y\}\). 
	\begin{remark}\rm{
		While the definition of FSR aligns with the directional FDR \citep{benjamini1993, Benjamini2005}, we underscore that our framework presents distinct perspectives compared to existing methodologies. Traditional approaches typically involve a two-stage process: first, individuals are selected using established FDR procedures, and second, adjustments are made to the signs of the selected units. In contrast, our framework begins with directional selections and subsequently collects additional information on the unselected units to enhance confidence in a sequential manner.}
	\end{remark}
	
	\begin{remark}\rm{
		A closely related error metric is the marginal FSR:  
		\begin{equation}\label{eq:mFSR}
			\mathrm{mFSR}_t=\dfrac{\mathbb{E}\left[\sum_{j:t^j_0\leq t} \mathbb{I}\{\delta^j_t\neq\theta^j, j\in\mathcal S_t\}\right]}{\mathbb{E}(|\mathcal S_t| \vee 1)}, \quad t\in\mathbb T.
		\end{equation}
		The FSR and mFSR are asymptotically equivalent (cf. \citealp{basu2018, CARS}) under independence or weak dependence but may differ substantially under strong dependence (cf. \citealp{Caoetal13, javanmard2018online}).}
	\end{remark}
	
	Our objective is to develop a class of real-time decision rules \(\pmb \delta = \{\pmb \delta_t\}_{t \in \mathbb{T}}\) that controls both \(\mathrm{FSR}_t\) and \(\mathrm{mFSR}_t\) below the nominal level $\alpha$ at all \(t \in \mathbb{T}\). In an online setting aimed at identifying promising candidates from a potentially extensive pool, our protocol for controlling the FSR and mFSR at all times ensures that resources are directed toward the most viable options. This approach closely aligns with the objectives in various practical contexts, where prudent resource allocation is essential for avoiding excessive spending and budget depletion.
	To compare the efficiency of different selection rules, we define the true selection rate (TSR), which is the expectation of the true selection proportion:
	\vskip-1cm
	\begin{equation*}
		\mathrm{TSP}_t(\pmb{\delta}_t) = 
		\dfrac{\sum_{j:t_0^j \leq t} \mathbb{I}\{\delta^j_t = \theta^j\}}{|\{j : t_0^j \leq t\}| \vee 1}.
	\end{equation*}

{
\begin{remark}\rm{
The current formulation treats the Drop action (Arm D) as a missed opportunity rather than a formal decision error. This design reflects a fundamental constraint in large-scale testing. In high-dimensional sparse inference, simultaneous tight control of the false discovery rate and the missed discovery rate is generally unattainable below the classification boundary \citep{CaiSun17}. This limitation similarly carries to doubly sequential experiments with finite per-task sampling budgets. The situation becomes even more involved in our doubly sequential framework. Beyond false discoveries and missed discoveries, we face an additional layer of complexity: a finite sampling budget per task, under which it is statistically impossible to simultaneously control both false positives and missed discoveries for all tasks. Consequently, we cannot formally treat Arm D as a decision error and control the drop error rate in the same manner as that of Arms A or B.  This underscores the precise role of Arm D in our framework: it serves as a safety valve that prevents the investigator from committing to an unreliable directional decision.

}

\end{remark}
}
	
	\section{The SAVA Algorithm}\label{sec:methodology}

	We propose a class of online inference procedures for doubly sequential experiments. Section \ref{subsec:avp} introduces the concept of always valid directional p-values. In Section \ref{subsec:int-rule}, we present an integrative rule for selective inference based on two directional p-values. Sections \ref{subsec:FSR-est} and \ref{subsec:test-level} discuss how to estimate the FSR and allocate task-specific test levels over time.  Finally, in Section \ref{subsec:pro-alg}, we introduce the SAVA algorithm and establish its anytime validity for online FSR control.

	\subsection{Always valid directional p-values}\label{subsec:avp}
	
	When comparing the effectiveness of two designs, a pre-specified null hypothesis or default arm may not be available. To address this issue, we perform simultaneous testing of two null hypotheses, \(H_{0,A}^j\) and \(H_{0,B}^j\), which correspond to selecting arm A and arm B as the default arm, respectively. Correspondingly, each task is associated with two directional p-values at a decision time. 
	
	The doubly sequential design involves the continuous monitoring of multiple data streams. The common practice, known as ``data peeking'', can significantly undermine the statistical validity of these sequential tests. To address these challenges, we build upon the work of \cite{Johetal22} and introduce the concept of \emph{always-valid directional p-values}, which remain valid for any data-adaptive stopping time $T$. This approach ensures \emph{safe testing} \citep{Gruetal24}, allowing practitioners to conduct sequential tests that are informed by prior decisions and newly acquired data without compromising statistical validity. 
	
	Let \((\Omega, \mathcal F)\) be a measurable space, and define the sigma-field \(\mathcal F^j_t = \sigma(X_s^j \colon s \leq t, s \in \mathcal T^j)\) for task \(j\), which encapsulates all information up to decision time $t$ related to the data stream \(\mathbf X^j\). For each \(j \in \mathbb J\), consider a (possibly infinite) stopping time \(T\) with respect to filtration \(\{\mathcal F^j_t\}_{t \in \mathbb T}\) on \((\Omega, \mathcal F)\). 
	
	\begin{definition}\label{def:avp}
		For each \(j \in \mathbb{J}\), \((p_t^{j,A}, p_t^{j,B})_{t \in \mathbb{T}}\) are always-valid directional p-values if for  any \(\alpha \in [0, 1]\) and any stopping time $T$, we have 
\vskip-0.8cm

		\begin{equation}\label{eq:avpdef}
			\mathrm{Pr}_{\theta^j = \mathrm{B}}(p_T^{j,A} \leq \alpha)\leq \alpha	\ \  \text{and} \ \  \mathrm{Pr}_{\theta^j = \mathrm{A}}(p_T^{j,B} \leq \alpha)\leq \alpha.
		\end{equation} 	
	\end{definition}
\vskip-0.5cm
	
	A systemic way of constructing always-valid directional p-values is via e-processes. 
	
	\begin{definition}\label{def:eprocess}
	A non-negative process $\{E^{j}_t\}_{t \in \mathbb{T}}$ 
	adapted to $\{\mathcal{F}^j_{t}\}$ is an 
	e-process with respect to $\theta^j=A$ if
		\begin{equation}\label{eq:eprocess-def}
			\sup_{\tau \in \mathcal{T}} \mathbb{E}_{\theta^j=A}[E^{j}_\tau] \leq 1,
		\end{equation}
		where $\mathcal{T}$ denotes the set of all stopping times with respect to $\{\mathcal{F}^{j}_t\}$.
	\end{definition}
	
	The connection between e-processes and always-valid p-values is given next.
	
	\begin{proposition}\label{prop:eprocess-to-pval}
		Suppose $\{E^{j,A}_t\}_{t \in \mathbb{T}}$ is an e-process with respect to $\theta^j = B$, and $\{E^{j,B}_t\}_{t \in \mathbb{T}}$ is an e-process respect to $\theta^j = A$. Define
		\vskip-1cm
		\begin{equation}\label{eq:pval-from-eprocess}
			p^{j,A}_t = \min\left\{1, \left(\max_{s \leq t, s \in \mathbb{T}} E^{j,A}_s\right)^{-1}\right\}, \quad
			p^{j,B}_t = \min\left\{1, \left(\max_{s \leq t, s \in \mathbb{T}} E^{j,B}_s\right)^{-1}\right\}.
		\end{equation}
	\vskip-0.5cm
Then $(p^{j,A}_t, p^{j,B}_t)_{t \in \mathbb{T}}$ are always-valid directional p-values.
	\end{proposition}

{	
Always-valid directional p-values can be readily derived from valid e-processes. The construction of e-processes is an active and rapidly developing research area, offering a rich set of tools that extend well beyond traditional parametric assumptions; in fact, many recent constructions are entirely model-free. The alpha-investing layer of our SAVA framework is completely modular and agnostic to the specific construction of these p-values, allowing deployment with diverse methods and model classes:

\begin{itemize}
    \item For data with bounded support, betting-based methods \citep{waudby2020estimating,Cle25} yield valid e-process constructions that are entirely model-free and require no parametric distributional assumptions.
    \item For one-parameter exponential families—including common data types such as Bernoulli (e.g., conversion rates in A/B testing), Poisson, and exponential distributions—mixture-martingale constructions \citep{Johetal22} can be applied.
    \item For a broad class of light-tailed distributions, martingale-based methods provide valid e-processes under mild assumptions such as sub-Gaussianity \citep{howard2021time}.
    \item The framework of universal e-processes provides a highly versatile construction with minimal distributional requirements \citep{ramdas2025hypothesis}.
\end{itemize}

Consequently, any new methodological or theoretical advances in e-process constructions can be directly leveraged to develop valid directional p-values, further broadening the applicability of the SAVA framework. We provide detailed examples of always-valid directional p-values in Section \ref{subsec:avp-cons} and Section \ref{sec:simulation} of the Supplement. 

We assume throughout this paper that \emph{always-valid directional p-values are non-increasing}. This is not a restrictive assumption. As justified by the following proposition, any valid sequence of directional p-values can be trivially transformed into a non-increasing sequence by taking the running minimum; this transformation is a costless algorithmic operation that preserves anytime validity without sacrificing statistical power.}

	\begin{proposition}\label{prop:avpconstruct}
		Suppose that $(\rho_t^{A}, \rho_t^{B})_{t\in\mathbb{T}}$  are always-valid directional p-values. For all $t\in\mathbb T$, define $p_t^{A} =\min\{\rho_r^{A}: r \le t, r\in \mathbb T\}$ and $p_t^{B} =\min\{\rho_r^{B}: r \le t, r \in \mathbb T\}$. Then $(p_t^{A}, p_t^{B})_{t\in\mathbb{T}}$  are also always-valid directional p-values.
	\end{proposition}

	\subsection{An integrative selection rule}\label{subsec:int-rule}
	
	Assuming we have calculated the directional p-values \((p^{j,A}_t, p^{j,B}_t)\) and allocated the corresponding test levels \((\alpha_t^{j,A}, \alpha_t^{j,B})\) for task \(j\) at time \(t\), the naive approach of directly comparing each directional p-value to its respective test level may initially appear straightforward. However, this method introduces significant complications in practice. A notable concern is the potential for conflicting decisions arising from the two directional tests. For example, it is possible that both p-values are significant; such situations can create ambiguity, making it challenging to draw meaningful conclusions.
	
\begin{figure}
		
		\centering{
			
			\includegraphics[width=5in,height=\textheight]{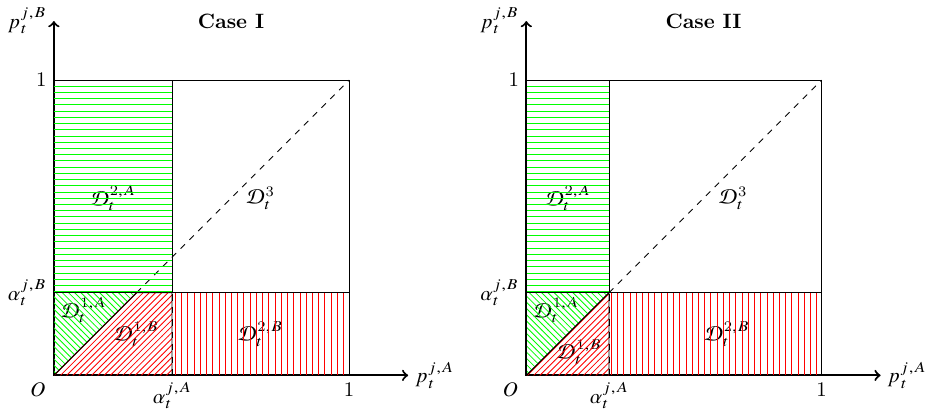}
		}
		\caption{\label{fig:ruleonline}\small An illustration of the partitioning of the p-value plane: the left panel represents the general case where \(\alpha_t^{j,A} \neq \alpha_t^{j,B}\), while the right panel shows the scenario where \(\alpha_t^{j,A} = \alpha_t^{j,B}\).}
	\end{figure}

	To mitigate these challenges, it is crucial to employ integrative approaches that consider both directional tests simultaneously, allowing for a coherent decision-making framework. The core idea of our proposal is to partition the p-value space \([0,1] \times [0,1]\) into the following non-overlapping regions:
	\begin{itemize}
		\tightlist
		\item \(\mathcal D^{1,A}_t\): where \(p^{j,A}_t \leq \alpha^{j,A}_t\), \(p^{j,B}_t \leq \alpha^{j,B}_t\), and \(p^{j,B}_t \geq p^{j,A}_t\);
		\item \(\mathcal D^{1,B}_t\): where \(p^{j,A}_t \leq \alpha^{j,A}_t\), \(p^{j,B}_t \leq \alpha^{j,B}_t\), and \(p^{j,B}_t < p^{j,A}_t\);
		\item \(\mathcal D^{2,A}_t\): where \(p^{j,A}_t \leq \alpha^{j,A}_t\) and \(p^{j,B}_t > \alpha^{j,B}_t\);
		\item \(\mathcal D^{2,B}_t\): where \(p^{j,B}_t \leq \alpha^{j,B}_t\) and \(p^{j,A}_t > \alpha^{j,A}_t\);
		\item \(\mathcal D^3_t\): where \(p^{j,A}_t > \alpha^{j,A}_t\) and \(p^{j,B}_t > \alpha^{j,B}_t\).
	\end{itemize}
	These regions are illustrated in Figure~\ref{fig:ruleonline}, with each area represented by distinct line styles and colors: \(\mathcal D^{1,A}_t\) and \(\mathcal D^{1,B}_t\) correspond to regions where both directional p-values are significant; \(\mathcal D^{2,A}_t\) and \(\mathcal D^{2,B}_t\) indicate areas where exactly one directional p-value is significant; and \(\mathcal D^3_t\) reflects the scenario in which p-values from neither direction are significant.

	In light of this partitioning, we propose the following decision rule:
	\vskip-1cm
	\begin{equation}\label{eq:decisionrule}
		\begin{aligned}
			\quad & \delta^j_t(p^{j,A}_t, p^{j,B}_t, b_j, t_0^j)\\
			= & \left\{
			\begin{array}{ll}
				\mathrm A, & \quad\text{ if } (p^{j,A}_t, p^{j,B}_t) \in \mathcal D^{1,A}_t \cup \mathcal D^{2,A}_t, \\
				\mathrm B,  & \quad \text{ if } (p^{j,A}_t, p^{j,B}_t) \in \mathcal D^{1,B}_t \cup \mathcal D^{2,B}_t,\\
				\mathrm C, &\quad\text{ if }(p^{j,A}_t, p^{j,B}_t)\in \mathcal D^3_t, \text{ and } t-t_0^j < b_j, \\
				\mathrm D, &\quad\text{ if }(p^{j,A}_t, p^{j,B}_t)\in \mathcal D^3_t, \text{ and } t-t_0^j \geq b_j.
			\end{array}
			\right.
		\end{aligned}
	\end{equation}

	The intuitions underlying \eqref{eq:decisionrule} are as follows: First, when both directional p-values are significant, the decision is made in favor of arm A (or B) if the corresponding pair of p-values falls within  \(\mathcal{D}_t^{1,A}\) (or \(\mathcal{D}_t^{1,B}\)). Second, if only one of the directional p-values is significant (\(\mathcal{D}_t^{2,A}\) or \(\mathcal{D}_t^{2,B}\)), the decision aligns with the arm associated with that particular direction. Finally, if neither directional p-value is significant, indicating that we are in region \(\mathcal{D}_{t}^3\), the decision hinges on whether \(t - t_0^j\) exceeds a pre-specified tolerance level \(b_j\), which signifies the maximum duration deemed acceptable for the experiment related to task \(j\). If \(t - t_0^j < b_j\), we continue data collection (C); otherwise, we drop the task (D). 
	
	The decision rule \eqref{eq:decisionrule}, which serves as the foundation of our later methodological developments, effectively addresses the issue of conflicting selections and provides coherent decisions for all possible configurations of directional p-values and test levels.  
	
	\subsection{Estimating the FSR in doubly sequential setups}\label{subsec:FSR-est}
	
	The following two subsections focus on the development of alpha-investing strategies \citep{foster2008alpha}. A critical component of our method is a novel formulation of the FSR estimate at time \(t \in \mathbb{T}\):
	\begin{equation}
		\label{eq:fsrhat}
		\widehat{\mathrm{FSR}}_t = \frac{\sum_{j\in\mathbb J: t_0^j \leq t} \bar{\alpha}_t^{j,A} \vee \bar{\alpha}_t^{j,B}}{|\mathcal{S}_t| \vee 1},
	\end{equation}
	where \(\mathcal{S}_t\) represents the index set of selected arms as defined in equation \eqref{eq:St}, and \(\bar{\alpha}_t^{j,A} = \max\{\alpha_s^{j,A}: s \leq t, s \in \mathbb{T}\}\) and \(\bar{\alpha}_t^{j,B} = \max\{\alpha_s^{j,B}: s \leq t, s \in \mathbb{T}\}\) serve as upper bounds for the alpha-wealth allocated to the respective directions when performing task \(j\) during the period leading up to time \(t\). 
	
	The FSR estimate \eqref{eq:fsrhat} has been carefully calibrated to address the complexities inherent in the doubly sequential setup (cf. left panel of Figure \ref{fig:comparison}). It extends and improves upon the FDR estimate provided by \cite{ramdas2017online} and \cite{javanmard2018online} for online multiple testing (cf. right panel of Figure \ref{fig:comparison}):
\vskip-0.8cm
	\begin{equation}
		\label{eq:ofdr}
		\widehat{\mathrm{FDR}}_t = \frac{\sum_{j \in \mathbb{J}: t_0^j \leq t} \alpha^j}{|\mathcal{R}_{t}| \vee 1}, \quad t\in\mathbb{T},
	\end{equation}
	where \(\alpha^j\) denotes the test level (or alpha-wealth) allocated for task \(j\) and \(\mathcal{R}_t\) represents the set of rejected null hypotheses up to time \(t\).

	To elucidate the differences between \eqref{eq:fsrhat} and \eqref{eq:ofdr}, we introduce the notion of active set at $t_k$, denoted as \(\mathcal{A}_{t_k}\), which includes the tasks for which new data have been collected between the current decision time and the previous decision time. Formally, let
\vskip-0.8cm
	\begin{equation}
		\label{eq:active-set}
		\mathcal{A}_{t_k} \coloneqq \{ j \in \mathbb{J} : \delta_{t_{k-1}}^j = \mathrm{C} \text{ or } t_{k-1} < t_0^j \leq t_k \}, \quad k \geq 2,
	\end{equation}
\vskip-0.5cm
with the initialization given by \(\mathcal{A}_{t_1} = \{ j \in \mathbb{J} : t_0^j \leq t_1 \}\). 
	Using this concept, our FSR estimate \eqref{eq:fsrhat} can be decomposed into two components: one accounting for alpha-wealth associated with completed tasks ($j\notin \mathcal{A}_{t}$) and the other for active tasks ($j\in\mathcal{A}_{t}$):
\vskip-0.8cm
	\begin{equation}
		\label{eq:fsrhat-twoparts}
		\widehat{\mathrm{FSR}}_t = \frac{\sum_{j:t_0^j\leq t, j\notin \mathcal{A}_t} \bar{\alpha}_t^{j,A} \vee \bar{\alpha}_t^{j,B}}{|\mathcal{S}_t| \vee 1} 
		+ \frac{\sum_{j:t_0^j\leq t, j\in \mathcal{A}_t} \bar{\alpha}_t^{j,A} \vee \bar{\alpha}_t^{j,B}}{|\mathcal{S}_t| \vee 1}, \quad t\in\mathbb T.
	\end{equation}
\vskip-0.5cm

We highlight several novel aspects of \eqref{eq:fsrhat-twoparts} compared to \eqref{eq:ofdr}.
	
	First, equation \eqref{eq:ofdr} considers only completed tasks up to $t$, corresponding to the first term in \eqref{eq:fsrhat-twoparts}. In online multiple testing, the experiments are conducted synchronously, meaning that a new experiment cannot commence until the current one concludes. Consequently, there is no active set of tasks, and the second term in \eqref{eq:fsrhat-twoparts} does not apply in this context. In contrast, our design necessitates the allocation of alpha wealth for tasks in the active set, which is represented by the second term in \eqref{eq:fsrhat-twoparts}. This term effectively captures the asynchronous structure inherent in doubly sequential designs. 
	
	Second, the FDR estimate \eqref{eq:ofdr} associates only one decision time with each task. In contrast, the FSR formulation allows each task to be linked to multiple decision times, reflecting the continuous evaluation of evidence as new data unfolds. Each task may thus be associated with multiple decision points with varied test levels, which are data-adaptive and aggregated using the maximum values \((\bar{\alpha}^{j,A}_t, \bar{\alpha}^{j,B}_t)\). This approach of taking the maximum facilitates continuous monitoring of the same data stream while ensuring that the alpha wealth allocated to a specific task is counted only once. 
	Furthermore, our FSR estimate \eqref{eq:fsrhat-twoparts} demonstrates that the agent can manage multiple tasks simultaneously at a single decision point. The test levels are allocated not only to tasks that have just concluded at time \(t\), but also to those that remain active at that time. 
	
	Finally, the test levels for the two opposing directions (A and B) associated with each task are combined using the maximum operator \(\vee\). Our FSR estimate effectively monitors the error rates in selective inference from both directions while preventing double counting of errors from opposing directions for the same task. This approach offers greater flexibility compared to online FDR analysis, which necessitates the specification of a default arm and considers only one direction when estimating the error rate.

	\subsection{Alpha-investing rules}\label{subsec:test-level}
	
	The FSR estimate \eqref{eq:fsrhat} offers valuable insights for designing alpha-investing rules for doubly sequential experiments, which require the allocation of test levels across multiple asynchronous tasks over time.
	
	Let \( T_{\mathrm{stop}}^j = \min\{t \in \mathbb{T} : \delta_t^j = \mathrm{A, B}, \text{ or } \mathrm{D}\} \) denote the stopping time at which task \( j \) is either selected or dropped. Additionally, let \( \mathcal{G}_t^j \) represent the $\sigma$-field generated by the observed samples \( \{X_s^j : s \leq (t \wedge T_{\mathrm{stop}}^j), s \in \mathcal{T}^j\} \) for task \( j \in \mathbb{J} \) at time \( t \in \mathbb{T} \). We first state two conditions for calibrating test levels, which serve as guiding principles for the design of alpha-investing rules.

	\begin{condition}\label{assump1}
		The set of test levels $\{(\alpha_t^{j,A},\alpha_t^{j,B}):j\in\mathbb J, t\in\mathbb T\}$ satisfies $\widehat{\mathrm{FSR}}_t\leq \alpha$ for all $j\in\mathbb J$ and $t\in\mathbb T$, where $\widehat{\mathrm{FSR}}_t$ is defined in \eqref{eq:fsrhat}.
	\end{condition}
	
	\begin{condition}\label{assump2}
		The test levels $\alpha_t^{j,A}$ and $\alpha_t^{j,B}$ are $\mathcal G_t^{1:(j - 1)}$-measurable, where $\mathcal G_t^{1:k}$ is the $\sigma$-field generated by $\{\mathcal G_t^{i}\}_{i=1}^{k}$.
	\end{condition}

	\begin{remark}\rm{
		The constraint \(\widehat{\mathrm{FSR}}_t \leq \alpha\) in Condition \ref{assump1} guarantees the anytime validity of online inference, provided that the FSR estimate is uniformly conservative for all \(j \in \mathbb{J}\) and \(t \in \mathbb{T}\). 
		The constraint imposed by the conservative FSR estimate is crucial. 
		Condition \ref{assump2} indicates that test levels allocated for task \( j \) should rely solely on information derived from data streams that have commenced prior to \( t_0^j \). This condition, which deliberately excludes the information from active streams that begin after \( t_0^j \), appears to be indispensable. In Sections \ref{sec:counterexample1} and \ref{sec:counterexample2}  of the Supplement, we provide counterexamples illustrating why the removal the conditions results in inflated FSR levels.}
	\end{remark}

	According to the constraint \(\widehat{\mathrm{FSR}}_t \leq \alpha\), each additional selection increases the denominator of \eqref{eq:fsrhat} by 1, thereby allowing an increase of \(\alpha\) in the numerator. Our basic strategy is that the test levels for task \( j \) are jointly determined by its arrival time \( (t_0^j) \) and the number of selections in its neighborhood with bandwidth \( k \in \mathbb{N} \). Initially, the test levels for tasks \( j \in \{1, \dots, k\} \) are assigned a value of \(\alpha/k\). Moving forward, the alpha-wealth acquired from each selection is equally distributed to the next \(k\) tasks, i.e., the selection of task $j$ increases the test levels for tasks $j+1,\dots,j+k$ by \(\alpha/k\). While \(k\) can be selected as any fixed integer, different values of \(k\) may impact power performance; this issue is investigated in Section \ref{subsec:bandwidth} of the online Supplement.
	
	Let $\mathcal{S}_{t,j-} = \{i\in\mathbb J: t_0^i<t_0^j, \delta_t^i = \mathrm{A}\text{ or }\mathrm{B}\}$ represent the index set of selected tasks arriving before task $j$ up to time $t$, and \(I_n(\mathcal{S}_{t,j-})\) denote the \(n\)-th smallest index in \(\mathcal{S}_{t,j-}\). The number of selected tasks in the (one-sided) neighborhood of $j$ at $t\in\mathbb T$ is given by: 
\vskip-1cm

	\begin{equation}\label{neighbor}
		N_{t,j-}(k) := |\{n\in\mathbb N: n\geq 2, j-k\leq I_n(\mathcal S_{t,j-})\leq j-1\}|.
	\end{equation} 
	
\vskip-0.5cm
	
	The test levels are given by:
\vskip-1cm

	\begin{equation}\label{eq:test-level}
		\alpha_t^{j,A} = \alpha_t^{j,B} = \frac{\alpha}{k}\left[\mathbb{I}\{j\leq k\} + N_{t,j-}(k)\right].
	\end{equation}
	
\vskip-1cm

	\begin{remark}\rm{
		We have deliberately excluded $I_1(\mathcal{S}_{t,j-})$ when calculating $N_{t,j-}(k)$. If included, the estimate \eqref{eq:fsrhat} could inflate beyond $\alpha$ at certain decision times, potentially violating Condition \ref{assump1}. Moreover, in our construction, we have set \(\alpha_t^{j,A} = \alpha_t^{j,B}\) for task \(j \in \mathcal{A}_t\). However, when separate constraints for arm-specific FSRs are imposed, it may be more appropriate to assign different values to \((\alpha_t^{j,A},\alpha_t^{j,B})\). Further details pertaining to this scenario can be found in Section \ref{subsec:dis-SAVA}. }
	\end{remark}
	
	Our alpha-investing strategy is inspired by the LORD method \citep{javanmard2018online}, but it features two key distinctions from the original LORD.
	First, our approach allocates alpha-wealth evenly across \(k\) tasks, rather than employing an infinite series converging to zero as done in \cite{javanmard2018online}. This significantly enhances the power of online inference. This benefit can be attributed to our innovative inference protocol, which incorporates an abstention option that facilitates continuous evidence collection, eventually leading to the rejection of p-values. Specifically, if a task is allowed to remain ongoing until any future time point, if properly constructed, one of the directional p-values must converge to 0, provided that $\mu_A^j\neq\mu_B^j$ \citep{robbins1970statistical, howard2021time, ramdas2021testing, Johetal22}. A definitive selection (A or B) will eventually be made for every task, and new ``alpha-wealth'' will be generated with that selection, thereby preventing the ``alpha-death'' issue commonly encountered in online multiple testing. 
	
	Second, our test levels are specifically calibrated to leverage the unique features of a doubly sequential design, where multiple tasks are handled simultaneously at each decision time. In contrast to traditional online FDR rules \citep{javanmard2018online, ramdas2017online, ramdas2018saffron, tian2019addis}, which rely exclusively on historical rejection data, our strategy evaluates all active tasks that began prior to \(t_0^j\). Specifically, we leverage concurrent selections at the decision time to enhance the available alpha-wealth, effectively increasing our test levels and, consequently, resulting in a greater number of selections.
	
	\subsection{The SAVA algorithm and its theoretical properties}\label{subsec:pro-alg}
	
	This section presents a general class of SAVA rules (summarized in Algorithm \ref{alg:SAVA} below). Let \(r_t(1) = \min \{j \in \mathbb{J} : j \in \mathcal{A}_t\}\) represent the smallest index in \(\mathcal{A}_t\). Further define \(r_t(k) = \min \{j \in \mathcal{A}_t : j > r_t(k-1)\}\) for \(2 \leq k \leq |\mathcal{A}_t|\).  
	At each decision time \(t_i \in \mathbb{T}\), we first calculate the always-valid directional p-values within the active set; some examples are provided  in Section \ref{subsec:avp-cons}. We then loop through the tasks from \(r_{t_i}(1)\) to \(r_{t_i}(|\mathcal{A}_{t_i}|)\) to: 
	\begin{description} 
		\item (a) determine the test levels adaptively as specified by \eqref{eq:test-level};
		\item (b) make decisions based on the selection rule given in \eqref{eq:decisionrule}.
	\end{description}
	Finally, we output the decisions for the active tasks in \(\mathcal{A}_{t_i}\): sampling is halted for tasks with decisions A, B, or D, while it continues for the remaining tasks. This process is continuously executed through all decision times.

	We first state an additional condition that has been widely adopted in online FDR rules \citep{javanmard2018online,ramdas2017online}. 
	
	\begin{condition}\label{assump3}
		The test levels $\alpha_t^{j,A}$ and $\alpha_t^{j,B}$ are non-decreasing functions of $(\mathbb I\{\delta_t^i = \mathrm{A} \text{ or }\mathrm{B}\})_{i=1}^{j-1}$ for all $t\in \mathbb T$.
	\end{condition}

{	
	\begin{remark}\rm{
		The monotonicity required by Condition 3 is both intuitively appropriate and technically necessary for doubly sequential inference. Intuitively, it reflects the core ``alpha-investing'' philosophy: a new discovery increases the FSR denominator, generating alpha-wealth that should strictly empower, rather than penalize, ongoing and future tasks. Technically, this monotonicity is essential for the ``leave-one-sequence-out'' technique used to prove Theorem \ref{thm:fsr}. It guarantees that forcing a counterfactual selection for task $j$
		will never decrease the test levels of other tasks. This predictable, one-directional bounding relationship is crucial to decouple complex asynchronous time-dependencies and rigorously control the overall FSR.}
	\end{remark}
}
	
	The next proposition shows that the test levels in SAVA fulfill Conditions \ref{assump1}-\ref{assump3}. 
	
	\begin{proposition}
		\label{prop:sava-lors}
		The proposed test levels in \eqref{eq:test-level} satisfy Conditions \ref{assump1} -- \ref{assump3}. 
	\end{proposition}

\singlespacing
	
\begin{algorithm}[htbp]
		\caption{The SAVA algorithm}\label{alg:SAVA}
		\vskip6pt
		\begin{algorithmic}
			\State \textbf{Input}: a grid of decision times $\mathbb T$, a target FSR level $\alpha$, a method for constructing always-valid directional p-values, $k$, and pre-specified tolerance durations $\{b_j\}_{j\in\mathbb J}$. 
			\State \textbf{For each} $t_i \in \mathbb{T}$, \textbf{do:}
			\State \quad \textbf{Step 1}: Update the index set of active tasks $\mathcal A_{t_{i}}$ defined in \eqref{eq:active-set}.
			\State \quad \textbf{Step 2}: Calculate always-valid directional p-values for tasks in $\mathcal A_{t_i}$ according to the given method.
			\State \quad \textbf{Step 3}: \textbf{For} $j=r_{t_{i}}(1),\dots, r_{t_{i}}(|\mathcal A_{t_{i}}|)$, \textbf{do}:
			
			\State \quad \quad \textbf{Step 3.1}: Calculate test levels by \eqref{eq:test-level}.
			\State \quad \quad \textbf{Step 3.2}: Obtain the decision $\delta_{t_i}^j$ according to decision rule \eqref{eq:decisionrule}.
			\State \quad \textbf{Step 4}: For tasks $j$ such that $\delta_{t_{i-1}}^j \neq \text{C}$, update $\delta_{t_i}^j = \delta_{t_{i-1}}^j$.
			\State \quad \textbf{Output}: decision-making states $\{\delta_{t_i}^{j}: j\in \mathcal A_{t_i}\}$. 
			\State \textbf{End for}
		\end{algorithmic}
	\end{algorithm}

\spacingset{1.9} 

	\begin{remark}\rm{
		In the proof of Proposition \ref{prop:sava-lors}, we  show that the test levels satisfying Conditions \ref{assump2}--\ref{assump3} are non-increasing in $t$. Consequently, the FSR estimate in Condition \ref{assump1} can be simplified as:
		$\widehat{\mathrm{FSR}}_t = \dfrac{\sum_{j: t_0^j \leq t} {\alpha}_t^{j,A} \vee {\alpha}_t^{j,B}}{|\mathcal{S}_t| \vee 1},
		$
		where only the test levels at the current decision time $t$ need to be considered. This simplification facilitates the concrete construction of test levels when applying Condition \ref{assump1}, although alternative approaches based on \eqref{eq:fsrhat} is possible.}
	\end{remark}

	The next theorem establishes the anytime validity for online selective inference.
	
	\begin{theorem}\label{thm:fsr}
		Assume that the samples from different data streams are independent.
		
		(a) If test levels $\alpha_t^{j,A}$ and $\alpha_t^{j,B}$ satisfy Conditions \ref{assump1}-\ref{assump3}, then $\mathrm{FSR}_t\leq \alpha$ for all $t\in \mathbb T$. 
		
		(b) If only Conditions \ref{assump1}-\ref{assump2} hold, then $\mathrm{mFSR}_t\leq \alpha$ for all $t \in \mathbb T$.  
	\end{theorem}

The proof for Theorem \ref{thm:fsr} presents substantial technical challenges that are not present in online FDR analysis. Specifically, we must navigate the complexities inherent in the doubly sequential structure, which includes (a) multiple decision times associated with a specific task and (b) multiple tasks being performed simultaneously at a single decision time. These complexities require a more careful analysis to upper-bound the true error rate using the novel FSR estimate \eqref{eq:fsrhat}.

	\section{Numerical Experiments}\label{sec:realdata}
	
		This section compares SAVA with online testing rules using Amazon Review Data \citep{Nietal2019}, which recorded items and real-time reviews released in 2014. The dataset contains multiple item categories, and we select the \textit{Amazon Fashion}, \textit{All Beauty}, and \textit{Luxury Beauty} categories to evaluate the performance of all methods. Within each category, items receive sequential customer reviews and ratings. Our objective is to identify items with consistently high ratings and those with consistently low ratings. The ratings follow a scale from 1 to 5, where higher scores indicate more attractive and higher-quality items. 
	This analysis underscores the practical relevance of SAVA to modern e-commerce, simulating a critical business scenario: the continuous, automated curation of inventory based on asynchronous user feedback. The problem aligns precisely with the selective inference framework in Sections \ref{subsec:SIframework} and \ref{subsec:avp}, requiring the simultaneous detection of directional signals (high vs. low quality) without pre-specified null hypothesis. By enabling continuous monitoring, SAVA effectively transforms high-volume data streams into actionable business insights, balancing the need for rapid identification with rigorous control over decision risk.

	To enhance the reliability of our analysis, we filter the dataset to include only items with more than 50 reviews, reflecting both popularity and sufficient rating data. 
	For each item, we define the true state as $\theta^j = \mathrm{A}$ if the average rating exceeds 3, and $\theta^j = \mathrm{B}$ otherwise. This simple and intuitive definition is used for demonstration purposes, though alternative thresholds or criteria could also be applied. For example, one could define $\theta^j = \mathrm{A}$ if averaged rating exceeds a pre-specified level, and $\theta^j = \mathrm{B}$ if it falls below a certain level.

	\subsection{Constructing always-valid directional p-values}\label{subsec:avp-cons}
	In this real-data application, the observations are naturally bounded. Customer ratings follow a discrete scale from 1 to 5, and we center the data by subtracting 3 (the neutral midpoint). Consequently, the transformed observations $X^{j}_t - 3$ are naturally bounded in $[-K, K]$ with $K = 2$. This boundedness property allows us to construct valid e-processes without requiring any parametric distributional assumptions.
	
	Following the betting-based e-process framework of \citet{waudby2020estimating}, we construct two directional e-processes for each item $j$:
		\vskip-1cm
	\begin{equation*}
	E^{j,A}_t = \prod_{t_0^j \leq i \leq t, i \in \mathcal{T}^j} \exp\left(\frac{\lambda_i (X_i^j-3)}{2K} - \frac{\lambda_i^2}{8}\right), 
	E^{j,B}_t = \prod_{t_0^j \leq i \leq t, i \in \mathcal{T}^j} \exp\left(\frac{-\lambda_i (X_i^j-3)}{2K} - \frac{\lambda_i^2}{8}\right),
\end{equation*}
	\vskip-0.5cm
where $\lambda_i = \left\{8\log(2/\alpha)/(r_i \log(r_i+1))\right\}^{1/2} \wedge 1$, $\alpha$ is the target FSR level ,and $r_i = |\{k \in \mathcal{T}_j : t_j^0 \leq k \leq i\}|$ denotes the cumulative sample count for item $j$ up to time $i$. By \citet{waudby2020estimating}, $E^{j,A}_t$ and $E^{j,B}_t$ are valid e-processes with respect to $\theta^j = B$ and $\theta^j = A$, respectively.
	 By Proposition \ref{prop:eprocess-to-pval} the corresponding always-valid directional p-values are:
	\vskip-0.8cm
\begin{equation}\label{eq:apval-realdata}
		p^{j,A}_t = \min\left\{1, \left(\max_{t_j^0 \leq s \leq t} 	E^{j,A}_s \right)^{-1}\right\}, \quad
		p^{j,B}_t = \min\left\{1, \left(\max_{t_j^0 \leq s \leq t} E^{j,B}_s\right)^{-1}\right\}.
	\end{equation}
	
	
		\vskip-1cm

	\subsection{Experiment design}\label{subsec:online-rule2}
\begin{figure}[ht]
			
			\centering{
				
				\includegraphics[width=6in,height=2in]{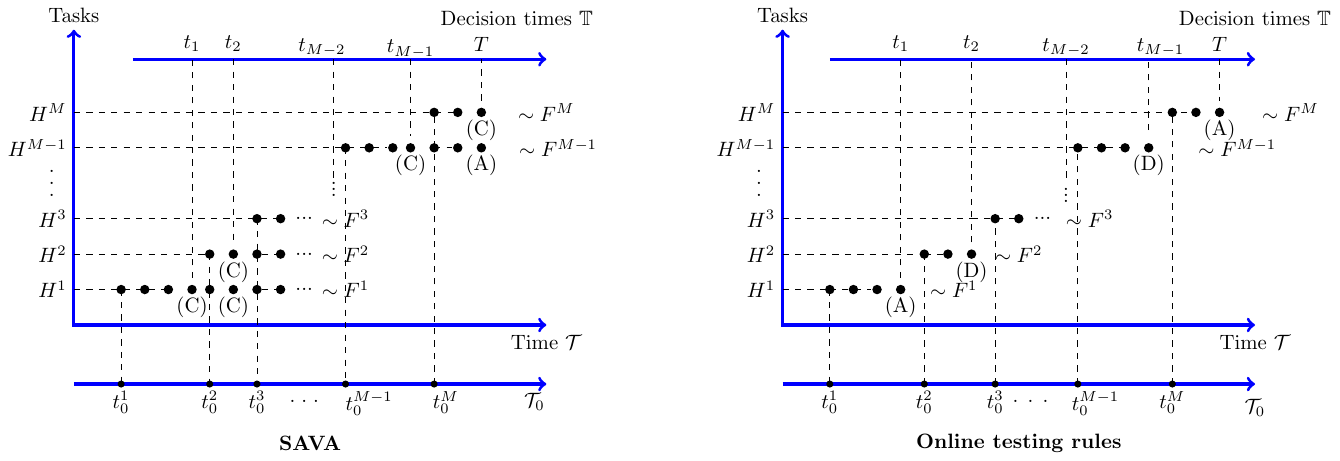}
				
			}
			
			\caption{\label{fig:illus-realdata}\small Decision times are synchronized across the online FDR and doubly sequential setups. The observations are represented by solid points, while decisions are denoted by (A, B, C, D).}
			
		\end{figure}

		Let $\mathcal{T}$ denote the complete set of review timestamps. We define $\mathcal{T}_0$ as the chronologically ordered set of the first review times for each item, and let $T$ represent the final timestamp across all items. To ensure a valid comparison between the online FDR methods and the SAVA framework, we set the decision times as $\mathbb{T} = \{t -1: t\in \mathcal T_0, t\geq t_0^2 \} \cup \{T\}$. 
			An illustration of the setup is shown in Figure~\ref{fig:illus-realdata}, where $M$ represent the total number of items.
			
		 This structure ensures that upon the arrival of a new task, the agent can immediately make a decision based on the collected data, thereby rendering the evaluation of different methods meaningful. For simplicity, we assume infinite resources for each task so SAVA never drops a task. 
		At each decision time $t_i \in \mathbb{T}$, the evaluation includes all cumulative reviews submitted prior to $t_i$. For the SAVA algorithm, the sequence of test levels is determined via Equation \eqref{eq:test-level} with a window size of $k = 100$.
	For the online FDR rules, p-values are calculated using the Wilcoxon signed-rank test based solely on the incremental samples collected between consecutive decision times. Further implementation details for the online FDR methods are provided in Supplement \ref{subsec:im-online}. The target FSR level is set to $\alpha = 0.2$.

	\subsection{Experiment result}\label{sec:realdata2}

\begin{figure}[ht]
		\centering{
			\includegraphics[width=5in,height=2in]{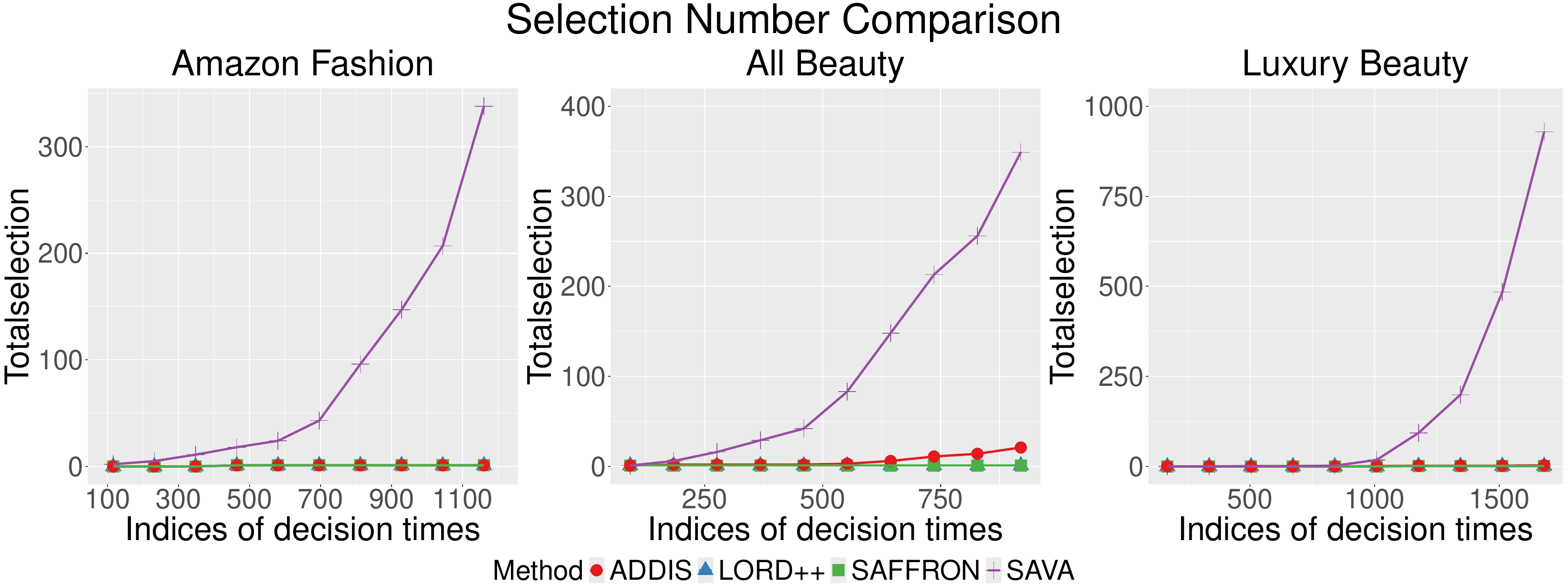}
		}
		\caption{\label{fig:realdata}Comparison of the cumulative number of selected items over time across three distinct datasets: Amazon Fashion, All Beauty, and Luxury Beauty.}
		
			\centering{
			\includegraphics[width=6in,height=2in]{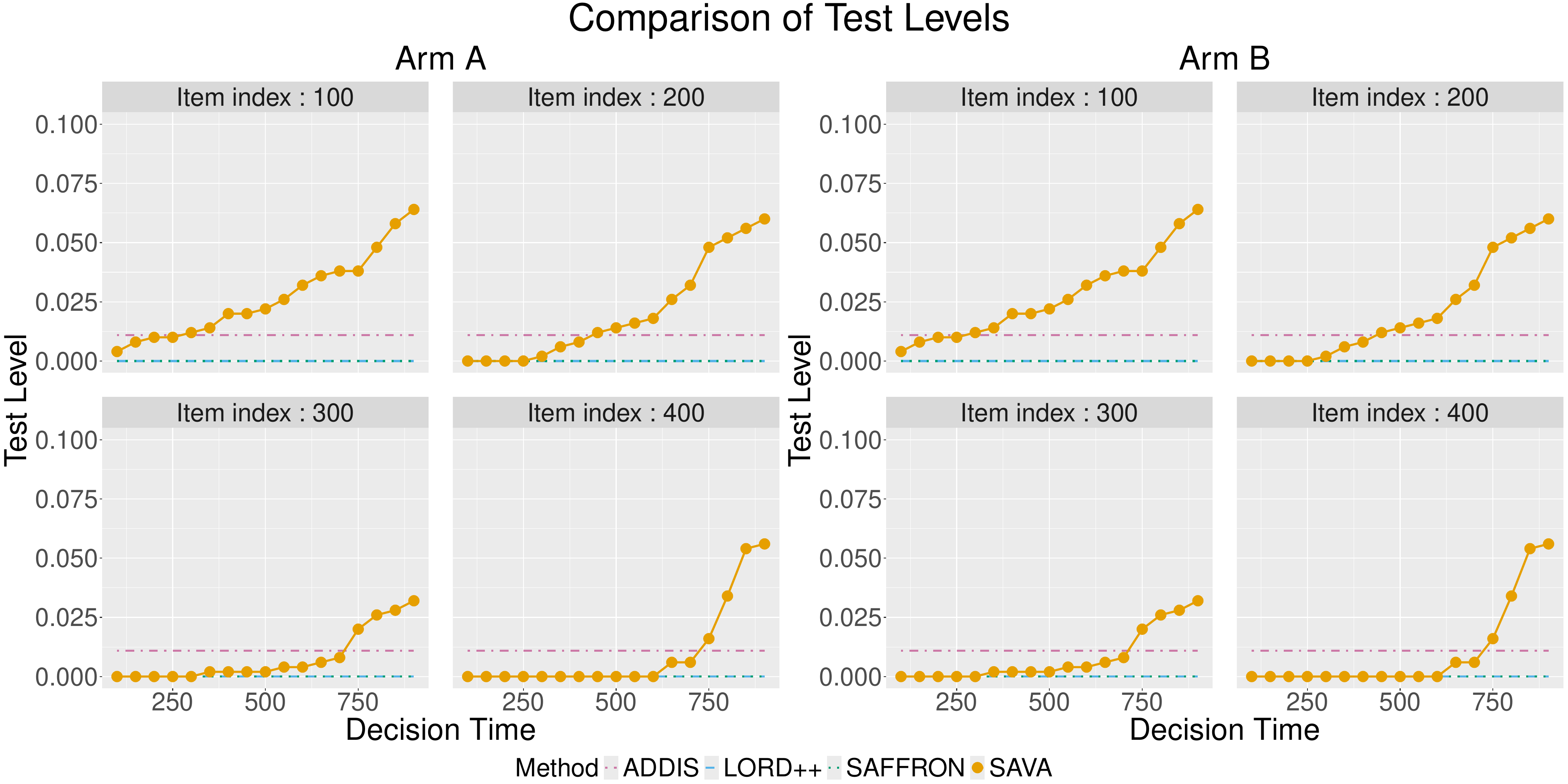}
		}
		
		\caption{\label{fig:realdata-testlevel}Temporal evolution of allocated test levels ($\alpha_t$) for four representative items (indices 100, 200, 300, and 400) within the Amazon Fashion dataset.}
	\end{figure}

	Figure~\ref{fig:realdata} illustrates the cumulative number of selections as a function of decision time for each category. We observe that the SAVA algorithm consistently yields a higher number of selections compared to online testing procedures. This performance advantage is largely attributable to the mitigation of the ``alpha-death'' phenomenon inherent in existing online FDR methods (e.g., LORD++, SAFFRON, and ADDIS). These online testing protocols necessitate an immediate, final decision upon the arrival of each new item, causing the available alpha wealth to deplete rapidly as the stream progresses.
	
	In contrast, the SAVA framework permits continuous sampling, allowing for the accumulation of evidence for a specific item over extended periods. Our alpha-investing strategy is designed to ensure that the test levels allocated to each task are non-decreasing. As data accumulates throughout the experiment, this approach effectively counters alpha depletion. To visualize this mechanism, Figure~\ref{fig:realdata-testlevel} tracks the evolution of test levels for representative items (indices 100, 200, 300, and 400) within the Amazon Fashion dataset. While the test levels for the online FDR benchmarks decay rapidly due to the high frequency of item arrivals, SAVA's test levels exhibit an upward trajectory. By leveraging the abstention option to defer judgment until sufficient evidence is gathered, SAVA preserves statistical power and avoids the premature exhaustion of the error budget. This results in higher selection yields of quality items, directly enhancing recommendation systems, inventory management, and targeted marketing campaigns.

{

\subsection{A summary of supplementary numerical results }
		In addition to the real-data analysis, we conduct extensive numerical experiments to evaluate the finite-sample performance of SAVA using simulated data. Full details are provided in Section~\ref{sec:simulation} of the Supplementary Material. We summarize the key findings below.
		
		\begin{enumerate}
			\item \textbf{FSR control and power under various models.} Under Gaussian, truncated Gaussian, Gamma, Beta, and sub-Gaussian models, SAVA maintains FSR control at the nominal level while achieving substantially higher true selection rates than LORD++, SAFFRON, and ADDIS (Section~\ref{subsec:ep-gaussian} to \ref{subsec:sub-gauss-model} ). 
			
			\item \textbf{Robustness to dependence.} When data streams exhibit moderate cross-stream dependence induced by preceding observations, SAVA continues to exhibit stable empirical FSR control and competitive power, suggesting robustness beyond the independence assumption required by the theory (Section~\ref{subsec:simu-dep}).
			
			\item \textbf{Robustness to model misspecification.} When constructing e-processes for bounded data, SAVA maintains valid FSR control even when the  bound is underestimated. Performance degradation occurs only under severe misspecification, indicating the procedure is conservative in practice (Section~\ref{subsec:simu-mis}).
		\end{enumerate}
		
}

{\subsection{Data and Code Availability}

The Amazon Review Data \citep{Nietal2019} used for the numerical experiments can be accessed at \url{https://nijianmo.github.io/amazon/index.html}. The code to reproduce all numerical results is publicly available at \url{https://github.com/melodies33/SAVA.git}.
}

\section{Conclusion and Discussion}

In this paper, we introduce the SAVA framework for doubly sequential online inference. By combining always-valid directional p-values with a dynamic alpha-investing rule, SAVA supports an ``active investigator'' setting, allowing continuous monitoring, adaptive resource allocation, and mitigation of alpha depletion, while rigorously controlling the FSR at all decision times.

{We conclude this article by discussing several promising directions for future research. First, in certain high-stakes applications where data collection is relatively inexpensive and decisions are not time-sensitive, mistakenly dropping a promising task (Arm D) incurs a genuine operational cost. In such settings, it becomes desirable to simultaneously control both the FSR and the missed discovery rate (MDR). This stands in contrast to the fast-paced, dynamic environments for which SAVA is primarily designed. To address simultaneous FSR and MDR control, one could formulate a compound sequential decision-theoretic framework \citep{SMART} that aims to minimize the total expected sampling cost subject to both FSR and MDR constraints. Such an approach would require carefully balancing exploration and exploitation. Deriving the optimal decision rule under this compound framework while preserving finite-sample FSR guarantees is highly non-trivial; it would necessitate the development of new alpha-investing strategies and corresponding e-processes capable of emulating the optimal rule. We view this as an important and challenging direction for future investigation. Second, the current SAVA framework exhibits notable conservativeness in FSR control, stemming in part from the conservative nature of the FSR estimator and the e-process constructions employed. Improving power through more adaptive FSR estimators and more powerful e-processes represents a valuable research direction. Third, extending the theoretical guarantees of SAVA to accommodate arbitrary cross-stream dependence remains an important open problem. Finally, although SAVA differs fundamentally from multi-armed bandit formulations (see Appendix~\ref{sec:mab} for a detailed discussion), integrating efficient tools and optimization techniques from that literature into the SAVA framework is another particularly promising avenue for enhancing its practical performance.}

	\section*{Acknowledgments}

The authors thank associate editor and two referees for their helpful comments that have resulted in significant improvements of the article. The authors declare that they have no conflicts of interest.

\medskip

\spacingset{1.0}
\small 

	\bibliography{paper-ref.bib}
	\newpage
	\phantomsection\label{supplementary-material}
	\bigskip

	\spacingset{1.9} 
	
	\normalsize
	\newpage
	\begin{center}
		\appendix
		{\Large\bf Supplements to \\ \vskip 0.5cm \LARGE ``Safe, Always-Valid Alpha-Investing Rules for Doubly Sequential Online Inference''}
	\end{center}
	
	This Supplement is organized as follows.  Section \ref{sec:notations} summarizes the main notations. Section \ref{sec:simulation} provides numerical illustrations of SAVA under different models and scenarios. Section \ref{supp:methods} discusses the adaptation of the SAVA algorithm for doubly sequential testing under the classical hypothesis testing setup, considering both overall and arm-specific constraints. Section \ref{sec:prf} provides the technical proofs for all theories presented. In Section \ref{sec:ctexam}, we analyze two counterexamples to underscore the necessity of the principles outlined in Section \ref{subsec:test-level}.  In Section \ref{sec:mab}, we discuss distinctions between SAVA and multi-armed bandits.

		\renewcommand\thetable{A.\arabic{table}}
	\setcounter{table}{0}
	\section{Summary of Notations}\label{sec:notations}
	Table~\ref{tbl-notation} provides a summary of the key notations used throughout the doubly sequential framework.
	
	\begin{longtable}{l p{0.75\textwidth}}
		\caption{Summary of key notations.}\label{tbl-notation}\\
		\toprule
		\textbf{Notation} & \textbf{Meaning} \\
		\midrule
		\endfirsthead
		
		\multicolumn{2}{c}%
		{{\bfseries Table \thetable\ continued from previous page}} \\
		\toprule
		\textbf{Notation} & \textbf{Meaning} \\
		\midrule
		\endhead
		
		\midrule
		\multicolumn{2}{r}{{Continued on next page}} \\
		\endfoot
		
		\bottomrule
		\endlastfoot
		
		$\mathcal T$ & Temporal domain within which all potential tasks operate \\
		$\mathcal T^j$ & Time interval during which data for task $j$ are collected \\
		$\mathbb T$ & Countable set of global decision times \\
		$\mathbb J$ & Index set for all potential tasks \\
		$t_0^j$ & Initiation time (arrival time) of task $j$ \\
		$\mathbf{X}^j$ & Data stream collected for task $j$ \\
		$\mu^j$ & Unknown parameter (e.g., effect size) associated with task $j$ \\
		$\theta^j \in \{\text{A, B}\}$ & True state of nature (superior arm) corresponding to $\mu^j$ \\
		$\pmb{ \delta}_t$ & Concurrent decisions for all tasks commenced prior to time $t$ \\
		$\delta_t^j \in \{\text{A, B, C, D}\}$ & Decision for task $j$ at time $t$ (Select A, Select B, Continue, or Drop) \\
		$\mathcal S_t$ & Index set of selected candidates (arms A or B) up to time $t$ \\
		$\mathcal A_t$ & Index set of active tasks (currently being sampled and monitored) at time $t$ \\
		$p_t^{j,A}, p_t^{j,B}$ & Always-valid directional p-values for task $j$ at time $t$ \\
		$\alpha_t^{j,A}, \alpha_t^{j,B}$ & Allocated test levels (alpha-wealth) for task $j$ at time $t$ \\
		$\text{FSR}_t, \text{mFSR}_t$ & False selection rate and marginal false selection rate at time $t$ \\
		$\widehat{\text{FSR}}_t$ &  Estimate of the false selection rate at time $t$ \\
	$ \mathcal{G}_t^j $ & The $\sigma$-field generated by the observed samples \( \{X_s^j : s \leq (t \wedge T_{\mathrm{stop}}^j), s \in \mathcal{T}^j\} \) for task \( j \in \mathbb{J} \) at time \( t \in \mathbb{T} \).
		
	\end{longtable}
	
	\setcounter{figure}{0}
		\renewcommand{\thefigure}{B.\arabic{figure}}
	\setcounter{table}{0}
\renewcommand{\thetable}{B.\arabic{table}}
		\setcounter{algorithm}{0}
		\renewcommand\thealgorithm{B.\arabic{algorithm}}
		
		\setcounter{theorem}{0}
		\renewcommand\thetheorem{B.\arabic{theorem}}
		\setcounter{proposition}{0}
		\renewcommand\theproposition{B.\arabic{proposition}}
		\setcounter{corollary}{0}
		\renewcommand\thecorollary{B.\arabic{corollary}}
		\setcounter{condition}{0}
		\renewcommand\thecondition{B.\arabic{condition}}
		\setcounter{algorithm}{0}
		\renewcommand\thealgorithm{B.\arabic{algorithm}}
		\setcounter{equation}{0}
		\renewcommand\theequation{B.\arabic{equation}}
		
		\setcounter{lemma}{0}
		\renewcommand\thelemma{B.\arabic{lemma}}

		\section{Experiments with Synthetic Data}\label{sec:simulation}

	This section begins by elaborating on the design of simulation studies (Section \ref{subsec:online-rule}), then presents the implementation details of online testing rules (Section \ref{subsec:online-implement}).
	We then present simulation results across a broad range of model classes (Sections \ref{subsec:ep-gaussian}--\ref{subsec:sub-gauss-model}). Next, we examine the performance of SAVA under dependent data streams (Section \ref{subsec:simu-dep}). Finally, we illustrate the behavior of SAVA when under model-misspecification (Section \ref{subsec:simu-mis}).
	
	\subsection{General considerations in the design of simulation studies}\label{subsec:online-rule}
	
	To facilitate a meaningful and informative comparison across various methods, we carefully design the simulation setup and adapt the online FDR methods within a doubly sequential context. Let \(\mathcal{T} = \{1, 2, \ldots, T\}\), \(T \in \mathbb{N}\), denote the temporal domain. A new task arrives at each \(t \in \mathcal{T}\) with probability \(p \in (0, 1)\). For convenience,  assume the first task is initiated at \(t = 1\). Subsequently, we generate independent Bernoulli variables \(\{\mathrm{Ber}_t(p)\}_{t=2}^T\), selecting the arrival times for the task stream as \(\mathcal{T}_0 = \{t \in \mathcal{T}: \mathrm{Ber}_t(p) = 1, t \geq 2\} \cup \{1\}\). 
	
\begin{figure}
				
		\centering{
			
			\includegraphics[width=6in,height=2in]{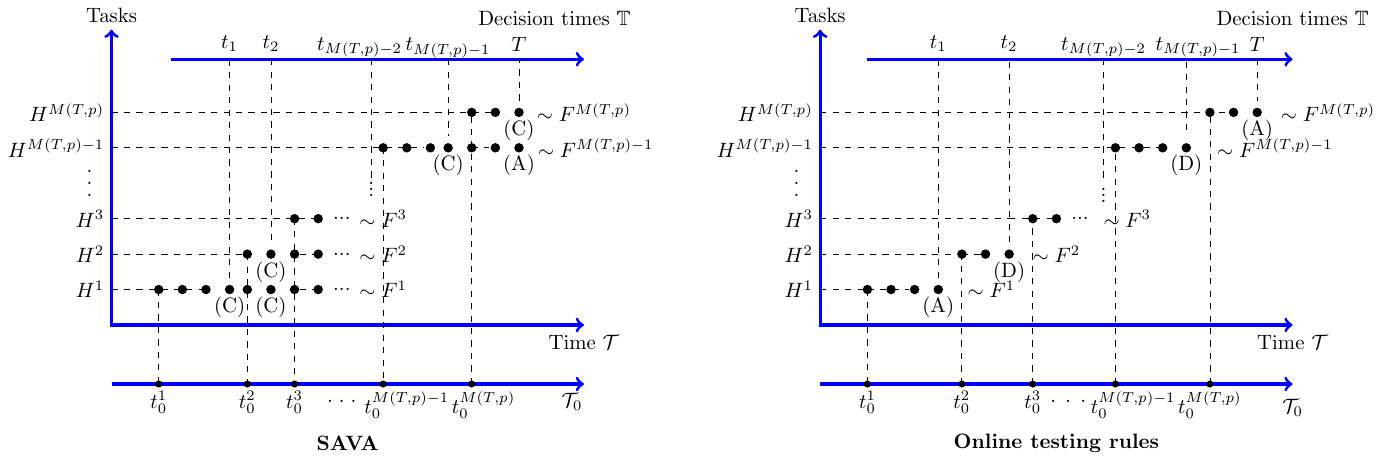}
			
		}
		
		\caption{\label{fig:illus-simu}Decision times are synchronized across the online FDR and doubly sequential setups. The observations are represented by solid points, while decisions are denoted by (A, B, C, D).}
		
	\end{figure}

	Let \(M(T, p)\) denote the total number of tasks and define \(\mathbb{J} = \{1, 2, \ldots, M(T, p)\}\). To align existing online FDR rules with the SAVA framework, we deliberately set the decision points for the doubly sequential experiments as \(\mathbb{T} = \{t - 1: t \in \mathcal{T}, \mathrm{Ber}_t(p) = 1, t \geq 2\} \cup \{T\}\). For simplicity, we assume infinite resources for each task so SAVA never drops a task. An illustration of the setup is shown in Figure~\ref{fig:illus-simu}. 
	
	For active tasks, we collect new samples \(X^j_t \sim F^j\) across two consecutive decision times, \(t_{i-1}\) and \(t_i\). At \(t_i\), the total number of newly collected samples for each active task is determined as follows: (a) If \(t_{i-1} < t_0^j \leq t_i\), the number of newly collected samples is \(t_i - t_0^j + 1\); (b) If \(t_0^j \leq t_{i-1}\), the number of newly collected samples is given by \(t_i - t_{i-1}\). This is the same setup as in Section \ref{sec:realdata}.

		\subsection{Implementation details of online testing rules}
	\label{subsec:im-online}\label{subsec:online-implement}
	
	This section presents specific designs of test levels in online FDR rules and outlines the operation of these rules within the online selective inference setup.
	
	\begin{itemize}
		\item \textit{LORD++} \citep{ramdas2017online}. For each direction, we set the test levels as $\alpha_j = w_0\gamma_j + (\alpha - w_0)\gamma_{j - \tau_1} + \alpha \sum_{i\colon \tau_i<j,\,i\geq 2} \gamma_{j - \tau_i}$, where $\tau_i$ denotes the index of $i$th selection. Furthermore, we set $w_0 = \alpha/10$ and choose $\gamma_j = 0.0722\log(j\vee 2)/(j\exp[{\{\log(j)\}^{1/2}}])$ as recommended in \cite{ramdas2017online}.
		\item \textit{SAFFRON}  \citep{ramdas2018saffron}. For each direction we set the test levels as follows. Let $p_j$ denote the p-value for $j$th hypothesis. For $j = 1$ set $\alpha_j = \min\{(1-\lambda)\gamma_1w_0, \lambda\}$. For $j>1$ set $	\alpha_j = \min\{\lambda, (1-\lambda)[w_0\gamma_{j - C_{0+}(j)} + (\alpha-w_0)\gamma_{j - \tau_1 - C_{1+}(j)}] + \alpha\sum_{i\colon \tau_i<j,\,i\geq 2}\gamma_{j - \tau_i - C_{j+}(j)}\}$, where $C_{i+}(j) = \sum_{k=\tau_i+1}^{j-1}\mathbb{I}\{p_k \leq \alpha_k\}$. The parameters are chosen as follows, $\gamma_j \propto (j+1)^{-1.6}$, $\sum_{j=1}^{\infty}\gamma_j=1$, $\lambda = 0.5$ and $w_0 = \alpha/2$, as recommended in \cite{ramdas2018saffron}.
		\item \textit{ADDIS} \citep{tian2019addis}. Let $p_j$ denote the p-value for the $j$th hypothesis. The test levels for each direction are set to be $\alpha_j = \min\{\lambda, \hat{\alpha}_j\}$, where $\hat{\alpha}_j = (\tau - \lambda)\Big(w_0\gamma_{S_j - C_{0+}(j)} + (\alpha-w_0)\gamma_{S_j -\kappa_1^*-C_{1+}(j)} + \alpha\sum_{i\geq 2}\gamma_{S_j - \kappa_j^* - C_{i+}(j)}\Big)$.
		Here $S_j = \sum_{i<j}\mathbb{I}\{p_i\leq\tau\}$, $C_{i+}(j) = \sum_{k = \kappa_i + 1}^{j-1}\mathbb{I}\{p_k\leq \lambda\}$, $\kappa_i = \min\{k\in\{1,\ldots,j-1\}\colon \sum_{l\leq k}\mathbb{I}\{p_l\leq\alpha_l\} \geq i\}$, $\kappa_i^* = \sum_{k\leq \kappa_i}\mathbb{I}\{p_k\leq \tau\}$.
		Furthermore, the parameters are set as $\lambda = 0.25$, $\tau = 0.5$, $w_0 = \alpha/2$ and $\gamma_j\propto (j+1)^{-1.6}$ as recommended by \cite{tian2019addis}.
	\end{itemize}
	
	Given the designs of the test levels, the operations of the online testing rules are as follows. At each decision time $t_i\in\mathbb T$, the active task is the task that arrived at $t_{i-1}+1$, which we denote as task $j$ (for ease of illustration), and we collect samples as described in Section \ref{subsec:online-rule}. We treat $H_{0,A}^j$ ($H_{0,B}^j$) as the null hypotheses and calculate valid p-values $p^{j,A}$ ($p^{j,B}$). Then  apply the given rules to calculate test levels for the two directions, by treating the sets $\{H_{0,A}^k\}_{k=1}^j$ ($\{H_{0,B}^k\}_{k=1}^j$) as null hypotheses respectively in the online testing rules. Finally, the rejection results are combined: if neither null hypothesis is rejected, the decision $\delta^j$ is set to D; if both null hypotheses are rejected, the decision is made in favor of the non-null arm in the hypothesis which generates a smaller p-value; and if one null hypothesis is rejected, the decision is made in favor of the non-null arm of the hypothesis.
	
	\subsection{Results for Gaussian model}\label{subsec:ep-gaussian}
	In this section, we provide the performance of SAVA under Gaussian model, to show that the SAVA framework is not conservative when the always-valid directional p-values are powerful. In addition, the choice of window size $k$ in test levels of SAVA \eqref{eq:test-level} is suggested under this model.

	\subsubsection{Constructing e-processes for Gaussian distributions}
	Assume that the distribution $F^j$ obeys $F^j = \frac{\mathrm{sign}(\mu^j)+1}{2}N(|\mu^j|,1) + \frac{\mathrm{sign}(-\mu^j)+1}{2}N(-|\mu^j|,1)$, where $\mathrm{sign}(x)$ represents the sign of $x$. Define $\theta^j = \mathrm A$ if $\mu^j >0 $ and $\theta^j = \mathrm B$ if $\mu^j \leq 0$. 
	
	Suppose that the absolute value of $\mu^j$ is known. Then the e-process $E_t$ can be constructed directly as the cumulative likelihood ratio for each true state. Concretely, define
	\begin{equation*}
		\begin{aligned}
			E^{j,A}_t &= \prod_{ t_0^j \leq i \leq t, i\in\mathcal T^j} \dfrac{\phi_{|\mu^j|}(X_i^{j})}{\phi_{-|\mu^j|}(X_i^j)}, \\
			E^{j,B}_t &= \prod_{ t_0^j \leq i \leq t, i\in\mathcal T^j} \dfrac{\phi_{-|\mu^j|}(X_i^j)}{\phi_{|\mu^j|}(X_i^j)},		
		\end{aligned}
	\end{equation*}
	where $\phi_{|\mu^j|}(\cdot)$ and $\phi_{-|\mu^j|}(\cdot)$ represent the density function of $N(|\mu^j|,1)$ and $N(-|\mu^j|,1)$, respectively. It is straightforward to verify that $E_t^{j,A}$ and $E_t^{j,B}$ are e-processes under $\theta^j = \mathrm B$ and $\theta^j = \mathrm A$, respectively. Therefore, the always-valid directional p-values are:
	\begin{equation}\label{eq:oraclepvalue}
		\begin{aligned}
			p_t^{j,A} &= \min\left\{1, \left(\max_{t_0^j \leq s\leq t}\prod_{i = t_0^j}^{s} \dfrac{\phi_{|\mu^j|}(X_i^{j})}{\phi_{-|\mu^j|}(X_i^j)}\right)^{-1}\right\},\\
			p_t^{j,B} &= \min\left\{1, \left(\max_{t_0^j \leq s\leq t}\prod_{i = t_0^j}^{s} \dfrac{\phi_{-|\mu^j|}(X_i^j)}{\phi_{|\mu^j|}(X_i^j)}\right)^{-1}\right\}.
		\end{aligned}
	\end{equation}
	
	\subsubsection{Simulation results for Gaussian model}\label{subsec:simugauss}
	Suppose $\theta^j = \mathrm A$ with probability $\pi^+$, and $\theta^j = \mathrm B$ with probability $1 - \pi^+$ for task $j$, where $\pi^+\in[0,1]$ represents the proportion of arm A. The process $X_t^j$ is generated according to model in Section \ref{subsec:ep-gaussian}. 
	The SAVA algorithm is implemented utilizing the always-valid directional p-values \eqref{eq:oraclepvalue}. 
	The sequence of test levels for SAVA is computed using \eqref{eq:test-level} with $k = 25$, guided by the analysis in Section \ref{subsec:bandwidth}.
	For the online FDR rules, p-values are calculated by the z-test method based on newly collected samples between two consecutive decision times for both arms. 
	
\begin{figure}[tbh!]
		\centering
		\includegraphics[width=0.90\linewidth]{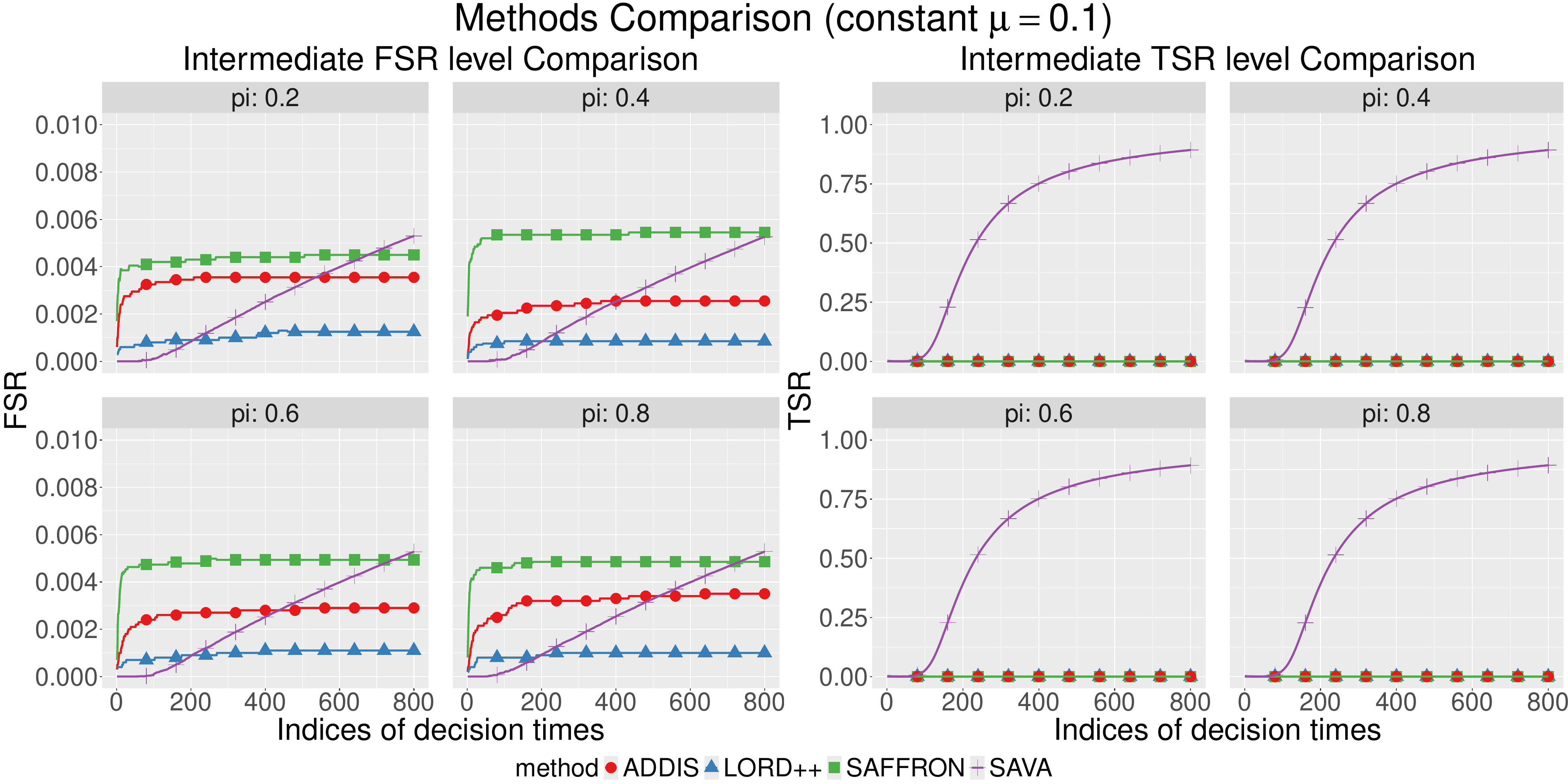}
		\caption{\small Comparisons of methods with different $\pi^{+}$ under the Gaussian model. }
		\label{fig:oracleplusratio}
	\end{figure}

\begin{figure}[tbh!]
		\centering
		\includegraphics[width=0.90\linewidth]{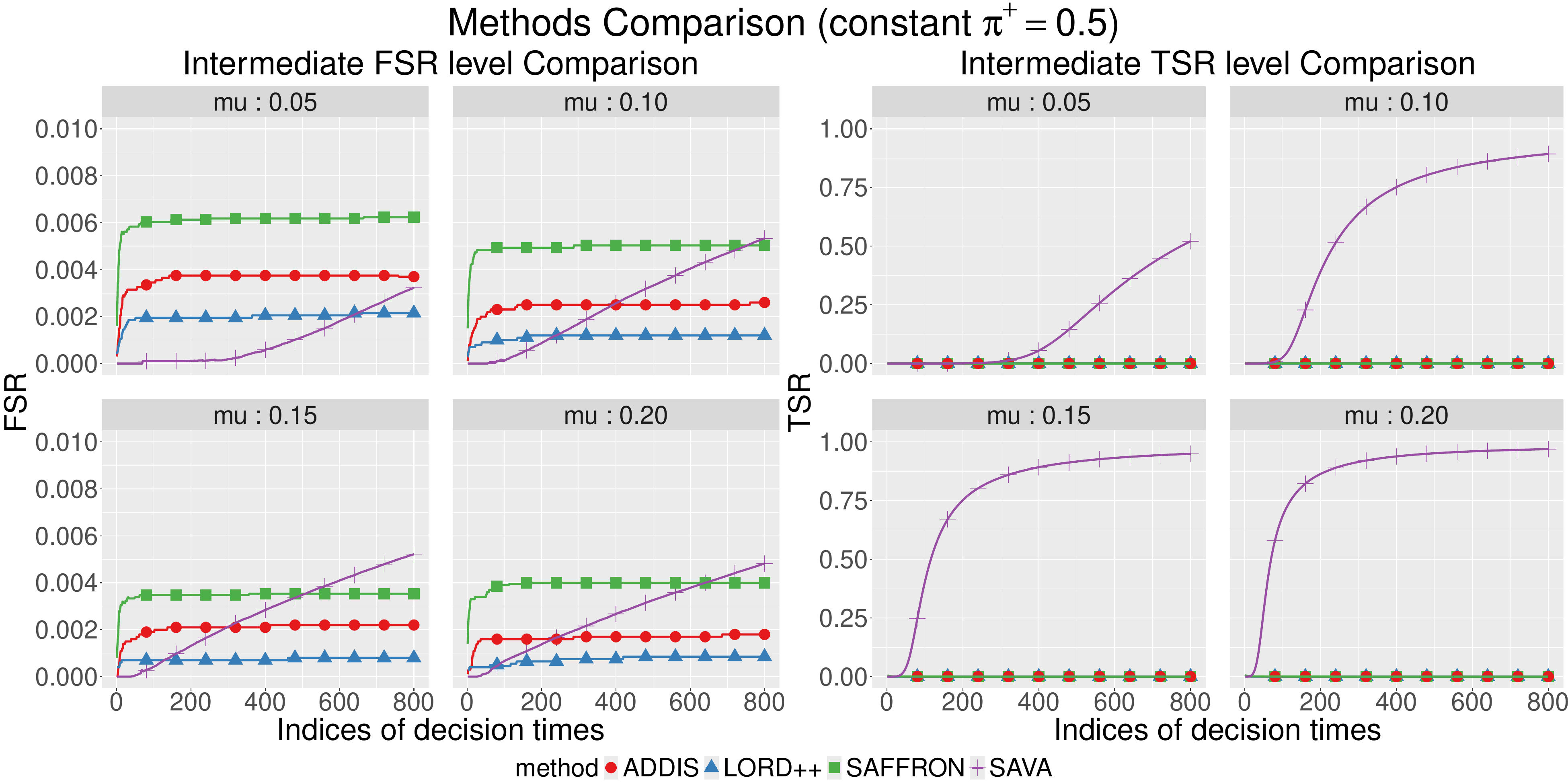}
		\caption{\small Comparisons of methods with different $\mu$ under the Gaussian model.}
		\label{fig:oraclemu}
	\end{figure}

	The target FSR level set is set to $\alpha = 0.05$, and set $T=3000$.
	The new task arrives with probability $p=1/3$.
	The following settings are considered: (i) Setting 1: Keep $\pi^{+} = 0.5$. Vary $\mu$ from 0.05 to 0.20 with step size 0.05; (ii) Setting 2: Keep $\mu = 0.1$. Vary values of $\pi^{+}$ from 0.2 to 0.8 with step size 0.2.
	For each setting, we repeat the experiments 1000 times and summarize the average FSP and TSP results at all decision times $t\in\mathbb T$ in Figure~\ref{fig:oracleplusratio} and Figure~\ref{fig:oraclemu}.
	Since the number of decision times may vary across multiple repetitions, we present results for the first 800 decision times.
	
	The results show that SAVA demonstrates considerable power improvements over online FDR rules. The conservativeness of online testing methods can largely be attributed to the ``alpha-death'' issue. By allowing the collection of  information and deferring decision-making to future decision times, the abstention option in the SAVA framework offers a significant advantage in scenarios where tasks arrive frequently and the strength of most signals is weak. This leads to higher power and helps avoid the alpha-death problem. Test levels for future tasks are increased due to more selections made from prior tasks, resulting in significantly higher power in the SAVA algorithm, as demonstrated in Figure~\ref{fig:oracleplusratio} and Figure~\ref{fig:oraclemu}, where the TSR increases following a plateau period.
	
	\subsubsection{Choosing the bandwidth of neighborhood}\label{subsec:bandwidth}
	
\begin{figure}[tb!]
		\centering
		\includegraphics[width=0.90\linewidth]{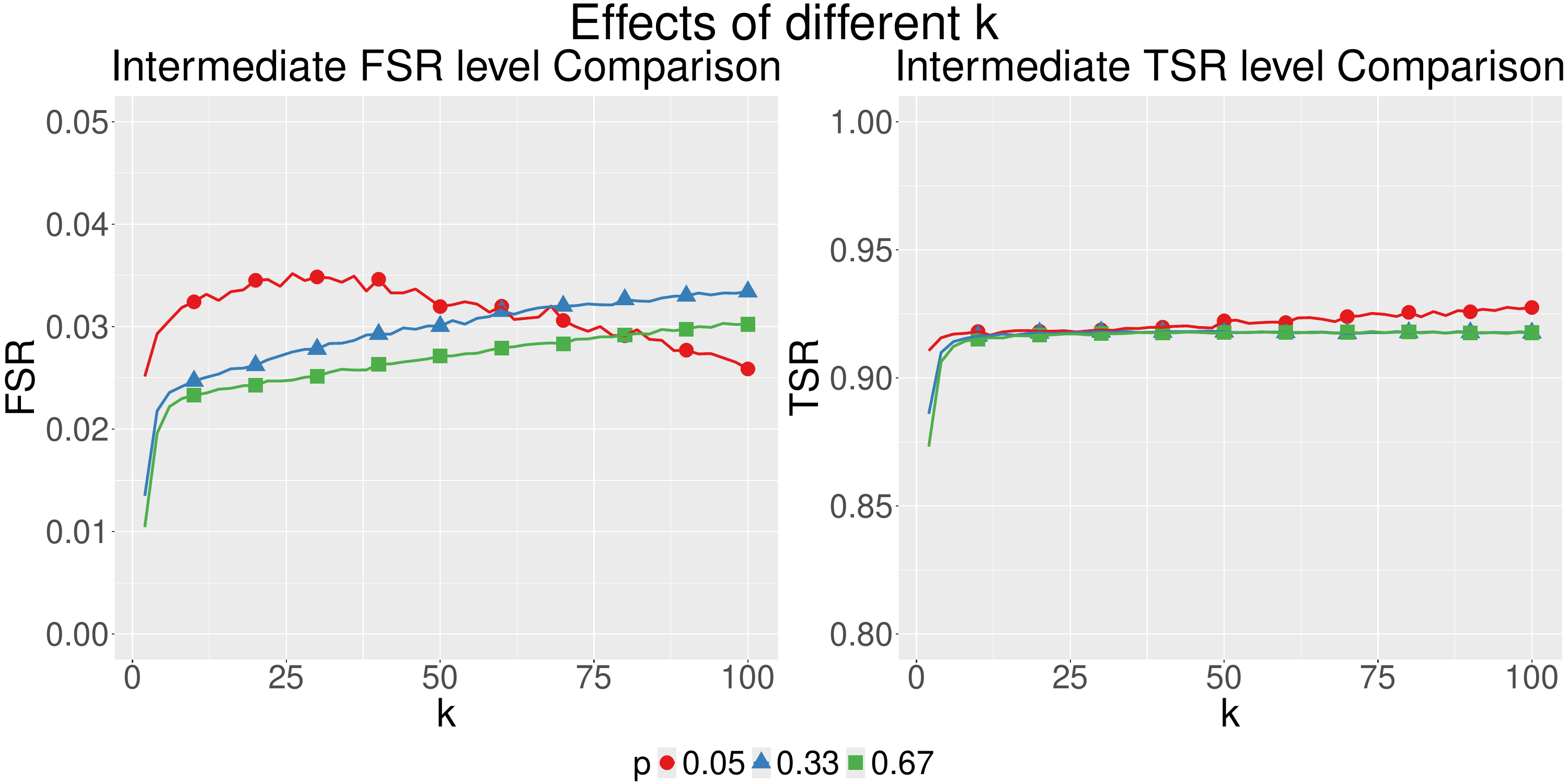}
		\caption{\small Effect of different $k$ in test levels under different arrival probability $p$. }
		\label{fig:comparisongamma}
	\end{figure}

	In order to determine the tuning parameter $k$ in test levels \eqref{eq:test-level}, we investigate how different choices of $k$ impact the performance of the SAVA under different arrival probability $p$. We consider the Gaussian model for illustration and use the same notations in Section \ref{subsec:simugauss}. The target FSR level set is set to $\alpha = 0.05$, and set $T=3000$. Fix $\pi^+ = 0.5$ and $\mu = 0.1$, and we vary $k$ from $2$ to $100$. The arrival probability $p$ is chosen from $\{1/20,1/3,2/3\}$. For each setting we repeat the experiments 1000 times and record the FSP and TSP at the final decision time $\max \mathbb T$. The average result is shown in Figure~\ref{fig:comparisongamma}. FSR is controlled for all choice of $k$ and $p$, and TSR varies little when $k>10$. Based on these results, as $k$ increases from a relatively small value to a larger one, more tasks are allocated alpha-wealth, increasing the likelihood of selections. However, when $k$ becomes sufficiently large, the alpha-wealth allocated to each task $(\alpha/k)$ becomes less sensitive to changes in $k$, and tasks receiving allocations may not arrive in time. Consequently, the overall selection performance becomes largely insensitive to the choice of $k$. Guided by these empirical observations and analysis, we fix $k=25$ for all experiments in our simulation studies.

	\subsection{Results for truncated Gaussian model}\label{sec:generalcdf}

	Suppose \(\theta^j = \mathrm{A}\) with probability \(\pi^+\), and \(\theta^j = \mathrm{B}\) with probability \(1 - \pi^+\), where \(\pi^+ \in [0, 1]\) represents the proportion of arm A. We consider the truncated Gaussian model for our comparison.
	Assume that the data are generated according to the following distribution: 
	\begin{equation}\label{Mod:data-gen}
		X^j_t|\theta^j\sim F^j = \mathbb I\{\theta^j = \mathrm A\} \mathrm{tr}N(\mu,1,-K,K) + \mathbb I\{\theta^j = \mathrm B\}\mathrm{tr}N(-\mu,1,-K,K),
	\end{equation}
	where \(\mu\) is an unknown positive parameter and \(\mathrm{tr}N(\mu,1,-K,K)\) denotes the truncated Gaussian distribution \(N(\mu,1)\) with \(K=2\).

	The SAVA algorithm is implemented utilizing the always-valid p-values constructed via the e-processes discussed in Section \ref{subsec:avp-cons} with $X^j_t-3$ replaced by $X^j_t$. The sequence of test levels for SAVA is computed using \eqref{eq:test-level} with \(k = 25\), guided by the analysis in Section \ref{subsec:bandwidth}. For online FDR rules, p-values are calculated using Wilcoxon's signed rank test based on newly collected samples between two consecutive decision times.

\begin{figure}[tbh!]
		\centering
		\includegraphics[width=0.90\linewidth]{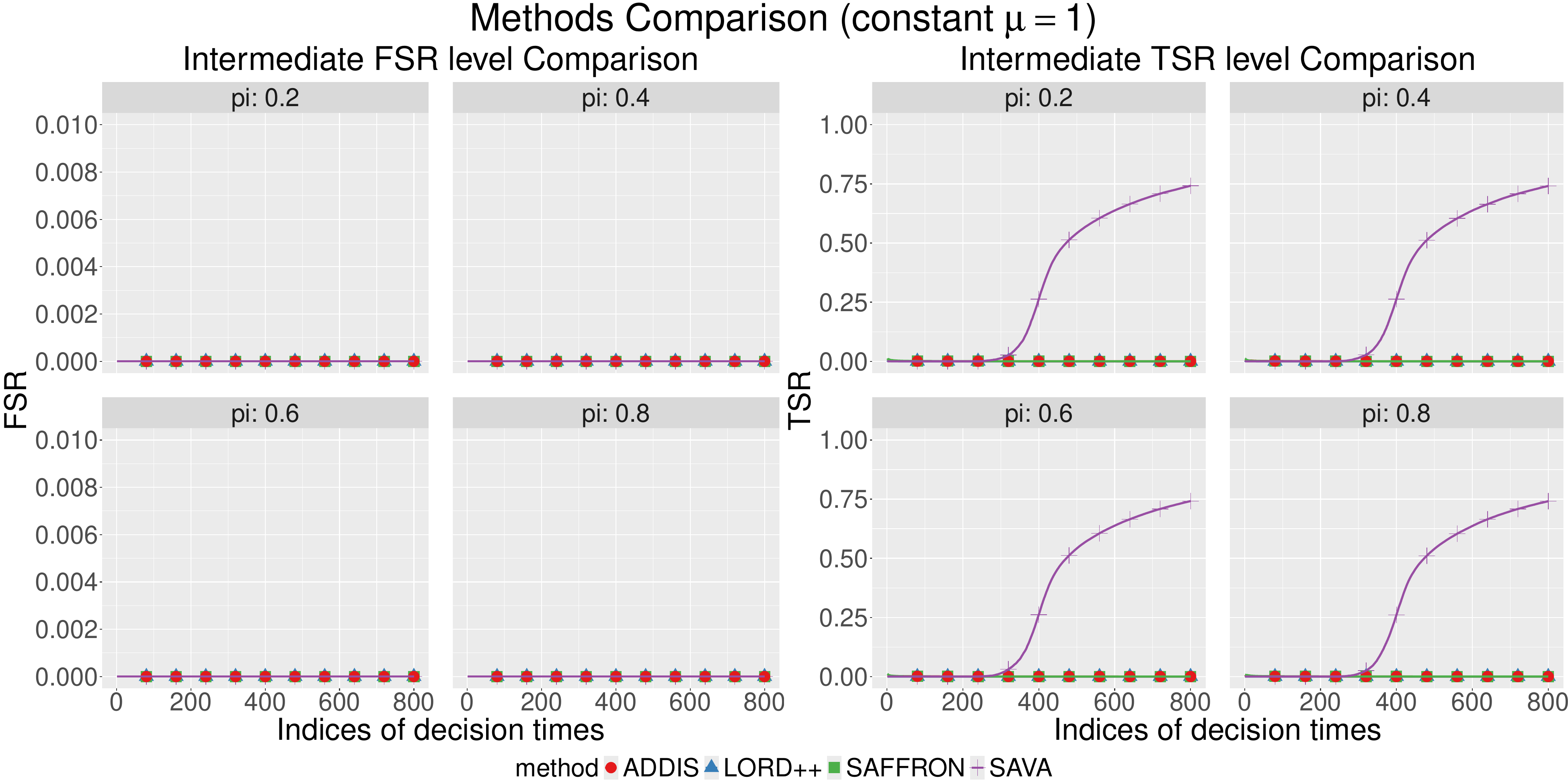}
		\caption{\small Comparisons of methods with different $\pi^{+}$ under the truncated Gaussian model.}
		\label{fig:comparisonmplratio}
	\end{figure}

\begin{figure}[tbh!]
		\centering
		\includegraphics[width=0.90\linewidth]{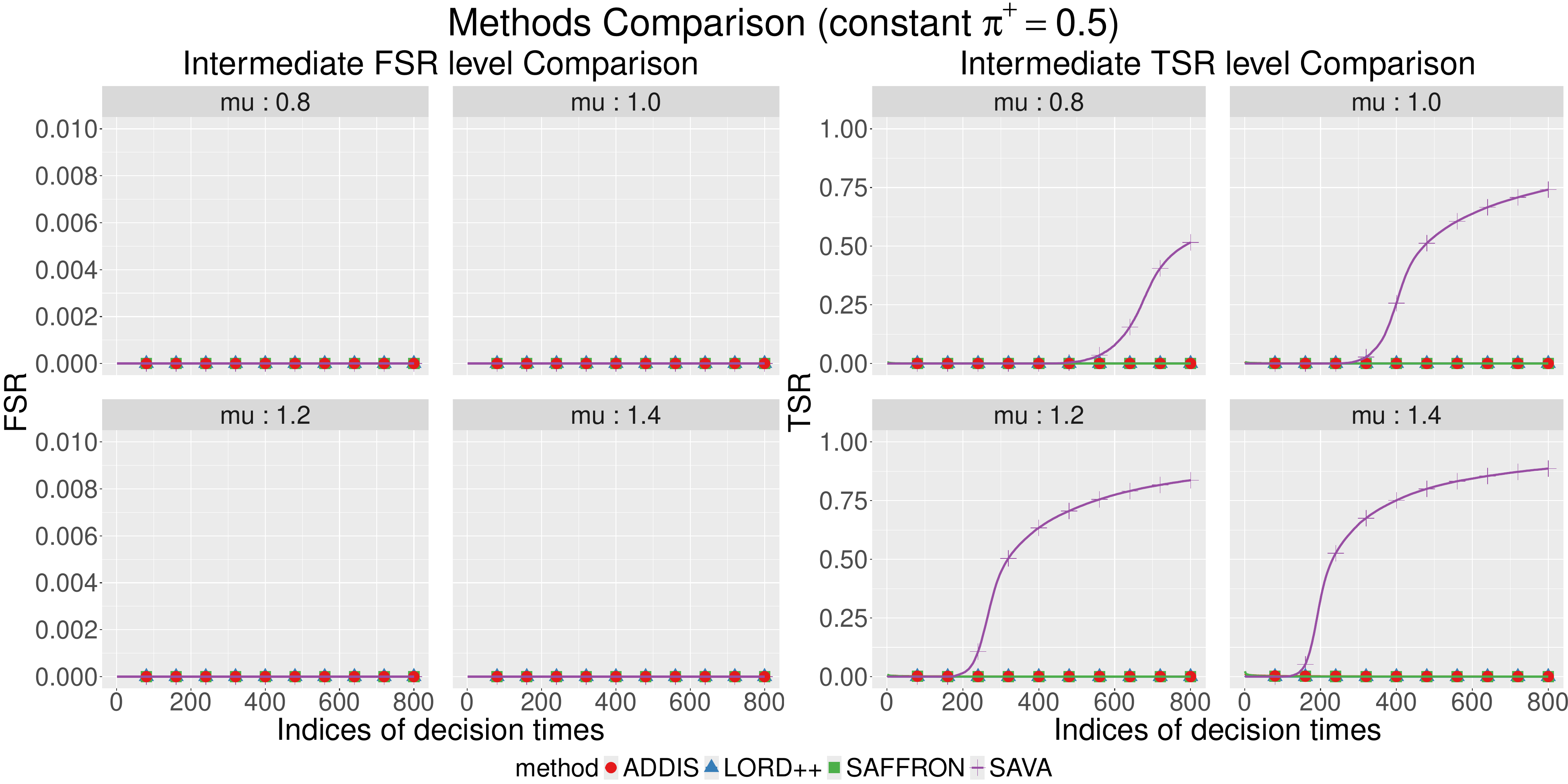}
		\caption{\small Comparisons of methods with different $\mu$ under the truncated Gaussian model.}
		\label{fig:comparisonmu}
	\end{figure}

	The target FSR level to be 0.05, and set $T = 3000$. The new task arrives with probability $p=1/3$.  The following settings are considered: (i) Setting 1: Keep $\pi^{+} = 0.5$. Vary $\mu$ from 0.8 to 1.4 with step size 0.2; (ii) Setting 2: Keep $\mu = 1$. Vary values of $\pi^{+}$ from 0.2 to 0.8 with step size 0.2. For each setting, we repeat the experiments 1000 times and summarize the average FSP and TSP results at each decision time $t\in\mathbb T$ in Figure~\ref{fig:comparisonmplratio} and Figure~\ref{fig:comparisonmu}.
	
	While all methods exhibit substantial conservativeness, SAVA demonstrates considerable power improvements over existing online testing rules. The conservativeness of SAVA primarily stems from the inefficiency of always-valid directional p-values. As illustrated in Section \ref{subsec:simugauss}, reducing the conservativeness of these directional p-values leads to less conservative FSR levels in the SAVA algorithm. Moreover, the conservativeness of online testing rules can largely be attributed to the ``alpha-death'' issue. The abstention option within the SAVA framework offers a significant advantage by allowing agents to continue sampling from active data streams even after the arrival of new tasks. This approach facilitates the accumulation of more evidence over time, resulting in more informative decisions and mitigating the alpha-death problem. As demonstrated in Figure~\ref{fig:comparisonmplratio} and Figure~\ref{fig:comparisonmu}, the TSR increases rapidly following a plateau period. As more selections are made from prior tasks, the test levels for future tasks are increased [cf. Equation \eqref{eq:test-level}], thereby further improving the power of the SAVA algorithm.

	\subsection{Results for Gamma model}\label{subsec:gamma-model}
	Suppose $\theta^j = \mathrm A$ with probability $\pi^+$ and $\theta^j = \mathrm B$ with probability $1 - \pi^+$. We consider a Gamma model for each task $j$, where $X_t^j \sim \mathrm{Gamma}(m,\sigma_j)$ with known shape parameter $m$ and unknown scale parameter $\sigma_j$. Let $\sigma_0>0$ be a pre-specified baseline parameter and define $\mu_0 = m\sigma_0$. The true state is determined by
	\begin{equation}
		\label{eq:true-state-setting}
		\theta^j =
		\begin{cases}
			\mathrm A, & \mu^j > \mu_0,\\
			\mathrm B, & \mu^j \leq \mu_0,
		\end{cases}
	\end{equation}
	where $\mu^j = m\sigma_j$. We assume the data are generated according to
	\begin{equation}
		\label{eq:simu-model-general}
		X_t^j|\theta^j\sim F^j=
		\mathbb I\{\theta^j=\mathrm A\}F_A^j
		+
		\mathbb I\{\theta^j=\mathrm B\}F_B^j,
	\end{equation}
	where \(F_A^j = \mathrm{Gamma}(m, \mu_0\mu_\delta /m),\) and \(F_B^j = \mathrm{Gamma}(m, \mu_0/(\mu_{\delta} m) )\), and $\mu_\delta>1$ controls the signal strength. In our experiments, we set $m=2$ and $\mu_0=4$.

	Following \cite{waudby2020estimating,howard2020time}, we construct e-processes using exponential supermartingales derived via Chernoff's method. Specifically, define
	\[
	E_{t,\mathrm{gam}}^j(\lambda)=
	\prod_{t_0^j\le i\le t}\exp\!\left(\lambda (X_i^j-\mu_0)-\phi_{\mathrm{gam}}(\lambda)\right),
	\]
	where
	\[
	\phi_{\mathrm{gam}}(\lambda)=-m\log(1-\sigma_0\lambda)-\lambda\mu_0,\quad |\lambda|<\sigma_0^{-1}.
	\]
	The directional e-processes are given by \(
	E^{j,A}_{t,\mathrm{gam}}(\lambda)=E_{t,\mathrm{gam}}^j(\lambda)\), 
	\(
	E^{j,B}_{t,\mathrm{gam}}(\lambda)=E_{t,\mathrm{gam}}^j(-\lambda)
	\).
	The corresponding always-valid directional p-values are defined as
	\begin{equation}
		\label{eq:e-pro-gamma}
		\begin{aligned}
			p_{t,\mathrm{gam}}^{j,A}&=
			\min\!\left\{1,\left(\max_{t_0^j\le s\le t} E^{j,A}_{s,\mathrm{gam}}(\lambda)\right)^{-1}\right\}, \\
			p_{t,\mathrm{gam}}^{j,B}&=
			\min\!\left\{1,\left(\max_{t_0^j\le s\le t}
			E^{j,B}_{s,\mathrm{gam}}(\lambda)\right)^{-1}\right\},
		\end{aligned}
	\end{equation}
	for any $\lambda\in(0, \sigma_0^{-1})$. In the subsequent simulations, we set $\lambda = 1/(2\sigma_0)$.
		
	The SAVA algorithm is implemented using the always-valid directional p-values in \eqref{eq:e-pro-gamma}. The sequence of test levels is computed via \eqref{eq:test-level} with $k=25$. The construction of p-values for online FDR rules is described in Section \ref{sec:online-pvalue-add}.

\begin{figure}[tbh!]
		\centering
		\includegraphics[width=0.90\linewidth]{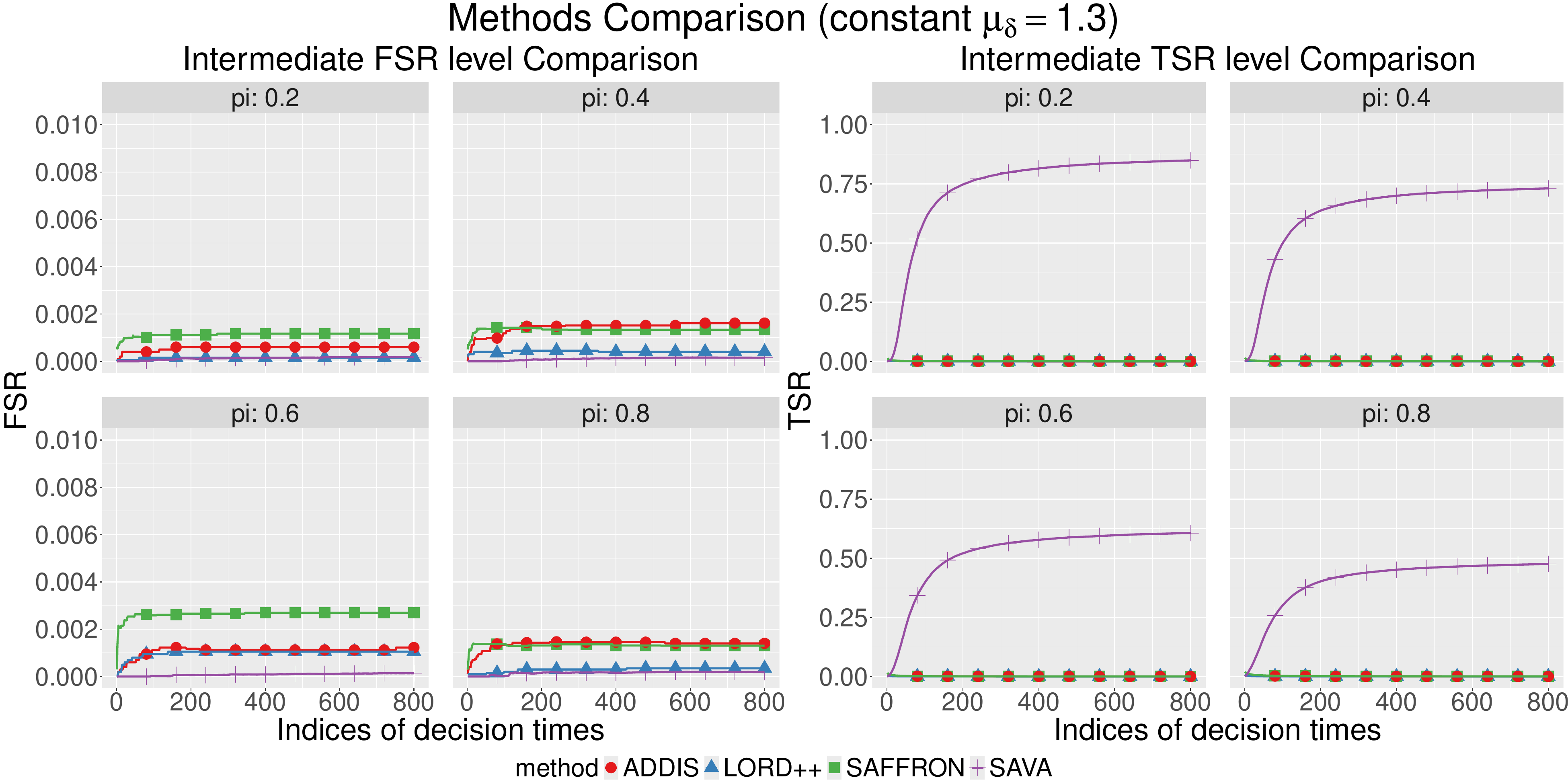}
		\caption{\small Methods with different $\pi^{+}$ under the Gamma model.}
		\label{fig:gamma-pi}
	\end{figure}

\begin{figure}[tbh!]
		\centering
		\includegraphics[width=0.90\linewidth]{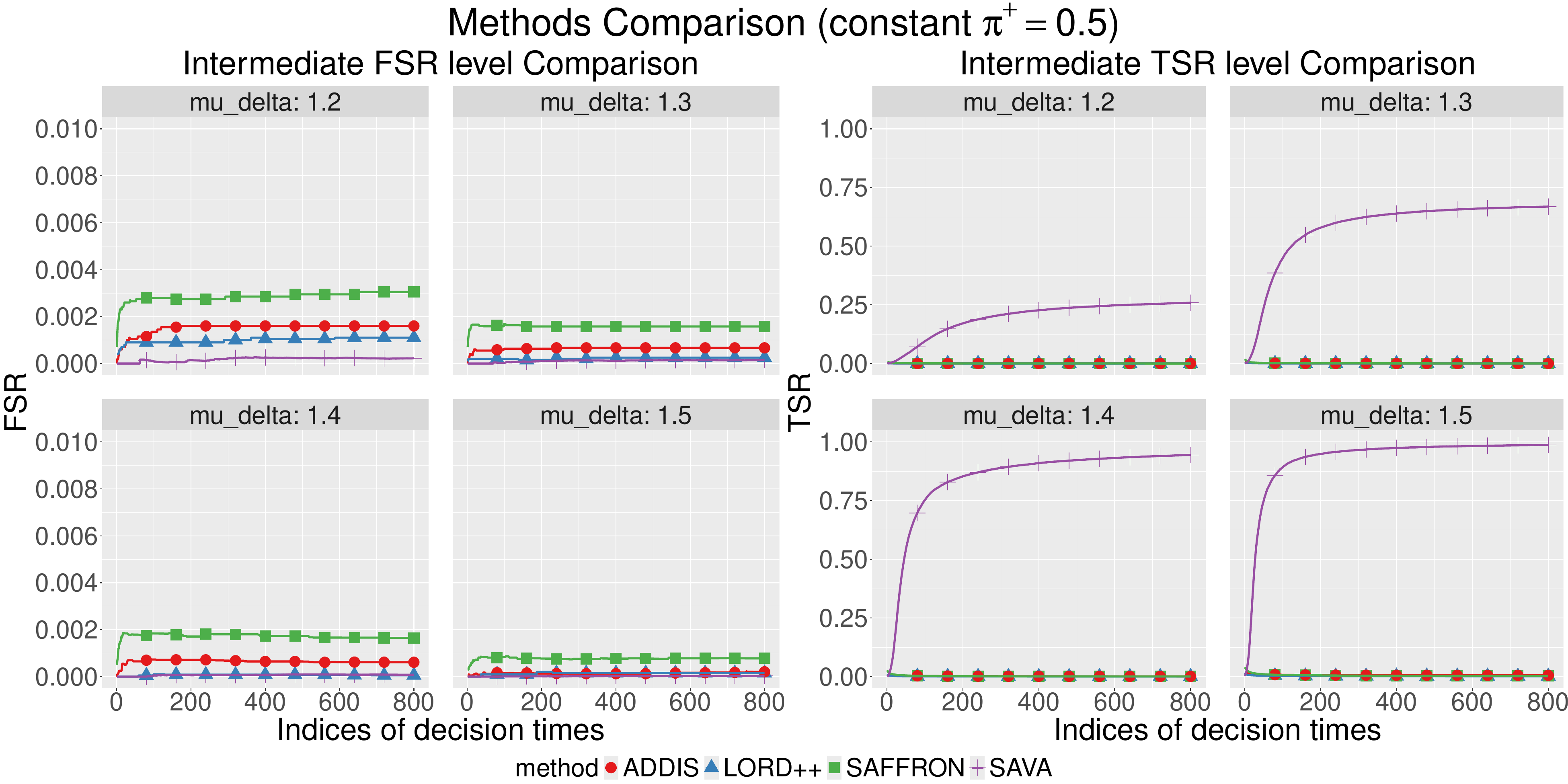}
		\caption{\small Methods with different $\mu_{\delta}$ under the Gamma model.}
		\label{fig:gamma-delta}
	\end{figure}

	We set the target FSR level to $0.05$ and set $T=3000$.  Assume that new tasks arrive with probability $p=1/3$. The following settings are considered: (i) Setting 1: fix $\pi^{+} = 0.5$ and vary $\mu_{\delta}$ from 1.2 to 1.5 with step size 0.1; (ii) Setting 2: fix $\mu_{\delta} = 1.3$ and vary $\pi^{+}$ from 0.2 to 0.8 with step size 0.2. Each experiment is repeated 1000 times. We report the average false selection proportion (FSP) and true selection proportion (TSP) at each decision time $t \in \mathbb T$ in Figure~\ref{fig:gamma-pi} and Figure~\ref{fig:gamma-delta}.
	
	While all methods exhibit substantial conservativeness, SAVA demonstrates notable power improvements over existing online testing rules. The conservativeness of both SAVA and the competing methods arises for reasons similar to those discussed in Section \ref{sec:generalcdf}. However, the SAVA framework allows continued sampling from active data streams, enabling the accumulation of stronger evidence over time and thereby improving its overall power.

\subsection{Results for Beta model}\label{subsec:beta-model}
Assume that $\theta^j = \mathrm A$ with probability $\pi^+$ and $\theta^j = \mathrm B$ with probability $1 - \pi^+$.
We consider the Beta distribution $F^j = \mathrm{Beta}(\alpha_j, \beta)$, where $\beta > 0$ is known and $\alpha_j > 0$ is unknown. 
The mean is
\(
\mu_j = \frac{\alpha_j}{\alpha_j + \beta}.
\)
Let $\alpha_0 > 0$ be a baseline parameter and define 
\(\mu_0 = \frac{\alpha_0}{\alpha_0 + \beta}.\)
The true state is defined as in \eqref{eq:true-state-setting}.
The data are generated according to \eqref{eq:simu-model-general} with
$F_A^j = \mathrm{Beta}(\beta_0(\mu_0 + \mu_{\delta})/(1-\mu_0-\mu_{\delta}), \beta_0)$, and $F_B^j = \mathrm{Beta}(\beta_0(\mu_0 - \mu_{\delta})/(1-\mu_0 + \mu_{\delta}), \beta_0)$, where $\mu_\delta > 0$ controls the signal strength. 
In the simulations, we set $\beta = 2$ and the baseline mean $\mu_0 = 1/2$, corresponding to $\alpha_0 = 2$.

To construct always-valid directional p-values, we first adopt a betting-based e-process motivated by \cite{waudby2020estimating}:
\[
	E_{t,\mathrm{beta}}^j(\lambda)=\prod_{t_0^j \le i \le t}\exp\!\left(\lambda (X_i^j - \mu_0)-\frac{\lambda^2}{8}\right),
	\]
	where $\lambda \in \mathbb{R}$ is a tuning parameter.
	This construction relies only on the boundedness of Beta random variables, and thus yields a valid e-process under the null via standard concentration arguments.
	Directional e-processes are defined as
	\[
	E^{j,A}_{t,\mathrm{beta}}(\lambda) = E_{t,\mathrm{beta}}^j(\lambda),
	\quad
	E^{j,B}_{t,\mathrm{beta}}(\lambda) = E_{t,\mathrm{beta}}^j(-\lambda).
	\]
	The corresponding always-valid directional p-values are defined as
	\begin{equation}
		\label{eq:e-pro-beta}
		\begin{aligned}
			p_{t,\mathrm{beta}}^{j,A}&=
			\min\!\left\{1,\left(\max_{t_0^j \le s \le t}
			E^{j,A}_{s,\mathrm{beta}}(\lambda)\right)^{-1}\right\}, \\
			p_{t,\mathrm{beta}}^{j,B}&=
			\min\!\left\{1,\left(\max_{t_0^j \le s \le t}
			E^{j,B}_{s,\mathrm{beta}}(\lambda)\right)^{-1}\right\},
		\end{aligned}
	\end{equation}
for any $\lambda>0$. In the simulations, we set $\lambda = 1$.
	
As an alternative, coin-betting methods \citep{Cle25} provide another class of e-processes with favorable optimality properties. 
Specifically, define
\begin{equation*}
	E_{t,\mathrm{coin}}^j(\lambda)=
	\prod_{t_0^j \le i \le t,\, i \in \mathcal{T}^j}\left(1 + \lambda (X_i^j - \mu_0)\right),
\end{equation*}
where \(
\lambda \in \left(0, \left(\max\{|\mu_0|, |1-\mu_0|\}\right)^{-1}\right) \).
This construction exploits the boundedness of $X_i^j \in [0,1]$ and yields a valid e-process under the null hypothesis.
Directional e-processes are defined analogously as
\[
E^{j,A}_{t,\mathrm{coin}}(\lambda) = E_{t,\mathrm{coin}}^j(\lambda),
\quad
E^{j,B}_{t,\mathrm{coin}}(\lambda) = E_{t,\mathrm{coin}}^j(-\lambda).
\] 
The corresponding always-valid directional p-values are defined as
\begin{equation}
	\label{eq:e-pro-coin}
	\begin{aligned}
		p_{t,\mathrm{coin}}^{j,A}&=
		\min\!\left\{1,\left(\max_{t_0^j \le s \le t}E^{j,A}_{s,\mathrm{coin}}(\lambda)\right)^{-1}\right\}, \\
		p_{t,\mathrm{coin}}^{j,B}&=
		\min\!\left\{1,\left(\max_{t_0^j \le s \le t}E^{j,B}_{s,\mathrm{coin}}(\lambda)\right)^{-1}\right\},
	\end{aligned}
\end{equation}
for any
\(
\lambda \in \left(0, \left(\max\{|\mu_0|, |1-\mu_0|\}\right)^{-1}\right) \). In the simulations, we set \(
\lambda = 0.9\left(\max\{|\mu_0|, |1-\mu_0|\}\right)^{-1} \).

For Beta model experiment 1, the SAVA algorithm is implemented using the always-valid directional p-values in \eqref{eq:e-pro-beta}.
 For Beta model experiment 2, the SAVA algorithm is implemented using the always-valid directional p-values in \eqref{eq:e-pro-coin}.
The sequence of test levels is computed via \eqref{eq:test-level} with $k=25$. 
The construction of p-values for online FDR rules is described in Section \ref{sec:online-pvalue-add}.

The target FSR level is $0.05$, the horizon is $T=3000$, and new tasks arrive with probability $p=1/3$.
The following settings are considered:  
\begin{enumerate}
	\item Beta model experiment 1.
	(i) Setting 1: Keep $\pi^{+} = 0.5$. Vary $\mu_{\delta}$ from 0.12 to 0.15 with step size 0.01; (ii) Setting 2: Keep $\mu_{\delta} = 0.14$. Vary values of $\pi^{+}$ from 0.2 to 0.8 with step size 0.2.
	\item Beta model experiment 2.
	(i) Setting 1: Keep $\pi^{+} = 0.5$. Vary $\mu_{\delta}$ from 0.04 to 0.07 with step size 0.01; (ii) Setting 2: Keep $\mu_{\delta} = 0.05$. Vary values of $\pi^{+}$ from 0.2 to 0.8 with step size 0.2.
\end{enumerate}
		
\begin{figure}[tbh!]
		\centering
		\includegraphics[width=0.90\linewidth]{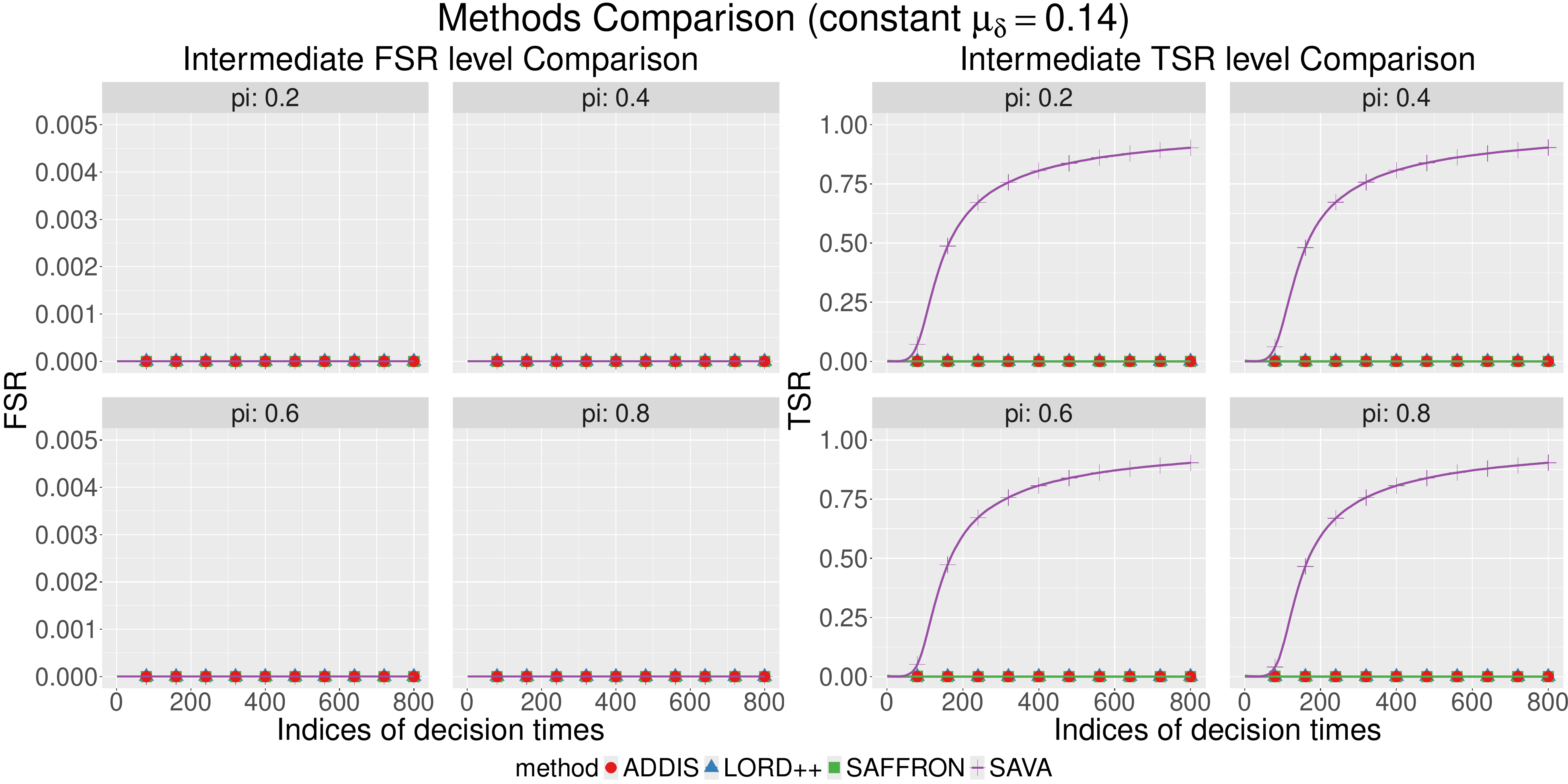}
		\caption{\small Methods with different $\pi^{+}$ under the Beta model experiment 1.}
		\label{fig:beta-pi}
	\end{figure}

\begin{figure}[tbh!]
		\centering
		\includegraphics[width=0.90\linewidth]{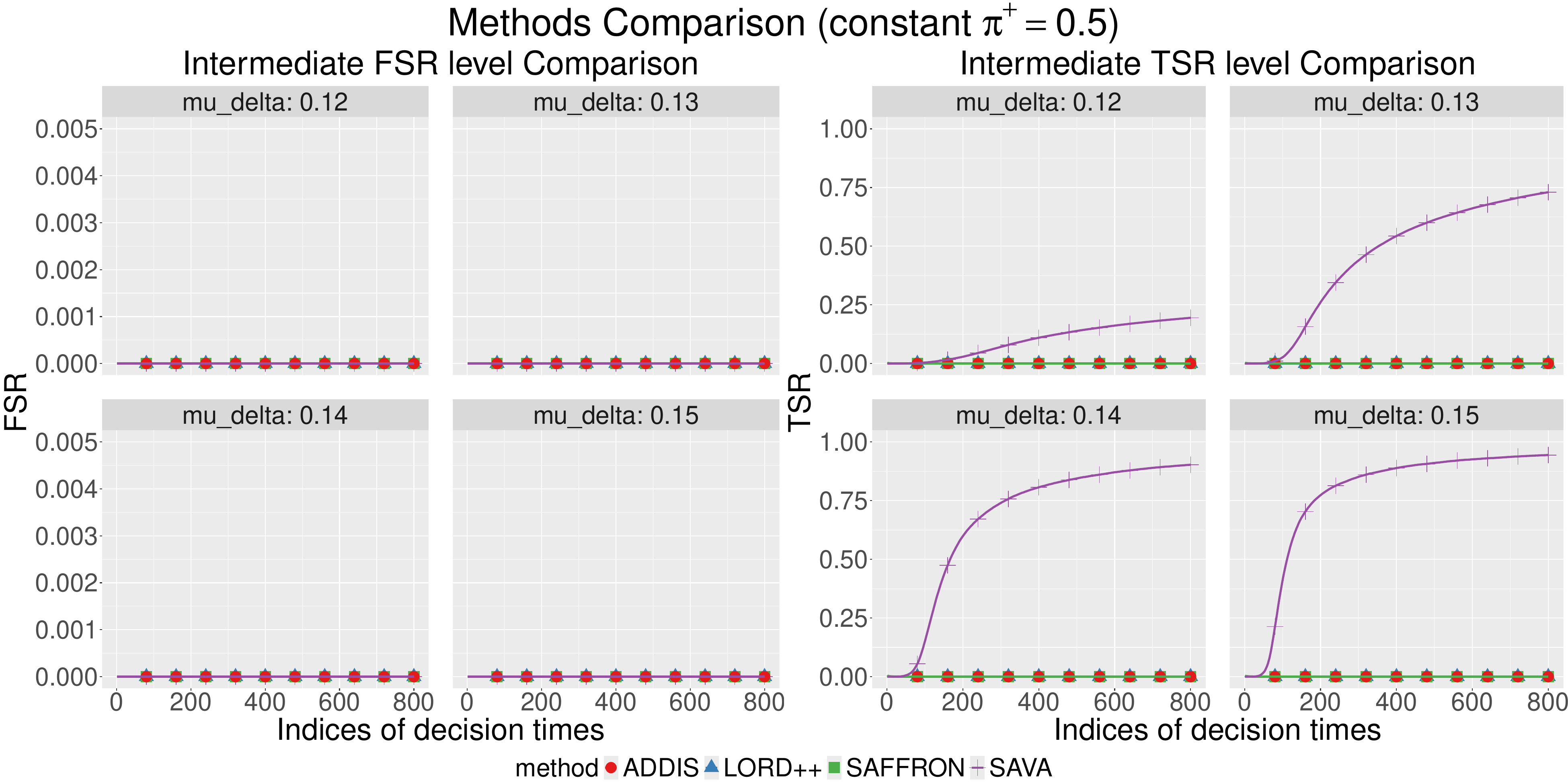}
		\caption{\small Methods with different $\mu_{\delta}$ under the Beta model experiment 1.}
		\label{fig:beta-delta}
	\end{figure}

\begin{figure}[tbh!]
		\centering
		\includegraphics[width=0.90\linewidth]{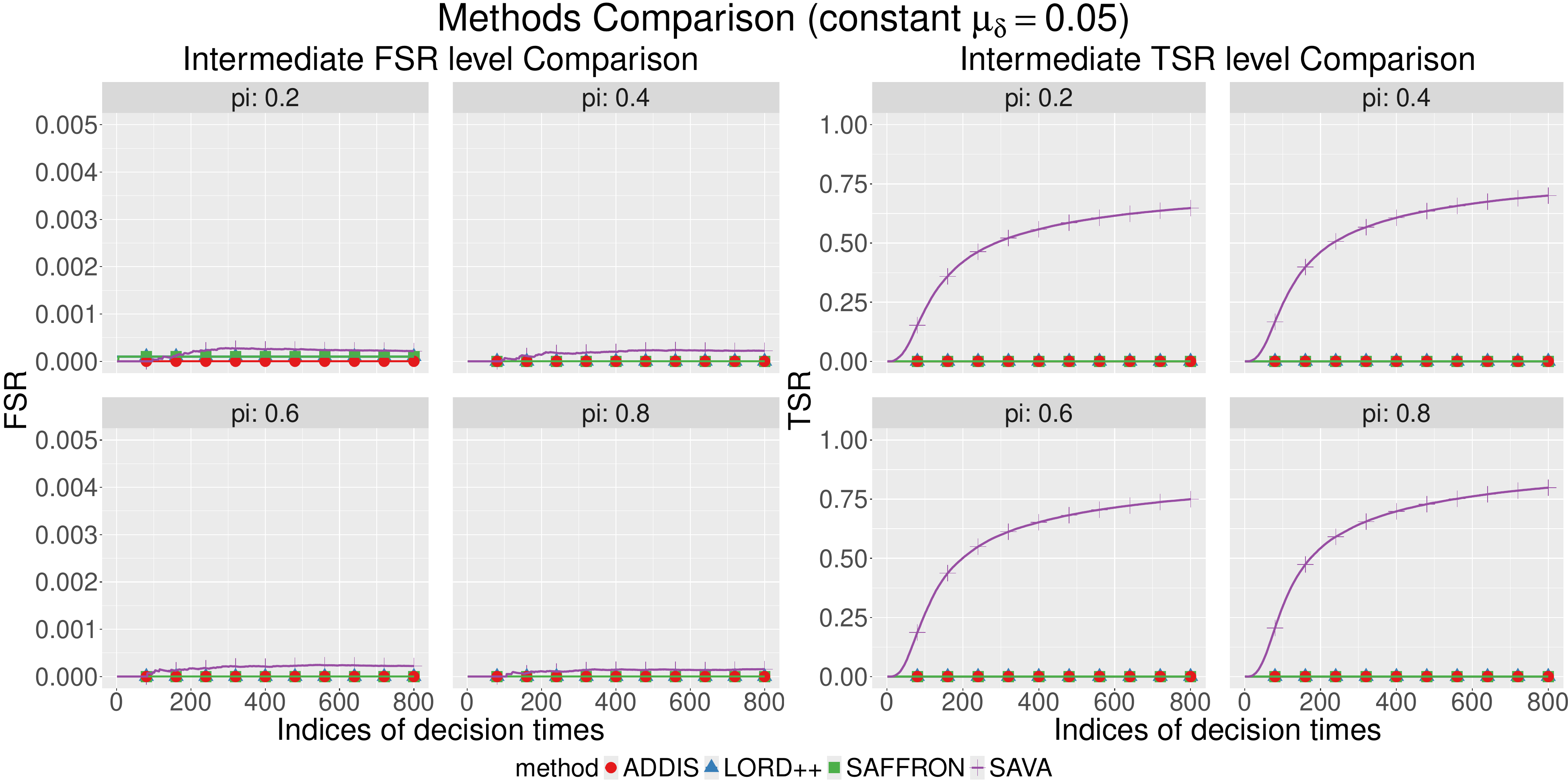}
		\caption{\small Methods with different $\pi^{+}$ under the Beta model experiment 2.}
		\label{fig:beta-pi-coin}
	\end{figure}

\begin{figure}[tbh!]
		\centering
		\includegraphics[width=0.90\linewidth]{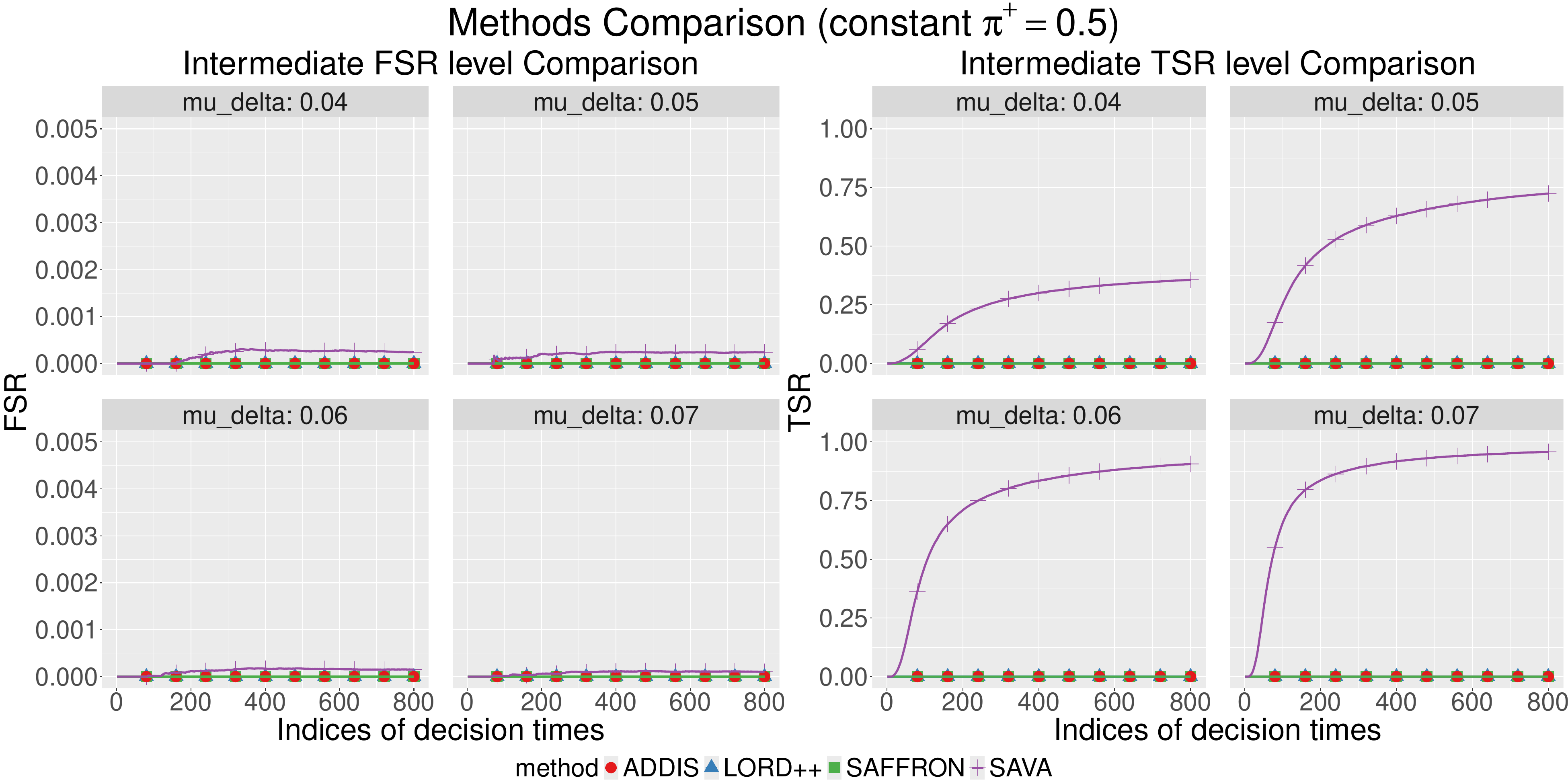}
		\caption{\small Methods with different $\mu_{\delta}$ under the Beta model experiment 2.}
		\label{fig:beta-delta-coin}
	\end{figure}

Each experiment is repeated 1000 times. We report the average false selection proportion (FSP) and true selection proportion (TSP) at each decision time for all methods in Figure~\ref{fig:beta-pi}, Figure~\ref{fig:beta-delta}, Figure~\ref{fig:beta-pi-coin} and Figure~\ref{fig:beta-delta-coin}. 

Across all settings, the procedures exhibit a notable degree of conservativeness, while SAVA consistently attains higher power than existing online testing rules. This behavior can largely be attributed to the same factors discussed in Section \ref{sec:generalcdf}. In addition, the results from Experiment 2 indicate that the choice of e-process construction can have a tangible impact on the power of the SAVA algorithm.

	\subsection{Results for sub-Gaussian distributions}\label{subsec:sub-gauss-model}
	Assume that $\theta^j = \mathrm A$ with probability $\pi^+$ and $\theta^j = \mathrm B$ with probability $1 - \pi^+$. 
	We consider a broad class of sub-Gaussian data streams such that
	\begin{equation}
		\label{eq:sub-gauss}
		\mathbb E \exp(\lambda(X_i^j-\mu_j))\leq 
		\exp\!\left(\frac{\lambda^2\sigma^2}{2}\right),
	\end{equation}
	for some $\sigma^2\leq\sigma_0^2$, where $\sigma_0$ is known. Let $\mu_0 > 0$ be a baseline parameter, and the true state is defined as \eqref{eq:true-state-setting}.	
	
	Given pre-specified $\mu_0 = 0$ and $\sigma_0 = 1$, the data are assumed to be generated according to \eqref{eq:simu-model-general} with following different distributions:
	\begin{itemize}
		\item Gaussian: \(
		F_A^j = N(\mu_0 + \mu_\delta, 1),\) and \(
		F_B^j = N(\mu_0 - \mu_\delta,1).
		\)
		\item Uniform: \(
		F_A^j = \mathrm{Unif}(\mu_0 + \mu_\delta - 1, \mu_0 + \mu_\delta + 1),\) and 
		\(
		F_B^j = \mathrm{Unif}(\mu_0 - \mu_\delta -1,\mu_0 - \mu_\delta+ 1).
		\)
		\item Drifted Bernoulli: Given $p_0 = 0.3$, \(F_A^j = \mu_0 + \mu_{\delta} + \mathrm{Ber}(p_0) - p_0\) and \(F_B^j = \mu_0 - \mu_{\delta} + \mathrm{Ber}(p_0) - p_0\).
	\end{itemize}
	It is easy to show that these distributions are sub-Gaussian with $\sigma\leq \sigma_0$.

	E-processes are constructed using the sub-Gaussian exponential supermartingale
	\[
	E_{t,\mathrm{sub}}^j(\lambda)=\prod_{t_0^j\le i\le t}\exp\!\left(\lambda(X_i^j-\mu_0)-\frac{\lambda^2\sigma_0^2}{2}\right).
	\]
	Directional e-processes are defined analogously
	\[
	E^{j,A}_{t,\mathrm{sub}}(\lambda)=E_{t,\mathrm{sub}}^j(\lambda),
	\quad
	E^{j,B}_{t,\mathrm{sub}}(\lambda)=E_{t,\mathrm{sub}}^j(-\lambda).
	\]
	In the same way, the desired always-valid directional p-values can be constructed as 
	\begin{equation}\label{eq:e-pro-sub}
		\begin{aligned}
			p_{t,\rm{sub}}^{j,A} &= \min\left\{1, \left(\max_{t_0^j \leq s \leq t} E^{j,A}_{s,\rm{sub}}(\lambda)\right)^{-1}\right\},\\
			p_{t,\rm{sub}}^{j,B} &= \min\left\{1, \left(\max_{t_0^j \leq s \leq t}  E^{j,B}_{s,\rm{sub}}(\lambda)\right)^{-1}\right\},
		\end{aligned}
	\end{equation}
	for any $\lambda>0$. In the simulations, we set $\lambda = 1$.
		
	The SAVA algorithm is implemented untilizing always-valid directional p-values \eqref{eq:e-pro-sub} with $\sigma_0 = 1$. The sequence of test levels for SAVA are computed via \eqref{eq:test-level} with $k=25$. The way of constructions of p-values for online FDR rules are presented in Section \ref{sec:online-pvalue-add}.
	The target FSR level is $0.05$, the horizon is $T=3000$, and new tasks arrive with probability $p=1/3$.
	The following settings are considered:  
		\begin{itemize}
			\item Gaussian model: 
			(i) Setting 1: Keep $\pi^{+} = 0.5$. Vary $\mu_{\delta}$ from 0.4 to 0.55 with step size 0.05; (ii) Setting 2: Keep $\mu_{\delta} = 0.5$. Vary values of $\pi^{+}$ from 0.2 to 0.8 with step size 0.2.
			
			\item Uniform model: 
			(i) Setting 1: Keep $\pi^{+} = 0.5$. Vary $\mu_{\delta}$ from 0.45 to 0.6 with step size 0.05; (ii) Setting 2: Keep $\mu_{\delta} = 0.5$. Vary values of $\pi^{+}$ from 0.2 to 0.8 with step size 0.2.

			\item Drifted Bernoulli model: (i) Setting 1: Keep $\pi^{+} = 0.5$. Vary $\mu_{\delta}$ from 0.5 to 0.8 with step size 0.1; (ii) Setting 2: Keep $\mu_{\delta} = 0.5$. Vary values of $\pi^{+}$ from 0.2 to 0.8 with step size 0.2.
		\end{itemize}
	
\begin{figure}[tbh!]
		\centering
		\includegraphics[width=0.90\linewidth]{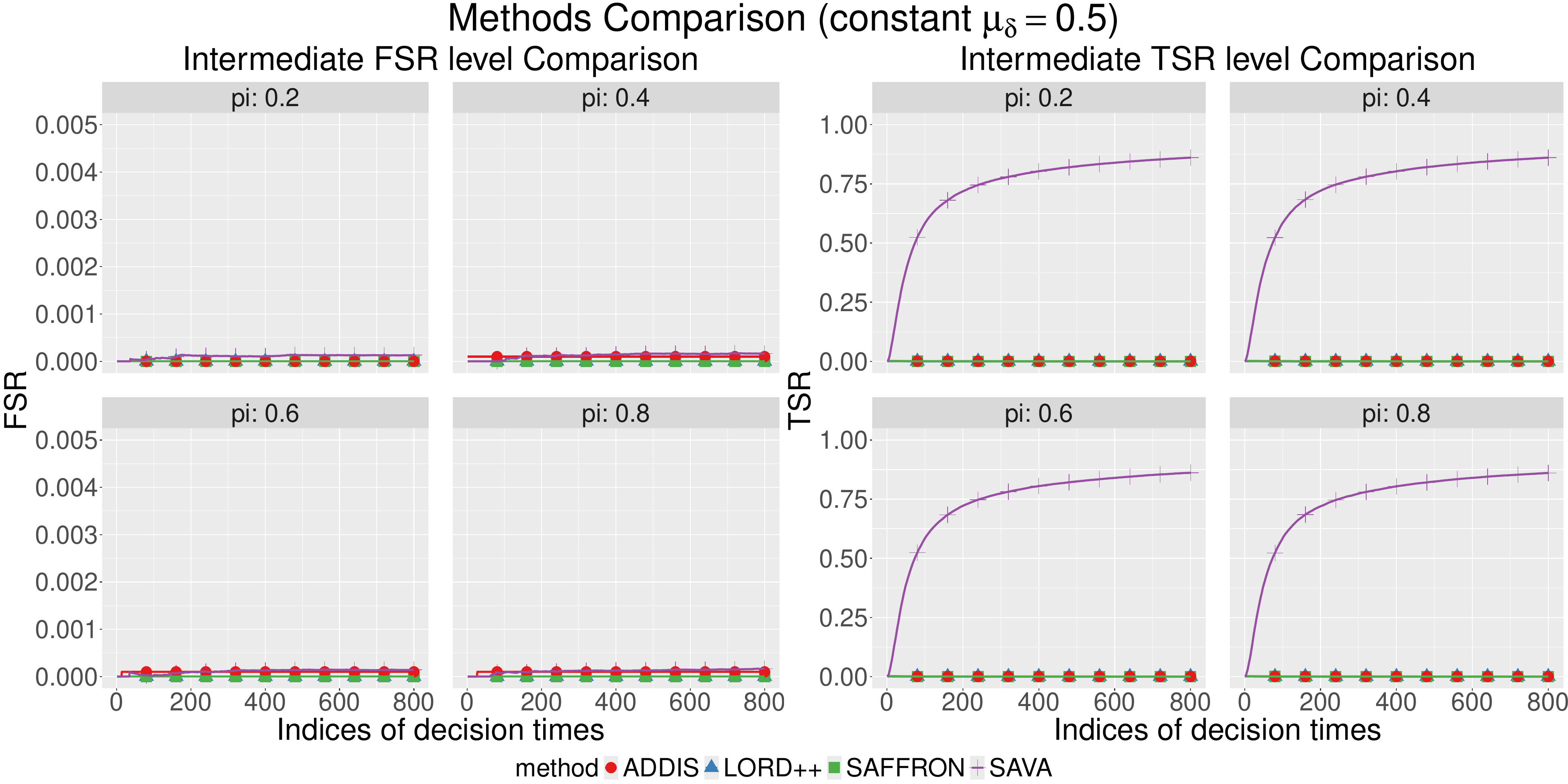}
		\caption{\small Methods with different $\pi^{+}$ under  Gaussian model.}
		\label{fig:sub-gauss-pi}
	\end{figure}

\begin{figure}[tbh!]
		\centering
		\includegraphics[width=0.90\linewidth]{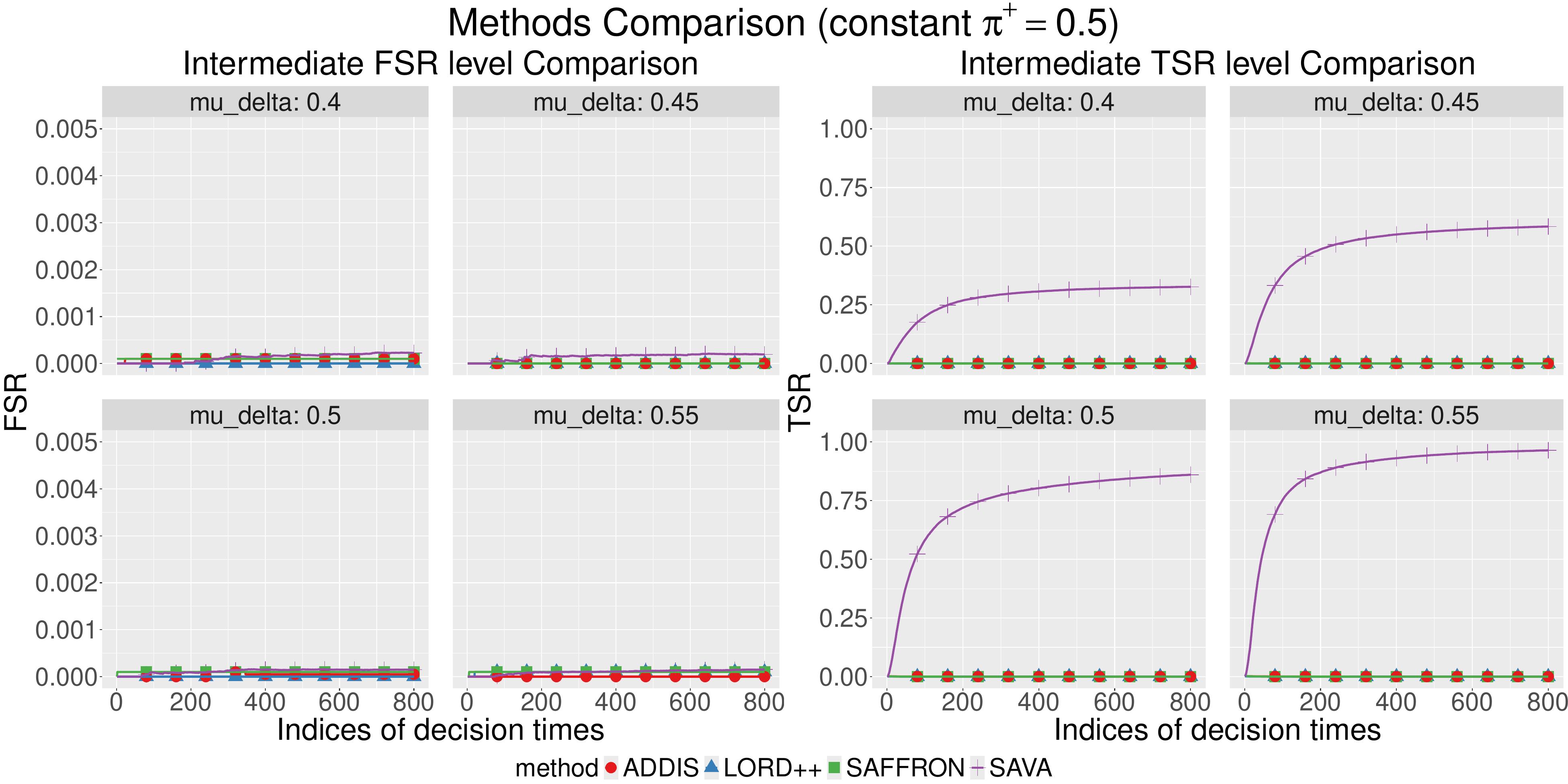}
		\caption{\small Methods with different $\mu_{\delta}$ under the Gaussian model.}
		\label{fig:sub-gauss-delta}
	\end{figure}

\begin{figure}[tbh!]
		\centering
		\includegraphics[width=0.90\linewidth]{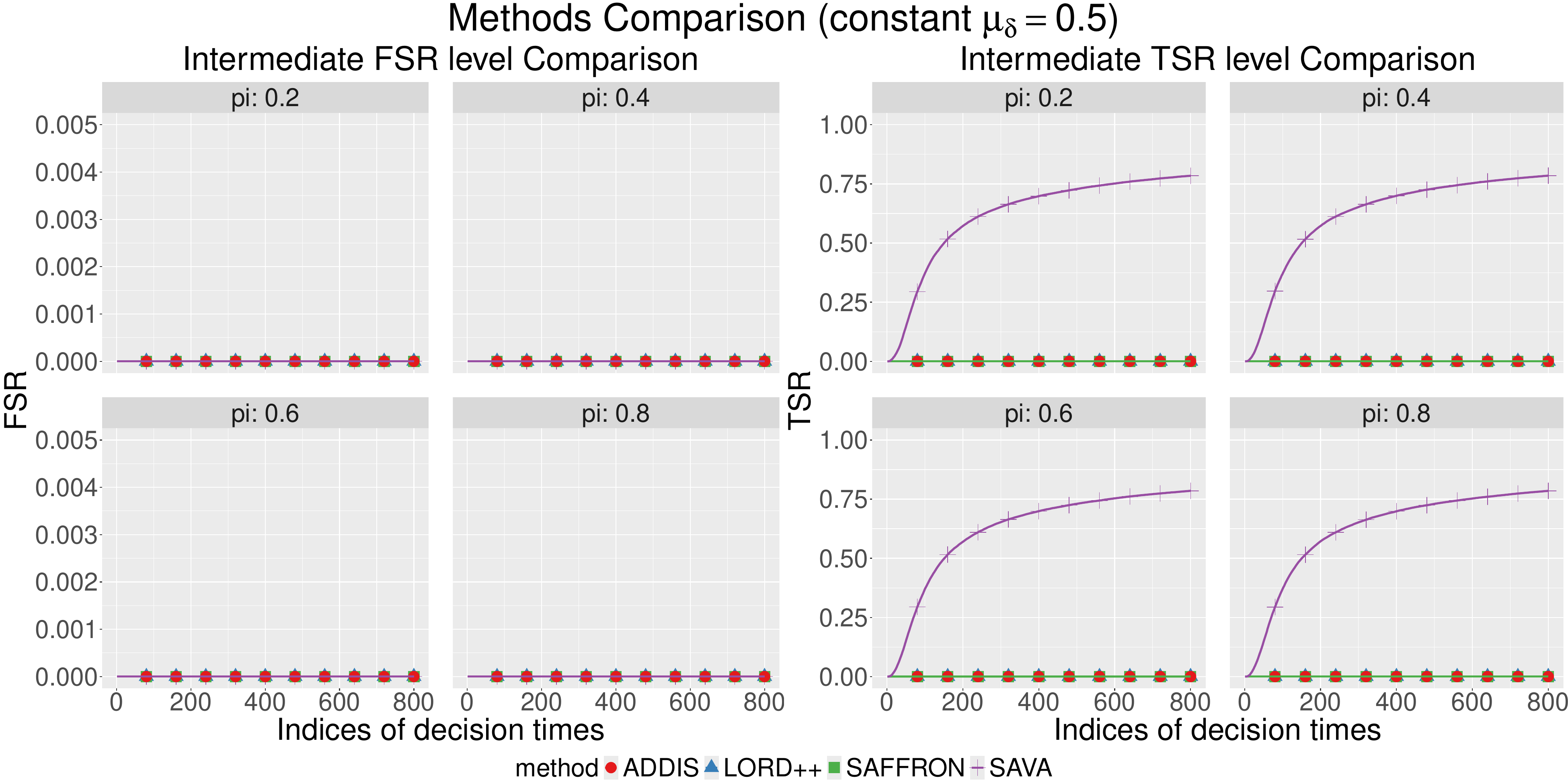}
		\caption{\small Methods with different $\pi^{+}$ under  Uniform model.}
		\label{fig:sub-unif-pi}
	\end{figure}

\begin{figure}[tbh!]
		\centering
		\includegraphics[width=0.90\linewidth]{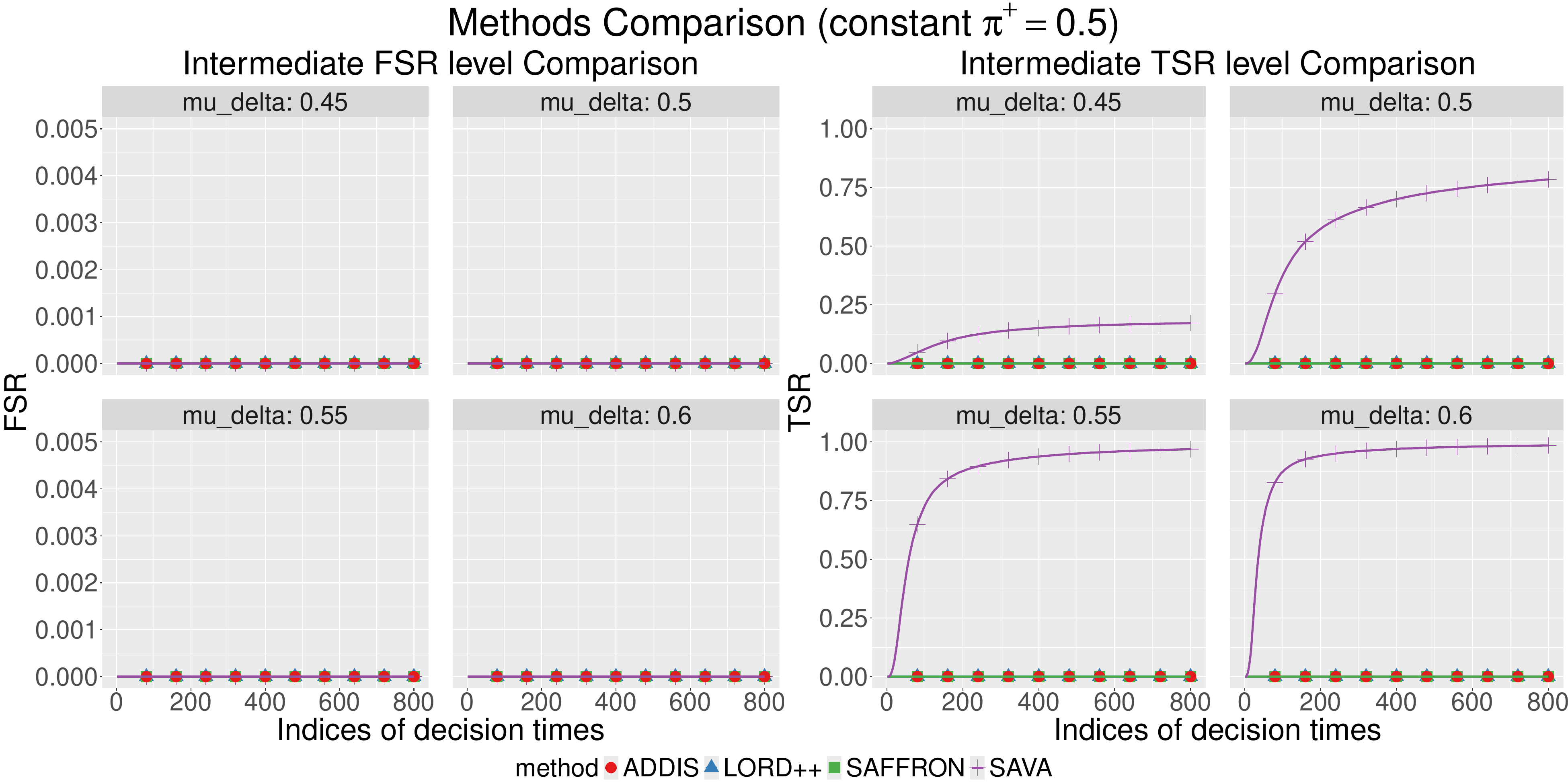}
		\caption{\small Methods with different $\mu_{\delta}$ under the Uniform model.}
		\label{fig:sub-unif-delta}
	\end{figure}

\begin{figure}[tbh!]
		\centering
		\includegraphics[width=0.90\linewidth]{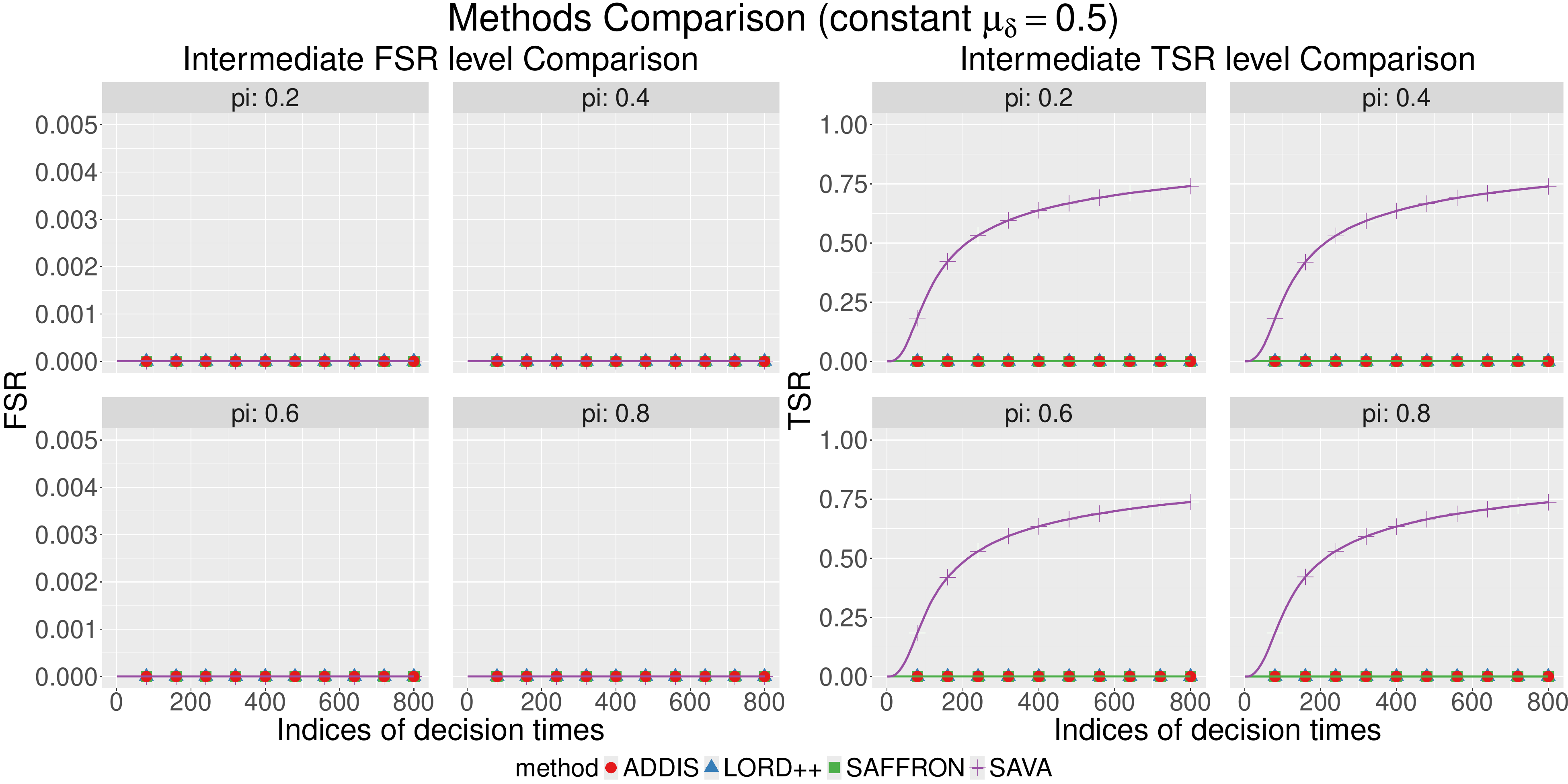}
		\caption{\small Methods with different $\pi^{+}$ under the drifted Bernoulli model.}
		\label{fig:sub-ber-pi}
	\end{figure}

\begin{figure}[tbh!]
		\centering
		\includegraphics[width=0.90\linewidth]{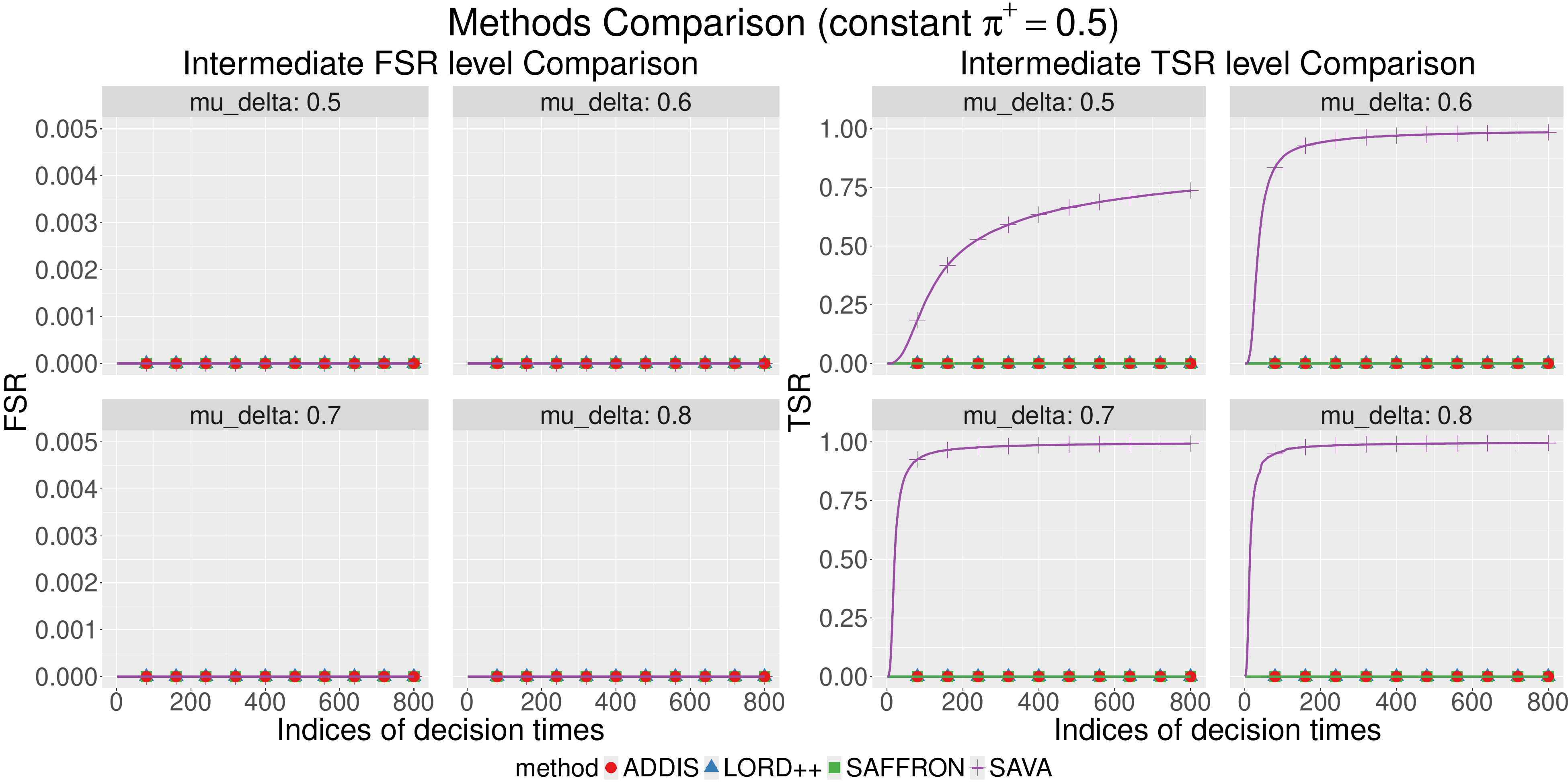}
		\caption{\small Methods with different $\mu_{\delta}$ under the drifted Bernoulli model.}
		\label{fig:sub-ber-delta}
	\end{figure}

	Each experiment is repeated 1000 times. We report the average false selection proportion (FSP) and true selection proportion (TSP) at each decision time in Figures~\ref{fig:gamma-pi}--\ref{fig:sub-ber-delta}. 
	
	While all methods exhibit substantial conservativeness, SAVA consistently achieves notable power improvements over existing online testing rules across all settings. The results demonstrate that SAVA can be effectively applied to a broad class of distributions even when the underlying distribution is unknown, provided that appropriate e-processes are employed. The conservativeness of both SAVA and the competing methods arises for reasons similar to those discussed in Section \ref{sec:generalcdf}. Nevertheless, the SAVA framework enables continued evidence accumulation over time, which in turn enhances its overall power.

	\subsection{Simulation results under dependent data streams}
	\label{subsec:simu-dep}
	
	We further investigate the empirical performances of SAVA and online FDR rules under dependent data streams. While the theoretical guarantees currently rely on independence across streams for SAVA, this experiment assesses the robustness of SAVA framework to moderate violations of this assumption.
	
	The simulation setup follows Section~\ref{subsec:online-rule}. The model is given by
	\[
	X_t^j|\theta^j\sim
	F^j=
	\mathbb I\{\theta^j=\mathrm A\}F_A^j
	+
	\mathbb I\{\theta^j=\mathrm B\}F_B^j.
	\]
	To induce dependence across streams, we let the mean of stream $j$ depend on the first observation of the preceding stream. Specifically, the following models are considered:
	\begin{itemize}
		
		\item Gaussian model
		\[
		F_A^j=N(\mu_j,1),
		\quad
		F_B^j=N(-\mu_j,1),
		\]
		where \(
	\mu_j=\rho \left|X_{t_0^{j-1}}^{j-1}\right| \).
		\item Gamma model
		\[
		F_A^j=\mathrm{Gamma}(2,2\mu_{j,\delta}),
		\quad
		F_B^j=\mathrm{Gamma}(2,2/\mu_{j,\delta}),
		\]
		where $\mu_{j,\delta}=1+\rho\left|X_{t_0^{j-1}}^{j-1}\right|$.
	\end{itemize}
	
	The remaining parameters follow the same configuration as the independent setting. The corresponding construction of always-valid directional p-values (for SAVA) and p-values (for online FDR rules) are summarized in Table.~\ref{tbl-pvalue-depend}.

	\begin{longtable}[]{@{}ccc@{}}
		\caption{Choices of test statistics for SAVA and online FDR rules.}\label{tbl-pvalue-depend}\tabularnewline
		\toprule\noalign{}
		Model & Method & Construction \\
		\midrule\noalign{}
		\endfirsthead
		\toprule\noalign{}
		Model & Method& Construction \\
		\midrule\noalign{}
		\endhead
		\bottomrule\noalign{}
		\endlastfoot
		Gamma & SAVA & \eqref{eq:e-pro-gamma} \\
		Gamma & Online FDR Rules & \eqref{eq:p-gamma} \\
		Gaussian & SAVA & \eqref{eq:e-pro-sub} \\
		Gaussian &  Online FDR Rules & \eqref{eq:pvalue-subgauss} \\
	\end{longtable}
	
	The sequence of test levels for SAVA is computed using \eqref{eq:test-level} with \(k = 25\). The implementations of online FDR rules follow in the same way as Section \ref{subsec:online-implement}.
	The target FSR level to be 0.05, and set $T = 3000$. The new task arrives with probability $p=1/3$. The following settings are considered:
	\begin{itemize}
		\item Gaussian model.
		(i) Setting 1: Keep $\pi^{+} = 0.5$. Vary $\rho$ from 0.2 to 0.5 with step size 0.1; (ii) Setting 2: Keep $\rho = 0.5$. Vary values of $\pi^{+}$ from 0.2 to 0.8 with step size 0.2.
		
		\item Gamma model.
		(i) Setting 1: Keep $\pi^{+} = 0.5$. Vary $\rho$ from 0.2 to 0.8 with step size 0.2; (ii) Setting 2: Keep $\rho = 0.5$. Vary values of $\pi^{+}$ from 0.2 to 0.8 with step size 0.2.
	\end{itemize}
	
\begin{figure}[tbh!]
		\centering
		\includegraphics[width=0.90\linewidth]{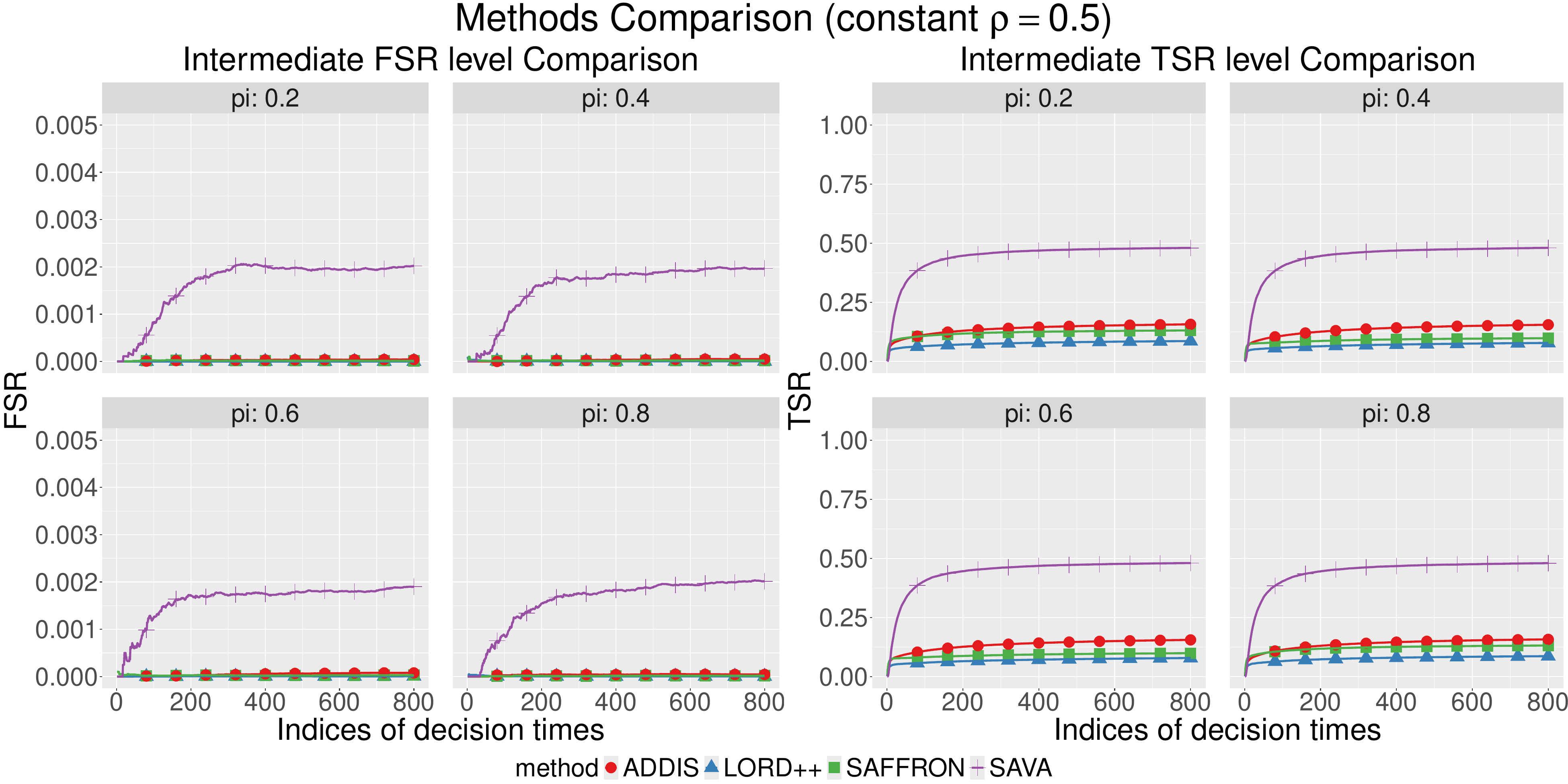}
		\caption{\small Methods with different $\pi^{+}$ under the Gaussian model with dependent data streams.}
		\label{fig:depend-gauss-pi}
	\end{figure}

\begin{figure}[tbh!]
		\centering
		\includegraphics[width=0.90\linewidth]{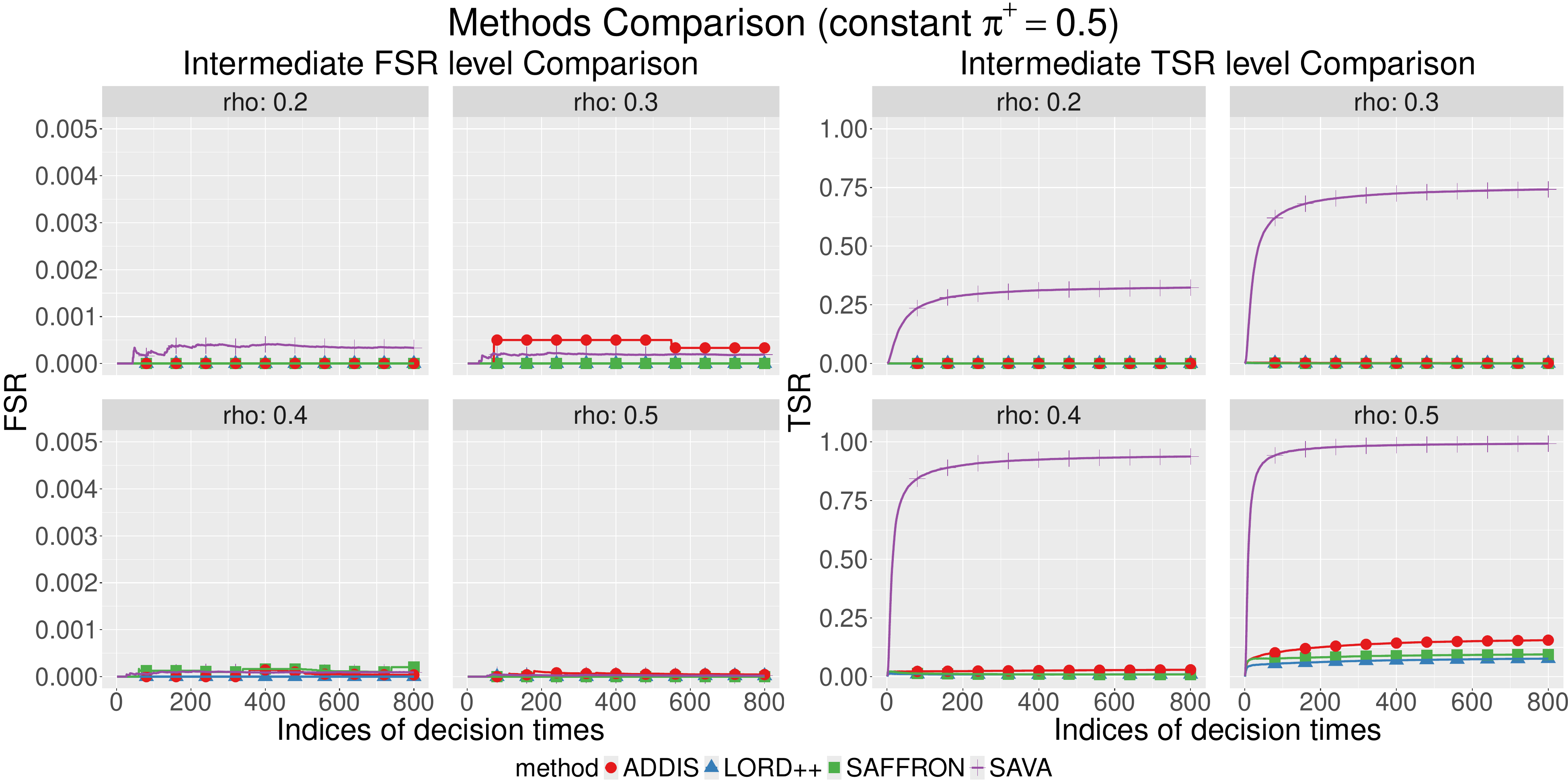}
		\caption{\small Methods with different $\rho$ under the Gaussian model with dependent data streams.}
		\label{fig:depend-gauss-delta}
	\end{figure}

\begin{figure}[tbh!]
		\centering
		\includegraphics[width=0.90\linewidth]{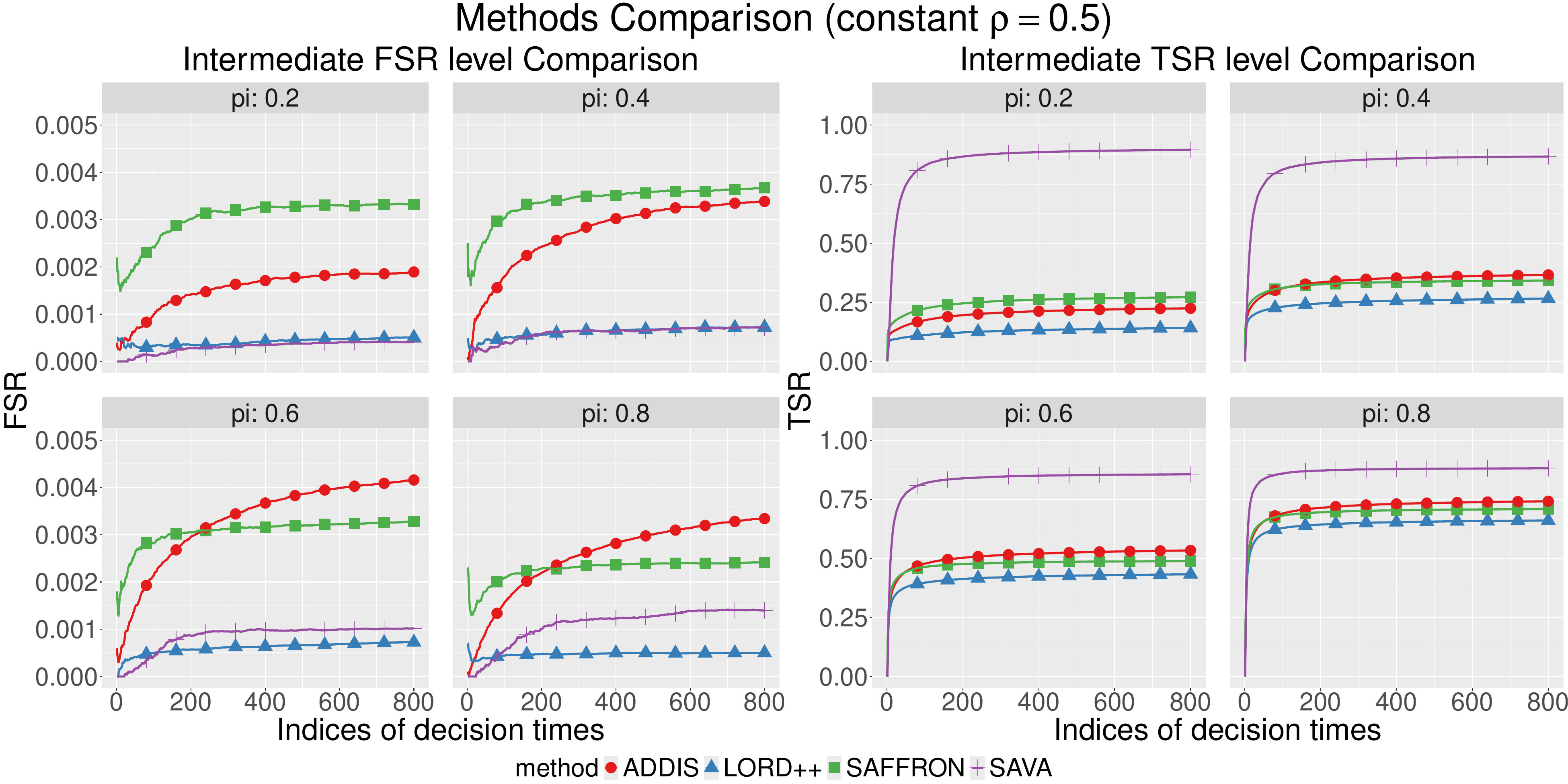}
		\caption{\small Methods with different $\pi^{+}$ under the Gamma model with dependent data streams.}
		\label{fig:depend-gamma-pi}
	\end{figure}

\begin{figure}[tbh!]
		\centering
		\includegraphics[width=0.90\linewidth]{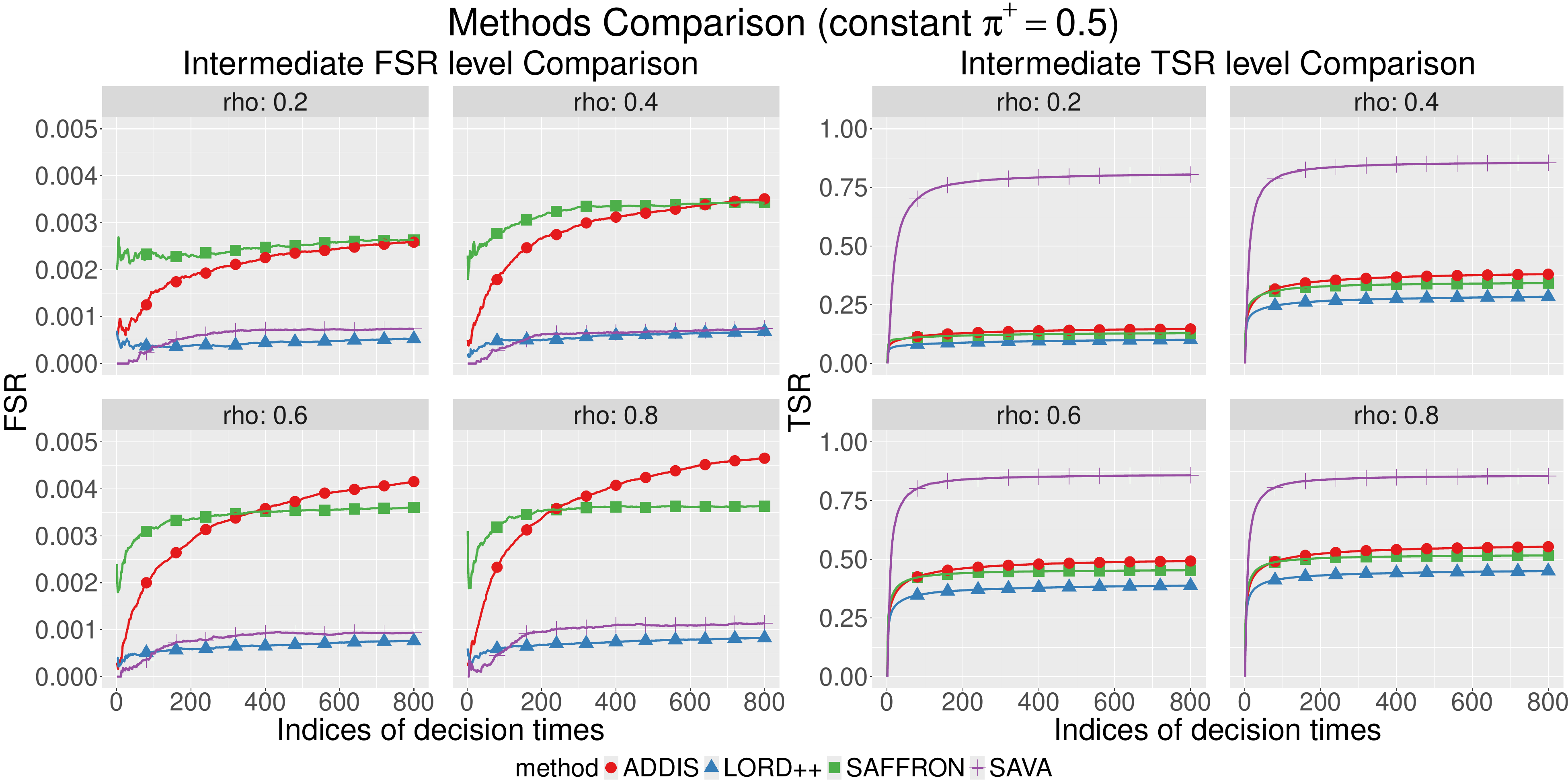}
		\caption{\small Methods with different $\rho$ under the Gamma model with dependent data streams.}
		\label{fig:depend-gamma-delta}
	\end{figure}

	For each configuration, we conduct 1000 simulation runs. The average FSP and TSP for all methods are summarized over time in Figure.~\ref{fig:depend-gauss-pi},  Figure.~\ref{fig:depend-gauss-delta},  Figure.~\ref{fig:depend-gamma-pi} and Figure.~\ref{fig:depend-gamma-delta}. 
	The results indicate that SAVA continues to exhibit stable empirical FSR control and strong power even in the presence of moderate dependence between streams. These findings suggest that the SAVA framework is robust beyond the independence setting, although establishing formal theoretical guarantees under dependence remains an important direction for future work.

	\subsection{Simulation results under model misspecification}
	\label{subsec:simu-mis}
	In this section, we investigate the empirical performance of SAVA when the e-processes are constructed under misspecified models. 
	We continue to adopt the data-generating model in \eqref{eq:simu-model-general}, where $\theta^j = \mathrm A$ with probability $\pi^+$ and $\theta^j = \mathrm B$ with probability $1 - \pi^+$.
	The following two representative model classes are considered:
	\begin{itemize}
		\item Uniform model. We consider a model with bounded support controlled by $a>0$:
		$F_A^j = \mathrm{Unif}(\mu_0+\mu_{\delta} - a, \mu_0+\mu_{\delta}  + a)$, and $F_B^j = \mathrm{Unif}(\mu_0 - \mu_{\delta} -a, \mu_0 - \mu_{\delta} +a)$. 

		\item Truncated Gaussian model. We consider
		$F_A^j = \mathrm{trN}(\mu_0+\mu_{\delta} ,1,-K, K)$ and $F_B^j = \mathrm{trN}(\mu_0 - \mu_{\delta} ,1,-K, K)$, where $K>0$ controls the truncation level.
	\end{itemize}
	
	To assess the impact of misspecification in the construction of always-valid directional p-values, we deliberately use incorrect parameters in the e-process construction. Specifically, for the Uniform model we implement \eqref{eq:e-pro-sub} with a misspecified sub-Gaussian parameter $\sigma_0$ that can be smaller than the true parameter. For the truncated Gaussian model, we use the construction in Section~\ref{subsec:avp-cons} with a misspecified truncation bound $K_0$ that is smaller than the true bound. Under these choices, the resulting e-processes are no longer guaranteed to be valid. We therefore investigate how the discrepancy between the true parameters (i.e., $\sigma$ in the Uniform model and $K$ in the truncated Gaussian model) and their misspecified counterparts (i.e., $\sigma_0$ and $K_0$) affects the performance of SAVA.
	
\begin{figure}[tbh!]
		\centering
		\includegraphics[width=0.99\linewidth]{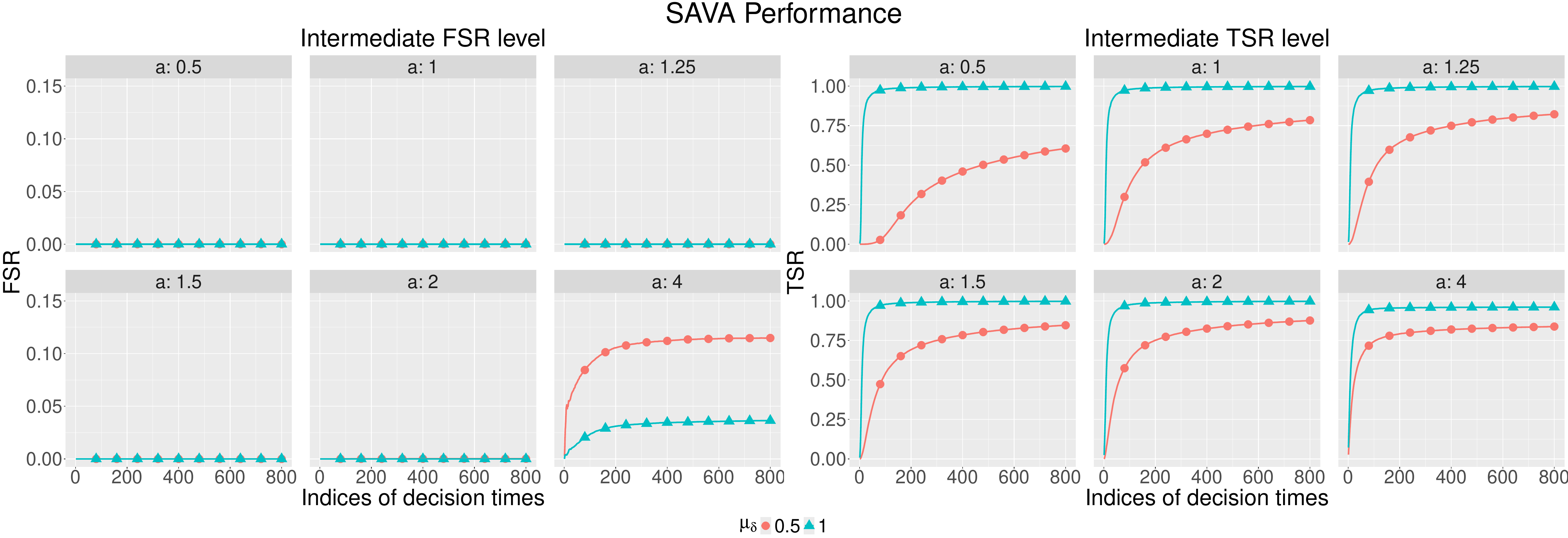}
		\caption{\small SAVA with different $a$ under the Uniform model.}
		\label{fig:unif-a}
	\end{figure}

\begin{figure}[tbh!]
		\centering
		\includegraphics[width=0.99\linewidth]{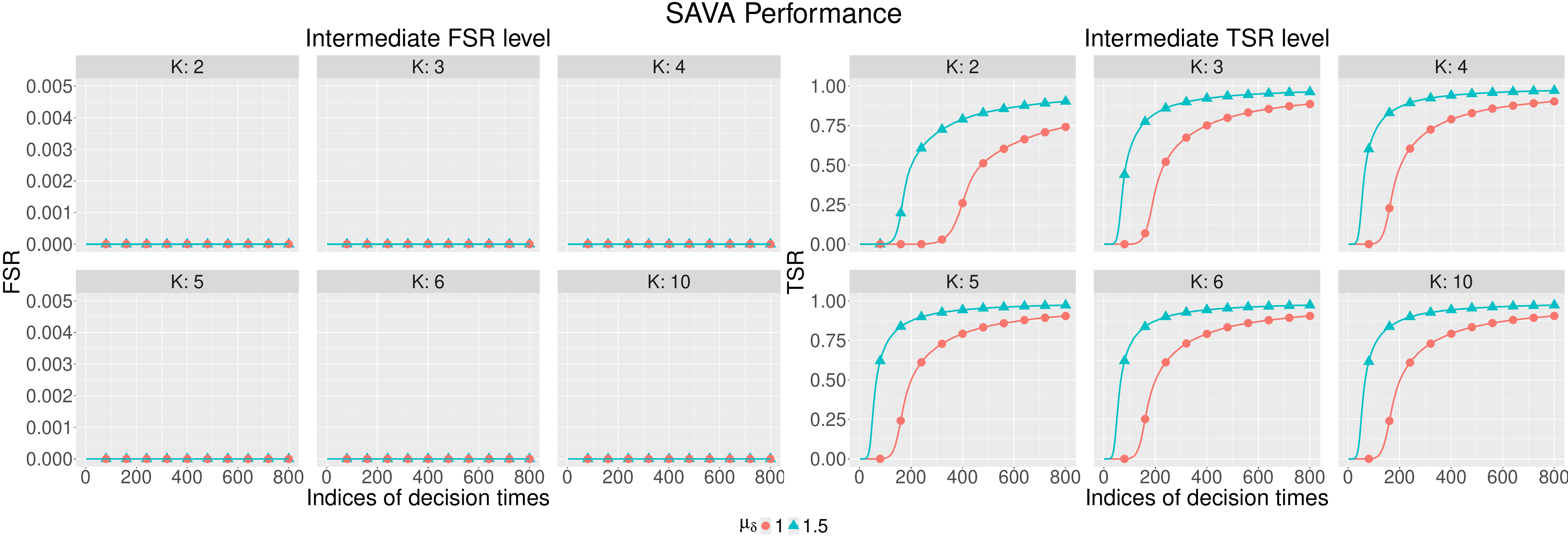}
		\caption{\small SAVA with different $K$ under the truncated Gaussian model.}
		\label{fig:truncgauss-k}
	\end{figure}

	When SAVA is implemented, the sequence of test levels is computed via \eqref{eq:test-level} with $k=25$. The target FSR level is set to $0.05$, the horizon is $T=3000$, and new tasks arrive independently with probability $p=1/3$. The detailed simulation settings are as follows:
	\begin{itemize}
		\item Uniform model. Set $\mu_0 = 0$ and $\pi^{+} = 0.5$. We fix $\sigma_0 = 3^{-1/2}$ in \eqref{eq:e-pro-sub}. Vary $a\in \{0.5,1,1.25,1.5,2,4\}$ and $\mu_{\delta}\in \{0.5,1\}$.
		
		\item Truncated Gaussian model. Set $\mu_0 = 0$ and $\pi^{+} = 0.5$. We fix $K_0 = 2$ in the construction of Section~\ref{subsec:avp-cons}. Vary $K\in \{2,3,4,5,6,10\}$ and $\mu_{\delta}\in \{1, 1.5\}$.
	\end{itemize}
	Each experiment is repeated 1000 times. We report the average FSP and TSP at each decision time in Figure~\ref{fig:unif-a} and Figure~\ref{fig:truncgauss-k}. 
	
	Overall, SAVA demonstrates strong empirical robustness under model misspecification. To better understand this behavior, we examine the impact of misspecification in specific models.
	
	In the Uniform model, note that for the distribution $\mathrm{Unif}(\mu - a, \mu + a)$ (for any $\mu\in\mathbb R$), a standard calculation shows that the sub-Gaussian parameter satisfying \eqref{eq:sub-gauss} is $\sigma = 3^{-1/2}a$. Hence, the choice $\sigma_0 = 3^{-1/2}$ in the Uniform model is valid only when the true parameter $a \leq 1$. When $a>1$, the resulting always-valid directional p-values are no longer theoretically valid. 
	This observation is consistent with the simulation results for the Uniform model: when the true parameter $a$ exceeds the assumed value, the resulting misspecified e-processes lead to a gradual increase in the FSP as $a$ grows. Notably, a pronounced inflation is observed only under severe misspecification (e.g., when $a$ is several times larger than the assumed value). 
	
	A similar analysis applies to the truncated Gaussian model. When the true parameter $K > K_0$, we find that the FSP remains stable across all configurations, and SAVA continues to exhibit valid behavior empirically. This robustness is primarily due to the inherent conservativeness of the construction in Section~\ref{subsec:avp-cons}, as well as the conservativeness of the estimator $\widehat{\mathrm{FSR}}_t$ used in computing test levels.
	
	In summary, these results suggest that SAVA is reasonably robust to moderate model misspecification, with degradation in performance occurring only under substantial deviations from the assumed model.

	\subsection{Construction of p-values for online FDR rules under different models}\label{sec:online-pvalue-add}
	In this section, we summarize the construction of p-values used in the online FDR rules implemented in previous sections. These constructions are based on classical test statistics under the models described earlier. 
	
	For each task $j \in \mathbb{J}$, recall that $t_0^j$ and $t_j$ denote the initial time and decision time for task $j$, respectively, as illustrated in the simulation setup in Section \ref{subsec:online-rule}. Define $T_j = \sum_{i = t_0^j}^{t_j} X_i^j$ as the cumulative sum of observations for task $j$, and let $n_j$ denote the corresponding sample size. The p-value constructions are given as follows.
	
	\begin{itemize}
		\item Gamma model.  
		Consider the setting in Section \ref{subsec:gamma-model}. By properties of the exponential family \citep{CasBer24}, the statistic $2T_j/\sigma$ follows a $\chi^2_{2k n_j}$ distribution. The p-values under $\theta^j = \mathrm{A}$ and $\theta^j = \mathrm{B}$ are defined as
		\begin{equation}
			\label{eq:p-gamma}
			\begin{aligned}
				p^{j,A}_{\mathrm{gam}} &= \mathrm{Pr}(\chi^2_{2kn_j}\geq 2t_{j,\mathrm{obs}}/\sigma_0), \\
				p^{j,B}_{\mathrm{gam}} &= \mathrm{Pr}(\chi^2_{2kn_j}\leq 2t_{j,\mathrm{obs}}/\sigma_0),
			\end{aligned}
		\end{equation}
		where $t_{j,\mathrm{obs}}$ denotes the observed value of $T_j$.
		\item Beta model.
		Consider the setting in Section \ref{subsec:beta-model}. We use Wilcoxon's signed rank test based on newly collected samples between two consecutive decision times for each task.
		\item Sub-Gaussian model.
		Consider the setting in Section \ref{subsec:sub-gauss-model}. Based on standard tail bounds for sub-Gaussian random variables \citep{Ver18}, we construct p-values as
		\begin{equation}
			\label{eq:pvalue-subgauss}
			\begin{aligned}
				p^{j,A}_{\mathrm{sub}} &= \begin{cases}
					\exp\left(-\frac{n_j(\bar{\pmb{X}}^j - \mu_0)^2}{2\sigma^2}\right), & \bar{\pmb{X}}^j > \mu_0, \\
					1, & \bar{\pmb{X}}^j \leq \mu_0.
				\end{cases} \\
				p^{j,B}_{\mathrm{sub}} &= \begin{cases}
					\exp\left(-\frac{n_j( \bar{\pmb{X}}^j - \mu_0)^2}{2\sigma^2}\right), & \bar{\pmb{X}}^j < \mu_0, \\
					1, & \bar{\pmb{X}}^j \geq \mu_0,
				\end{cases} 
			\end{aligned}
		\end{equation}
		where $ \bar{\pmb{X}}^j = T_j/n_j$ is the sample mean for task $j$.
	\end{itemize}

	\section{SAVA Algorithms for Alternative Setups}\label{supp:methods}
	
	\setcounter{algorithm}{0}
	\renewcommand\thealgorithm{C.\arabic{algorithm}}
	\setcounter{equation}{0}
	\renewcommand\theequation{C.\arabic{equation}}
	\setcounter{theorem}{0}
	\renewcommand\thetheorem{C.\arabic{theorem}}
	\setcounter{proposition}{0}
	\renewcommand\theproposition{C.\arabic{proposition}}
	\setcounter{figure}{0}
	\renewcommand\thefigure{C.\arabic{figure}}
	
	\setcounter{condition}{0}
	\renewcommand\thecondition{C.\arabic{condition}}
	\setcounter{remark}{0}
	\renewcommand\theremark{C.\arabic{remark}}

	\subsection{Conventional hypothesis testing setup}\label{subsec:classical-setup}

	We present a simplified version of the SAVA algorithm designed for the conventional setup where a specific direction is specified as the null hypothesis. Let \(\theta^j\) denote whether a new treatment outperforms a pre-specified benchmark \(c_j \geq 0\):
	\begin{equation*}
		\theta^j=\left\{\begin{array}{ll}
			\text{A},& \text{ if }\mu^j>c_j\\
			\text{B},& \text{ if }\mu^j\leq c_j\\
		\end{array}\right..
	\end{equation*}
	Assuming A is the arm of interest and B is the default arm corresponding to the null hypothesis, with \(\mu_A^j\) and \(\mu_B^j\) denoting their effect sizes. The conventional A/B testing problem can be recovered by setting \(\mu^j = \mu_A^j - \mu_B^j\) and \(c_j = 0\), with the goal of identifying \(\theta^j = \text{A}\). Using notations in Section \ref{subsec:SIframework}, the decision rule is given by:
	\begin{equation}
		\label{eq:decision-classical}
		\delta_t^j(p_t^{j,A}, b_j, t_0^j) = \left\{ \begin{array}{ll}
			\text{A}, & \text{if } p_t^{j,A} \leq \alpha_t^{j,A};\\
			\text{C}, & \text{if } p_t^{j,A} > \alpha_t^{j,A}, \text{ and } t - t_0^j < b_j, \\
			\text{D}, & \text{if } p_t^{j,A} > \alpha_t^{j,A}, \text{ and } t - t_0^j \geq b_j.
		\end{array}\right.
	\end{equation}
	
	In selective inference, we focus solely on the set of selected units up to \(t \in \mathbb{T}\):
	$
	\mathcal{S}_t^A= \{j \in \mathbb{J} : t_0^j \leq t, \delta_t^j = \mathrm{A}\}.
	$
	Define \(\mbox{FSR}^A\) as the expected proportion of the FSP:
	$
	\mathrm{FSR}_t^A = \mathbb{E}\{\mathrm{FSP}_t^A(\pmb{\delta}_t)\},
	$
	where 
	$\mathrm{FSP}_t^A(\pmb{\delta}_t) = \dfrac{\sum_{j: t_0^j \leq t} \mathbb{I}\{\delta_t^j \neq \theta^j, j \in \mathcal{S}_t^A\}}{|\mathcal{S}_t^A| \vee 1}, \quad t \in \mathbb{T}.
	$
	
	The SAVA algorithm employs an alpha-investing rule based on a conservative estimate of the FSR: 
	\begin{equation}
		\label{eq:fsrhatonesample}
		\widehat{\mathrm{FSR}}^{A}_t = \dfrac{\sum_{j: t_0^j \leq t} \bar{\alpha}_t^{j,A}}{|\mathcal{S}^{A}_t| \vee 1},
	\end{equation}
	where \(\bar{\alpha}_t^{j,A} = \max_{s \leq t, s \in \mathbb{T}}\{\alpha_s^{j,A}\}\). The following two conditions serve as guiding principles for designing alpha-investing rules.
	
	\begin{condition}
		\label{ass:classical1}
		The set of test levels $\{\alpha_t^{j,A}:j\in\mathbb J, t\in\mathbb T\}$ satisfies $\widehat{\mathrm{FSR}}^A_t \leq \alpha^A$ for all $j\in\mathbb J$ and $t\in\mathbb T$.
	\end{condition}
	
	\begin{condition}
		\label{ass:classical2}
		The test level $\alpha_t^{j,A}$ is $\mathcal G_t^{1:(j-1)}$-measurable.
	\end{condition} 
	
	The SAVA algorithm under the classical setup employs the following alpha-investing rule, with its operations summarized in Algorithm \ref{alg:SAVA-classical}:
	\begin{equation}
		\label{eq:level-classical}
		\alpha_t^{j,A} = \frac{\alpha}{k}\left[\mathbb{I}\{j\leq k\} + N_{t,j-}^A(k)\right],
	\end{equation}
	where $N_{t,j-}^A(k) \coloneqq |\{n\in\mathbb N: n\geq 2, j-k \leq I_n(\mathcal{S}^A_{t,j-}) \leq j-1 \}|$, $\mathcal{S}^A_{t,j-} = \{i\in\mathbb J: t_0^i<t_0^j, \delta_t^i = \text{A} \}$ and $I_n(\mathcal{S}^A_{t,j-})$ is the $n$-th smallest index in $\mathcal{S}^A_{t,j-}$. 
	
	\begin{algorithm}
		\caption{The SAVA framework in classical setup}\label{alg:SAVA-classical}
		\vskip6pt
		\begin{algorithmic}
			\State \textbf{Input}: a grid of decision time $\mathbb T$, a target FSR level $\alpha^A$, a method for computing always-valid directional p-values, a tuning parameter $k$, and pre-specified tolerance durations $\{b_j\}_{j\in\mathbb J}$.
			
			\State \textbf{For each} $t_i \in \mathbb{T}$, \textbf{do:}
			\State \quad \textbf{Step 1}: Update the index set of active tasks $\mathcal A_{t_{i}}$ defined in \eqref{eq:active-set}.
			\State \quad \textbf{Step 2}: Calculate always-valid directional p-values for tasks in $\mathcal A_{t_i}$ according to the given method.
			\State \quad \textbf{Step 3}: \textbf{For} $j=r_{t_{i}}(1),\dots, r_{t_{i}}(|\mathcal A_{t_{i}}|)$, \textbf{do}:
			\State \quad \quad \textbf{Step 3.1}: Calculate test levels by \eqref{eq:level-classical}.
			\State \quad \quad \textbf{Step 3.2}: Obtain the decision $\delta_{t_i}^j$ according to decision rule \eqref{eq:decision-classical}.
			\State \quad \textbf{Step 4}: For tasks $j$ such that $\delta_{t_{i-1}}^j \neq \text{C}$, update $\delta_{t_i}^j = \delta_{t_{i-1}}^j$.
			\State \quad \textbf{Output}: decision-making states $\{\delta_{t_i}^{j}: j\in \mathcal A_{t_i}\}$. 
			\State \textbf{End for}
		\end{algorithmic}
	\end{algorithm}

	\begin{condition}
		\label{ass:classical3} $\alpha_t^{j,A}$ is non-decreasing in $(\mathbb I\{\delta_t^i = \mathrm A\})_{i=1}^{j-1}$.
	\end{condition}
	
	The anytime validity of the SAVA algorithm in the classical setup can be established by integrating Proposition \ref{prop:test-levels} with Theorem \ref{thm:classical} below. The proofs, which are similar to those of the theorems in the main text, are therefore omitted. 
	
	\begin{proposition}\label{prop:test-levels}
		The proposed test levels in \eqref{eq:level-classical} satisfy Conditions \ref{ass:classical1}-\ref{ass:classical3}.
	\end{proposition}
	
	\begin{theorem}\label{thm:classical}
		Assume that the samples from different data streams are independent. 
		
		(a) If test levels $\alpha_t^{j,A}$ satisfy Conditions \ref{ass:classical1}-\ref{ass:classical3}, then $\mathrm{FSR}_t^A\leq \alpha^A$ for all $t\in\mathbb T$.  
		
		(b) If only Conditions \ref{ass:classical1}-\ref{ass:classical2} hold, then $\mathrm{mFSR}^A_t\leq \alpha^A$ for all $t\in\mathbb T$.
	\end{theorem}

	\subsection{Arm-specific error constraints}\label{subsec:dis-SAVA}
	
	In practical scenarios where it is advantageous to differentiate between the two arms with varying tolerance levels for error rates, we can impose two distinct constraints: \(\mathrm{FSR}^A_t \leq \alpha^A\) and \(\mathrm{FSR}^B_t \leq \alpha^B\). This section discusses how the SAVA framework may be adjusted to handle this new scenario. Consider the class of decision rules defined in \eqref{eq:decisionrule}. The SAVA algorithm employs the following conservative estimates of the arm-specific FSRs to bound the true FSR levels: 
	\begin{equation}\label{distinctfsrhat}
		\widehat{\mathrm{FSR}}^A_t = \dfrac{\sum_{j:t_0^j\leq t} \bar{\alpha}_t^{j,A}}{|\mathcal S_t^{A}| \vee 1},\quad \widehat{\mathrm{FSR}}^B_t = \dfrac{\sum_{j:t_0^j\leq t} \bar{\alpha}_t^{j,B}}{|\mathcal S_t^{B}| \vee 1},
	\end{equation}
	where $\mathcal S_t^A = \{j\in\mathbb J: \delta_t^j = \mathrm A\}$ and $\mathcal S_t^B = \{j\in\mathbb J: \delta_t^j = \mathrm B\}$, \(\bar{\alpha}_t^{j,A} = \max\{\alpha_s^{j,A}: s \leq t, s \in \mathbb{T}\}\), and \(\bar{\alpha}_t^{j,B} = \max\{\alpha_s^{j,B}: s \leq t, s \in \mathbb{T}\}\). We modify Condition \ref{assump1} as follows: 
	\begin{condition}\label{assump4}
		The set of test levels $\{(\alpha_t^{j,A},\alpha_t^{j,B}):j\in\mathbb J, t\in\mathbb T\}$ satisfies $\widehat{\mathrm{FSR}}_t^A\leq \alpha^A$ and $\widehat{\mathrm{FSR}}_t^B\leq \alpha^B$ for all $j\in\mathbb J$ and $t\in\mathbb T$.
	\end{condition}
	
	Let $\mathcal S_{t,j-}^A = \{i\in\mathbb J:t_0^i<t_0^j,\delta_t^i = \mathrm A\}$ and $\mathcal S_{t,j-}^B = \{i\in\mathbb J:t_0^i<t_0^j,\delta_t^i = \mathrm B\}$. The alpha-investing rule is given by:
	\begin{equation}\label{eq:testlevelposneg}
		\alpha_t^{j,A} = \dfrac{\alpha^A}{k^A}\left[\mathbb{I}\{j\leq k^A\}  + N^A_{t,j-}(k^A)\right],\quad
		\alpha_t^{j,B} = \dfrac{\alpha^B}{k^B}\left[\mathbb{I}\{j\leq k^B\}  + N^B_{t,j-}(k^B)\right],
	\end{equation}
	where $N^A_{t,j-}(k^A) = |\{n\in\mathbb N: n\geq 2, j-k^A\leq I_n(\mathcal S_{t, j-}^A)\leq j-1\}|$, $N^B_{t,j-}(k^B) = |\{n\in\mathbb N:n\geq 2, j-k^B\leq I_n(\mathcal S_{t, j-}^B)\leq  j-1\}|$, and $(k^A, k^B)$ are tuning parameters characterizing the neighborhood sizes. The modified operations are summarized in Algorithm \ref{alg:sava-arm-spec}.

	\begin{remark}\rm{
			There are two key differences between the two alpha-investing rules \eqref{eq:testlevelposneg} and \eqref{eq:test-level}: first, the sets of selections in the two directions differ; second, the test levels for the two arms are distinct in \eqref{eq:testlevelposneg}.}
	\end{remark} 
	
	\begin{algorithm}
		\caption{The SAVA algorithm with arm-specific error constraints}\label{alg:sava-arm-spec}
		\vskip6pt
		\begin{algorithmic}
			
			\State \textbf{Input}: a grid of decision time $\mathbb T$, target FSR levels $\alpha^A$ and $\alpha^B$, a method for computing always-valid directional p-values, two tuning parameters $k^A$ and $k^B$, and pre-specified tolerance durations $\{b_j\}_{j\in\mathbb J}$.
			
			\State \textbf{For each} $t_i \in \mathbb{T}$, \textbf{do:}
			\State \quad \textbf{Step 1}: Update the index set of active tasks $\mathcal A_{t_{i}}$ defined in \eqref{eq:active-set}.
			\State \quad \textbf{Step 2}: Calculate always-valid directional p-values for tasks in $\mathcal A_{t_i}$ according to the given method.
			\State \quad \textbf{Step 3}: \textbf{For} $j=r_{t_{i}}(1),\dots, r_{t_{i}}(|\mathcal A_{t_{i}}|)$, \textbf{do}:
			\State \quad \quad \textbf{Step 3.1}: Calculate test levels $\alpha_t^{j,A}$ and $\alpha_t^{j,B}$ by \eqref{eq:testlevelposneg}.
			\State \quad \quad \textbf{Step 3.2}: Obtain the decision $\delta_{t_i}^j$ according to decision rule \eqref{eq:decisionrule}.
			\State \quad \textbf{Step 4}: For tasks $j$ such that $\delta_{t_{i-1}}^j \neq \text{C}$, update $\delta_{t_i}^j = \delta_{t_{i-1}}^j$.
			\State \quad \textbf{Output}: decision-making states $\{\delta_{t_i}^{j}: j\in \mathcal A_{t_i}\}$. 
			\State \textbf{End for}
		\end{algorithmic}
	\end{algorithm}
	
	We present a condition that serves as a replacement for Condition \ref{assump3}.
	
	\begin{condition}\label{assump5}
		The test levels $\alpha_t^{j,A}$ and $\alpha_t^{j,B}$ are non-decreasing functions of $(\mathbb I\{\delta_t^i = \mathrm{A}\})_{i=1}^{j-1}$ and $(\mathbb I\{\delta_t^i = \mathrm{B}\})_{i=1}^{j-1}$, respectively, for all  $t\in \mathbb T$.
	\end{condition}

	The following proposition and theorem establish the anytime validity of SAVA under the new setup.
	\begin{proposition}
		\label{prop:level-arm-spec}
		The test levels in \eqref{eq:testlevelposneg} satisfy Conditions \ref{assump2}, \ref{assump4} and \ref{assump5}.
	\end{proposition}
	
	\begin{theorem}\label{thm:distinct}
		Assume that the samples from different data streams are independent. 
		
		(a) If test levels $\alpha_t^{j,A}$ and $\alpha_t^{j,B}$ satisfy Conditions \ref{assump2}, \ref{assump4} and \ref{assump5}, then $\mathrm{FSR}_t^{A}\leq \alpha^A$ and $\mathrm{FSR}_t^{B}\leq \alpha^B$ for all $t\in \mathbb T$.   
		
		(b) If only Conditions \ref{assump2} and \ref{assump4} hold, then $\mathrm{mFSR}_t^{A}\leq \alpha^A$ and $\mathrm{mFSR}_t^{B}\leq\alpha^B$ for all $t\in\mathbb T$.
	\end{theorem}
	
	The proofs of Proposition \ref{prop:level-arm-spec} and Theorem \ref{thm:distinct} are provided in Sections \ref{proof-prop-a.2} and \ref{proof-thm-a.2}, respectively.

	\section{Proofs}\label{sec:prf}

	\setcounter{algorithm}{0}
	\renewcommand\thealgorithm{D.\arabic{algorithm}}
	
	\setcounter{theorem}{0}
	\renewcommand\thetheorem{D.\arabic{theorem}}
	\setcounter{proposition}{0}
	\renewcommand\theproposition{D.\arabic{proposition}}
	\setcounter{corollary}{0}
	\renewcommand\thecorollary{D.\arabic{corollary}}
	\setcounter{condition}{0}
	\renewcommand\thecondition{D.\arabic{condition}}
	\setcounter{algorithm}{0}
	\renewcommand\thealgorithm{D.\arabic{algorithm}}
	\setcounter{equation}{0}
	\renewcommand\theequation{D.\arabic{equation}}
	\setcounter{lemma}{0}
	\renewcommand\thelemma{D.\arabic{lemma}}
	\setcounter{figure}{0}
	\renewcommand\thefigure{D.\arabic{figure}}

	\subsection{Proof of Proposition \ref{prop:eprocess-to-pval}}
    
	By Lemma 3 (a) in \cite{howard2021time}, $\{p_t^A\}_{t\in\mathbb{T}}$ and $\{p_t^B\}_{t\in\mathbb T}$ are always-valid directional p-values if they satisfy the following equivalent condition: For any $\alpha\in[0,1]$, 
	\begin{equation}\label{eq:equavp}
		\mathrm{Pr}_{\theta=\mathrm B}(\exists t \in\mathbb T: p_t^{A}\leq \alpha) \leq \alpha,\quad \mathrm{Pr}_{\theta = \mathrm A}(\exists t \in \mathbb T: p_t^B\leq \alpha)\leq \alpha.
	\end{equation}
	For the e-process $\{E_t^{j,A}\}_{t\in\mathbb T}$, Ville's inequality \citep{ville1939etude,howard2020time} implies that
    \[
    \begin{aligned}
        \mathrm{Pr}_{\theta^j = \mathrm{B}}(\exists t \in\mathbb T: p_t^{j,A}\leq \alpha) & = \mathrm{Pr}_{\theta^j = \mathrm{B}}(\exists t \in\mathbb T: \max_{s\leq t, s\in\mathbb T} E_s^{j,A}\geq 1/\alpha) \\
        & = \mathrm{Pr}_{\theta^j = \mathrm{B}}(\exists t \in\mathbb T: E_t^{j,A}\geq 1/\alpha) \leq \alpha,
    \end{aligned}
    \]
where the final inequality follows from Ville's inequality. An analogous bound holds for the e-process $\{E_{t}^{j,B}\}_{t\in\mathbb T}$.

	\subsection{Proof of Proposition \ref{prop:avpconstruct}}
	
	By the same argument as in the proof of Proposition~\ref{prop:eprocess-to-pval}, it suffices to show that, for any $\alpha\in[0,1]$, 
	\begin{equation*}
		\mathrm{Pr}_{\theta=\mathrm B}(\exists t \in\mathbb T: \rho_t^{A}\leq \alpha) \leq \alpha,\quad \mathrm{Pr}_{\theta = \mathrm A}(\exists t \in \mathbb T: \rho_t^B\leq \alpha)\leq \alpha.
	\end{equation*}
	Recall that \(t_1 \in \mathbb{T}\) represents the first decision time. The desired result can be established by observing that 
	\[
	\mathrm{Pr}_{\theta = \mathrm{B}}(\exists t \in \mathbb{T}: p_t^A \leq \alpha) = \mathrm{Pr}_{\theta = \mathrm{B}}(\exists t' \in \mathbb{T}: \min_{t_1 \leq t' \leq t} \rho_{t'}^A \leq \alpha) \leq \mathrm{Pr}_{\theta = \mathrm{B}}(\exists t \in \mathbb{T}: \rho_t^A \leq \alpha) \leq \alpha.
	\]
	Similarly, we can show that 
	\(
	\mathrm{Pr}_{\theta = \mathrm{A}}(\exists t \in \mathbb{T}: p_t^B \leq \alpha) \leq \alpha. \qedsymbol
	\)

	\subsection{Proof of Proposition \ref{prop:sava-lors}}
	
	We first demonstrate that the test levels $\alpha_t^{j,A}$ and $\alpha_t^{j,B}$ satisfy Conditions \ref{assump2}--\ref{assump3}, then show that the test levels also satisfy Condition \ref{assump1}.

	\medskip
	\noindent\textbf{Part 1. Verification of Condition 2. }
	
	Fix the index \(j \in \mathbb{J}\). It is straightforward to verify that the construction of the test levels at any \(t \in \mathbb{T}\) relies exclusively on the observed data streams from tasks that have commenced prior to task \(j\). Thus, the test levels are \(\mathcal{G}_t^{1:(j-1)}\)-measurable. 
	
	\medskip
	\noindent\textbf{Part 2. Verification of Condition 3. }
	
	Note that $\alpha_t^{j,A}$ and $\alpha_t^{j,B}$ are fully determined by the decisions $\mathbb I\{\delta_t^1 = \mathrm A \text{ or } \mathrm B\},\dots, \mathbb I\{\delta_t^{j-1} = \mathrm A \text{ or } \mathrm B\}$, we rewrite the test levels $\alpha_t^{j,A}$ and $\alpha_t^{j,B}$ as a function $f_t^j:\{0,1\}^{j-1}\rightarrow\mathbb{R}$ such that 
	$$
	\alpha_t^{j,A} = \alpha_t^{j,B}=f_t^j(\mathbb I\{\delta_t^1 = \mathrm A \text{ or } \mathrm B\},\dots,\mathbb I\{\delta_t^{j-1} = \mathrm A \text{ or } \mathrm B\}). 
	$$
	
	For any $i<j$, it suffices to show that 
	\begin{equation*}
		\begin{aligned}
			a_1 &= f_t^j(\mathbb I\{\delta_t^1 = \mathrm A \text{ or } \mathrm B\},\dots,\mathbb I\{\delta_t^{i-1} = \mathrm A \text{ or } \mathrm B\},0,\mathbb I\{\delta_t^{i+1} = \mathrm A \text{ or } \mathrm B\},\dots,\mathbb I\{\delta_t^{j-1} = \mathrm A \text{ or } \mathrm B\}) \\
			& \leq f_t^j(\mathbb I\{\delta_t^1 = \mathrm A \text{ or } \mathrm B\},\dots,\mathbb I\{\delta_t^{i-1} = \mathrm A \text{ or } \mathrm B\},1,\mathbb I\{\delta_t^{i+1} = \mathrm A \text{ or } \mathrm B\},\dots,\mathbb I\{\delta_t^{j-1} = \mathrm A \text{ or } \mathrm B\}) \\
			& =: a_2. 
		\end{aligned}
	\end{equation*}

	Recall that when computing $N_{t,j-}(k) = |\{n\in\mathbb N: n\geq 2, j-k\leq I_n(\mathcal S_{t,j-})\leq j-1\}|$ in \eqref{eq:test-level}, we have deliberately excluded the first selection. The proof of this inequality involves analyzing three possible cases: 
	
	\begin{itemize}
		\item Case I: There is no selection among tasks $1,2,\dots ,i-1,i+1,\dots,j-1$. This case is trivial to verify by noting that $a_2 = {\alpha}\mathbb{I}(j\leq k)/k = a_1$. 
		
		\item Case II: There is only one selection among tasks $1,2,\dots ,i-1,i+1,\dots,j-1$. Suppose the $m$-th task has been selected. We have 
		$a_2 = {\alpha}[\mathbb{I}(j\leq k)+\mathbb{I}(j\leq \min\{m, i\} + k)]/k> {\alpha}\mathbb{I}(j\leq k)/k = a_1.$
		
		\item Case III: There is more than one selection among tasks $1,2,\dots ,i-1,i+1,\dots,j-1$. Let $M \subset \{1,2,\dots,i-1,i+1,\dots,j-1\}$ be the index set of selected tasks. Define $l = \min_{m\in M} m$. Then $a_1 = {\alpha}[\mathbb{I}(j\leq k)/k + |\{m\in M\colon m>l, j\leq m+k\}|]$. It is easy to see that $a_2 = \alpha[\mathbb{I}(j\leq k) + |\{m\in M\cup\{i\}\colon m>\min (M\cup\{i\}), j\leq m+k\}|]/k>a_1$. 
		
	\end{itemize}
	
	Therefore, we conclude that $a_1\leq a_2$, and $f_t^j$ is non-decreasing in each coordinate.  
	
	\medskip
	
	\noindent\textbf{Part 3. Verification of Condition 1. }
	
	Write $\mathbf{x} = (x_1,\dots, x_n)\preceq \mathbf{y} = (y_1,\dots, y_n)$ if $x_i\leq y_i$ for all $i=1,\dots, n$. 
	We say that a function \( f: \mathbb{R}^n \rightarrow \mathbb{R} \) is componentwise non-increasing if, for any two vectors \( \mathbf{x}, \mathbf{y} \in \mathbb{R}^n \), the following holds: whenever \( \mathbf{x} \preceq \mathbf{y} \), we have \( f(\mathbf{x}) \geq f(\mathbf{y}) \). With this definition in place, we now state a claim that demonstrates the componentwise monotonicity of the test levels.

	\begin{lemma}\label{lem:monotonethre}
		
		For any $j\in\mathbb{J}$, the test levels $\alpha_t^{j,A}$ and $\alpha_t^{j,B}$ satisfying Conditions \ref{assump2}-\ref{assump3} are componentwise non-increasing in $(p_t^{1,A}, \dots, p_t^{(j-1),A},p_t^{1,B}, \dots, p_t^{(j-1),B})$, and thus are non-decreasing in $t\in \mathbb T$. 
	\end{lemma}
	
	Lemma \ref{lem:monotonethre} is proved in Section \ref{sec:monotoneproof}.
	According to Lemma \ref{lem:monotonethre}, $f_t^j$ is non-decreasing in $t$ for all $j$. It follows that the estimate for FSR in \eqref{eq:fsrhat} becomes   
	$$
	\widehat{\mathrm{FSR}}_t = \dfrac{\sum_{j: t_0^j \leq t} {\alpha}_t^{j,A} \vee {\alpha}_t^{j,B}}{|\mathcal S_t| \vee 1}.
	$$

	Fix the decision time point $t\in\mathbb T$. Let $l_j$ denote the smallest index in a non-empty $\mathcal{S}_{t,j-}$ and  $n_t= |\{j\in\mathbb J: t_0^j \leq t\}|$ the number of tasks having commenced prior to $t$. We consider the following cases: (i) $|\mathcal{S}_t|\leq 1$; (ii) $|\mathcal{S}_t| := s>1$. 
	For case (i), we have 
	\begin{equation*}
		\widehat{\mathrm{FSR}}_t = \sum_{j\colon t_0^j \leq t} \alpha_t^{j,A}\vee \alpha_t^{j,B} = \dfrac{\alpha}{k}\sum_{j\colon t_0^j\leq t}\mathbb{I}\{j\leq k\}\leq \alpha.
	\end{equation*}
	For case (ii), we have
	\begin{equation*}
		\begin{aligned}
			\sum_{j:t_0^j \leq t} \alpha_t^{j,A}\vee \alpha_t^{j,B}
			&= \dfrac{\alpha}{k}\sum_{j\colon t_0^j\leq t}\mathbb{I}\{j\leq k\} + \dfrac{\alpha}{k}\sum_{j:t_0^j\leq t}\sum_{i:i\in\mathcal{S}_{t,j-},l_j<i<j} \mathbb{I}\{j\leq i+k\} \\
			& = \dfrac{\alpha}{k}\sum_{j\colon t_0^j\leq t}\mathbb{I}\{j\leq k\} + \dfrac{\alpha}{k}\sum_{j:t_0^j\leq t}\sum_{i:i\in\mathcal{S}_{t},l_{n_t}<i<j} \mathbb{I}\{j\leq i+k\} \\
			& = \dfrac{\alpha}{k}\sum_{j\colon t_0^j\leq t}\mathbb{I}\{j\leq k\} + \dfrac{\alpha}{k}\sum_{i:i\in\mathcal{S}_t,i>l_{n_t}} \sum_{j:t_0^j\leq t}\mathbb{I}\{i<j\leq i+k\} \\
			&\leq \alpha + \alpha|\{i:i\in\mathcal{S}_t,i>l_{n_t}\}| =\alpha|\mathcal{S}_t|,
		\end{aligned}
	\end{equation*} 
	where the second equation follows from the fact that if $l_j < j$ then $l_{n_t} = l_{j}$ and thus $\mathcal{S}_{t,j-}\cap \{l_j,\dots, j-1\} = \mathcal{S}_{t}\cap \{l_j,\dots, j-1\}$.
	Therefore, in the second case we obtain
	\begin{equation*}
		\widehat{\mathrm{FSR}}_t = {\sum_{j:t_0^j \leq t} {\alpha}_t^{j,A}\vee {\alpha}_t^{j,B}}/{|\mathcal{S}_t|} \leq \alpha.
	\end{equation*}
	The desired result follows by combining the results for the two cases.  \qedsymbol

	\subsection{Proof of Lemma \ref{lem:monotonethre}}\label{sec:monotoneproof}
	
	Define the set $M = \{i\leq j-1: (p_t^{i,A}\wedge p_t^{i,B}) \leq (q_t^{i,A}\wedge q_t^{i,B})\}$. Let $M_{(1)}$ denote the smallest element in the set $M$. Since smaller directional p-values lead to a larger indicator $\mathbb I\{\delta_t^{M_{(1)}} = \text{A or B}\}$, the test levels for tasks $j>M_{(1)}$ under input $\mathbf{p}_1$ are larger than those under input $\mathbf{p}_2$, according to Condition \ref{assump3}. Therefore, under input $\mathbf p_1$, all tasks are more likely to be selected in the loop step of the SAVA algorithm, resulting in larger test levels for all tasks. The same argument applies to the remaining items in the set $M$, and thus it is clear that test levels $\alpha_t^{j,A}$ and $\alpha_t^{j,B}$ are larger under input $\mathbf{p}_1$.
	Hence, it follows that $\alpha_t^{j,A}$ and $\alpha_t^{j,B}$ are non-increasing in $(p_t^{1,A}, \dots, p_t^{j-1,A}, p_t^{1,B}, \dots, p_t^{j-1,B})$ with respect to the componentwise order.
	
	Note that the always-valid directional p-values are non-increasing in $t$, and thus each of $(\mathbb I\{\delta_t^i = \mathrm A \text{ or } \mathrm B\})_{i=1}^{j-1}$ is non-decreasing in $t$. The test levels $\alpha^{j,A}_t$ and $\alpha_t^{j,B}$ involve the number of selections according to \eqref{neighbor} and \eqref{eq:test-level}. Consequently, they are non-decreasing in $t \in \mathbb{T}$, which completes the proof. \qedsymbol

	\subsection{Proof of Theorem \ref{thm:fsr}} \label{proof:thm1}

	\textit{\textbf {Proof of part (a)}}. By Proposition \ref{prop:avpconstruct}, assume that the always-valid directional p-values are non-increasing in $t$ for all $j\in\mathbb J$. For any $t_k\in\mathbb T$, we have
	\begin{equation*}
		\mathrm{FSR}_{t_k}  = \mathbb E\left[\dfrac{\sum_{j:t_0^j\leq t_k, \theta^j = \mathrm B}\mathbb{I}\{\delta_{t_k}^j = \mathrm A\}}{|\mathcal{S}_{t_k}| \vee 1}\right] + \mathbb E\left[\dfrac{\sum_{j:t_0^j\leq t_k, \theta^j = \mathrm A}\mathbb{I}\{\delta_{t_k}^j = \mathrm B\}}{|\mathcal{S}_{t_k}| \vee 1}\right]
		\coloneqq I^A + I^B.
	\end{equation*}
	We employ a novel leave-one-out technique to show that $I^A$ and $I^B$ are bounded above by $\mathbb E\left[\dfrac{\sum_{j:t_0^j\leq t_k, \theta^j =\mathrm B}\bar{\alpha}_{t_k}^{j,A}}{|\mathcal{S}_{t_k}| \vee 1}\right]$ and $\mathbb E\left[\dfrac{\sum_{j:t_0^j\leq t_k, \theta^j =\mathrm A}\bar{\alpha}_{t_k}^{j,B}}{|\mathcal{S}_{t_k}| \vee 1}\right]$, respectively. 
	
	We summarize some notations before proceeding to the proof. Let $s^j = \min\{i: t_i \geq t^j_0\}$ represent the index of first decision time encountered by task $j$. For each $j\in\mathbb J$, the always-valid directional p-values $\{q_{t_i}^{j,A}\}_{i = s^j}^{+\infty}$ and $\{q_{t_i}^{j,B}\}_{i = s^j}^{+\infty}$ are 
	\begin{equation*}
		q_{t_k}^{j,A} = f^A((X_t^j)_{t_0^j \leq t \leq t_k, t\in \mathcal T^j}),\quad q_{t_k}^{j,B} = f^B((X_t^j)_{t_0^j \leq t \leq t_k, t\in \mathcal T^j}),\quad k = s^j, s^j+1,\dots,
	\end{equation*}
	where $f^A$ and $f^B$ are functions for generating always-valid directional p-values.
	Furthermore, define $T^{j,A} = \inf\{s \geq t_{s^j}: q_{s}^{j,A} \leq \alpha_s^{j,A}, s\in\mathbb T\}$, $T^{j,B} = \inf\{s\geq t_{s^j}: q_s^{j,A} \leq \alpha_s^{j,A}, s\in \mathbb T\} $, $T^j_{\rm{drop}} = \inf\{s \geq t_{s^j}: s - t_0^j \geq b_j, s\in\mathbb T\}$ and define $T^{j} = \min\{T^{j,A}, T^{j,B}, T^j_{\rm{drop}}\}$. Then the always-valid directional p-values observed by the agent can be written as $p_t^{j,A} = q_{t\wedge T^{j}}^{j,A}$ and $p_t^{j,B} = q_{t\wedge T^{j}}^{j,B}$, $t\in\mathbb T$. In addition, let
	$$\mathbf{p}_{t_k}^{j,A} \coloneqq (p_{t_{s^j}}^{j,A}, p_{t_{s^j+1}}^{j,A}, \dots, p_{t_k}^{j,A}), \quad \mathbf{p}_{t_k}^{j,B} \coloneqq (p_{t_{s^j}}^{j,B}, p_{t_{s^j+1}}^{j,B}, \dots, p_{t_k}^{j,B}),$$
	represent the sequence of always-valid directional p-values from $t_{s^j}$ to $t_k$ for task $j$.
	Given any decision time $t_k$, define $n_k=\max\{j:t_0^j\leq t_k\}$ as the index of the newest task. 
	
	Let $\pmb{\delta}_{t_k}=(\delta_{t_k}^1, \delta_{t_k}^2,\dots,\delta_{t_k}^{n_k})$ denote the vector of decisions for the $n_k$ tasks at time $t_k$. 
	For any $j \le n_k$, let 
	\begin{equation}\label{eq:input}
		\tilde{\pmb{\delta}}_{t_k}^{-j} = (\tilde{\delta}_{t_k}^{1,-j},\tilde{\delta}_{t_k}^{2,-j},\cdots,\tilde{\delta}_{t_k}^{n_k,-j}) 
	\end{equation}
	represent the vector of decisions when the decision rule \eqref{eq:decisionrule} is applied to 
	\begin{small}
		\begin{equation}\label{eq:leaveone}
			\mathbf{p}_{t_k}^{1,A}, \dots, \mathbf{p}_{t_k}^{(j-1),A}, \mathbf 0, \mathbf{p}_{t_k}^{(j+1),A}, \dots, \mathbf{p}_{t_k}^{n_k,A}\quad\text{and}\quad \mathbf{p}_{t_k}^{1,B}, \dots, \mathbf{p}_{t_k}^{(j-1),B}, \mathbf{0}, \mathbf{p}_{t_k}^{(j+1),B}, \dots, \mathbf{p}_{t_k}^{n_k,B}.
		\end{equation}
	\end{small}
	
	Consider $\pmb{\delta}_{t_k}$ and its modified version $\tilde{\pmb{\delta}}_{t_k}^{-j}$ defined in \eqref{eq:input}. Obviously $\delta_{t_k}^{i} = \tilde{\delta}_{t_k}^{i,-j}$ for all $i<j$. Moreover, since  the test levels with input \eqref{eq:leaveone} are always larger than or equal to original test levels (by Lemma \ref{lem:monotonethre}), we have 
	\begin{equation}
		\label{eq:decision-relation}
		\mathbb I\{\delta_{t_k}^{i} = \mathrm{A}\text{ or }\mathrm{B}\} \leq \mathbb I\{\tilde{\delta}_{t_k}^{i,-j} = \mathrm{A}\text{ or }\mathrm{B}\},\quad \text{ for all } i\geq j.
	\end{equation}
	
	We introduce a few notations. For each $n\in\mathbb{N}$, define $g:\{\mathrm{A,B,C,D}\}^n\rightarrow \mathbb{R}$ as $g(x_1,x_2,\dots,x_n)=(\sum_{i=1}^n \mathbb I\{x_i = \text{A or B}\}) \vee 1$. 
	By construction, for task $j$ such that $t^j_0 \leq t_k$ and $\theta^j = \mathrm B$, on the event $\{\delta_{t_k}^j = \mathrm A\} = \{T^{j,A} \leq t_k, T^{j,A}<\min\{ T^{j,B}, T^j_{\rm{drop}}\}\}$, we have $\mathbb I\{\delta_{t_k}^j = \mathrm{A} \text{ or }\mathrm{B}\} = \mathbb I\{\tilde{\delta}_{t_k}^{j,-j}= \mathrm{A} \text{ or }\mathrm{B}\} = 1$. 
	Likewise, for task $l$ such that $t_0^l\leq t_k$ and $\theta^l = \mathrm A$, we have $\mathbb I\{\delta_{t_k}^l= \mathrm{A} \text{ or }\mathrm{B}\} = \mathbb I\{\tilde{\delta}_{t_k}^{l,-j} = \mathrm{A} \text{ or }\mathrm{B}\} = 1$ on the event $\{\delta^l_{t_k} = \mathrm B\}$. 
	Furthermore, we have the following result:
	\begin{lemma}\label{lem:deltatilde}
		Consider \(\tilde{\delta}^{i,-j}_{t_k}\) as defined in \eqref{eq:input}. We have: (a) The equality \(\mathbb{I}\{\delta_{t_k}^i = \mathrm{A} \text{ or } \mathrm{B}\} = \mathbb{I}\{\tilde{\delta}^{i,-j}_{t_k} = \mathrm{A} \text{ or } \mathrm{B}\}\) holds on the event \(\{\delta_{t_k}^j = \mathrm{A}\}\) for each \(i = j + 1, \ldots, n_k\); (b) The equality \(\mathbb{I}\{\delta_{t_k}^i = \mathrm{A} \text{ or } \mathrm{B}\} = \mathbb{I}\{\tilde{\delta}^{i,-j}_{t_k} = \mathrm{A} \text{ or } \mathrm{B}\}\) holds on the event \(\{\delta_{t_k}^j = \mathrm{B}\}\) for each \(i = j + 1, \ldots, n_k\). \end{lemma}
	
	The proof for Lemma \ref{lem:deltatilde} is provided in Section \ref{proof-b.2}. It follows from Lemma \ref{lem:deltatilde} that
	\begin{equation*}
		\dfrac{\mathbb{I}\{\delta_{t_k}^j = \mathrm A\}}{g({\pmb \delta_{t_k}})} =\dfrac{\mathbb{I}\{\delta_{t_k}^j = \mathrm A\}}{g(\tilde{\pmb \delta}_{t_k}^{-j})}.
	\end{equation*}
	Meanwhile, it is worth noting that decisions for tasks that arrived before $t^j_0$ are independent of $\mathbf{X}^j$. By Condition \ref{assump2}, the test levels for task $j$ at time $t_k$ are non-random conditional on  $\mathcal{G}_{t_k}^{1:(j-1)}$. Moreover, $\tilde{\pmb{\delta}}_{t_k}^{-j}$ is  independent of the directional p-values $\mathbf{p}_{t_k}^{j,A}$ and $\mathbf{p}_{t_k}^{j,B}$ conditional on $\mathcal{G}_{t_k}^{1:(j-1)}$. It follows that $\tilde{\pmb{\delta}}_{t_k}^{-j}$ and $\delta_{t_k}^j$ are conditionally independent given $\mathcal{G}_{t_k}^{1:(j-1)}$. Therefore, for each $j$ such that $\theta^j = \mathrm B$, we have
	\begin{align}
		\mathbb E_{\theta^j=\mathrm B}\left[\left.\dfrac{\mathbb{I}\{\delta_{t_k}^j = \mathrm A\}}{g(\pmb{\delta}_{t_k})}\right| \mathcal{G}_{t_k}^{1:(j-1)}\right]
		& = \mathbb E_{\theta^j=\mathrm B}\left[\left.\dfrac{\mathbb{I}\{\delta_{t_k}^j = \mathrm A\}}{g(\tilde{\pmb{\delta}}_{t_k}^{-j})}\right| \mathcal{G}_{t_k}^{1:(j-1)}\right] \nonumber \\ 
		& = \mathrm{Pr}_{\theta^j=\mathrm B}\left(\left.\delta_{t_k}^j = \mathrm A\right| \mathcal{G}_{t_k}^{1:(j-1)}\right) \mathbb E_{\theta^j=\mathrm B}\left[\left.\dfrac{1}{g(\tilde{\pmb{\delta}}_{t_k}^{-j})}\right| \mathcal{G}_{t_k}^{1:(j-1)} \right] \nonumber \\
		& \leq \mathrm{Pr}_{\theta^j=\mathrm B}\left(\left.\delta_{t_k}^j = \mathrm A\right| \mathcal{G}_{t_k}^{1:(j-1)}\right) \mathbb E_{\theta^j=\mathrm B}\left[\left.\dfrac{1}{g(\pmb \delta_{t_k})}\right| \mathcal{G}_{t_k}^{1:(j-1)} \right] \label{eq:denomi},
	\end{align}
	where the inequality \eqref{eq:denomi} follows from the fact $g(\tilde{\pmb \delta}_{t_k}^{-j}) \geq g({\pmb \delta}_{t_k})$, which follows from  $\mathbb I\{\tilde{\delta}_{t_k}^{i,-j} =  \text{A or B}\}\geq \mathbb I\{ \delta_{t_k}^{i} = \text{A or B}\}$ for all $i \leq n_k$ according to \eqref{eq:decision-relation}. 
	
	To upper bound $\mathrm{Pr}_{\theta^j=\mathrm B}(\delta_{t_k}^j = \mathrm A|\mathcal{G}_{t_k}^{1:(j-1)})$, note that the test levels are monotone in time $t\in\mathbb T$ (by Lemma \ref{lem:monotonethre}). It follows that
	\begin{equation}
		\label{eq:denomi2}
		\begin{aligned}
			\mathrm{Pr}_{\theta^j=\mathrm B}\left(\left.\delta_{t_k}^j = \mathrm A\right|\mathcal{G}_{t_k}^{1:(j-1)}\right) & = \mathrm{Pr}_{\theta^j=\mathrm B}\left(\left.T^{j,A}\leq t_k, T^{j,A}<\min\{T^{j,B},T^j_{\rm{drop}}\}\right| \mathcal{G}_{t_k}^{1:(j-1)}\right) \\
			& \leq \mathrm{Pr}_{\theta^j=\mathrm B}\left(\left.T^{j,A}\leq t_k\right| \mathcal{G}_{t_k}^{1:(j-1)}\right)  \\
			& = \mathrm{Pr}_{\theta^j=\mathrm B}\left(\left.\exists t_i \in[t_0^j,t_k]: q_{t_i}^{j,A} \leq \alpha_{t_i}^{j,A} \right| \mathcal{G}_{t_k}^{1:(j-1)}\right) \\
			& \leq \mathrm{Pr}_{\theta^j=\mathrm B}\left(\left.\exists t_i \in[t_0^j,t_k]: q_{t_i}^{j,A} \leq \bar{\alpha}_{t_k}^{j,A} \right| \mathcal{G}_{t_k}^{1:(j-1)}\right) \leq \bar{\alpha}_{t_k}^{j,A}, 
		\end{aligned}
	\end{equation}
	where the last inequality follows from the independence of data streams and the definition of always-valid directional p-values. 
	Combining \eqref{eq:denomi} with \eqref{eq:denomi2} yields 
	\begin{equation*}
		\begin{aligned}
			I^A & = \mathbb E\left[\sum_{j:t_0^j\leq t_k, \theta^j = \mathrm B}\mathbb E_{\theta^j=\mathrm B}\left\{\left.\dfrac{\mathbb{I}\{\delta_{t_k}^j = \mathrm A\}}{|\mathcal{S}_{t_k}| \vee 1}\right|  \mathcal{G}_{t_k}^{1:(j-1)}\right\}\right] \\
			& \le \mathbb E\left[\sum_{j:t_0^j\leq t_k, \theta^j = \mathrm B}\mathbb E_{\theta^j=\mathrm B}\left\{\left.\dfrac{\bar{\alpha}_{t_k}^{j,A}}{|\mathcal{S}_{t_k}| \vee 1}\right|  \mathcal{G}_{t_k}^{1:(j-1)}\right\}\right] \\
			& = \mathbb E\left[\sum_{j:t_0^j\leq t_k, \theta^j = \mathrm B}\dfrac{\bar{\alpha}_{t_k}^{j,A}}{|\mathcal{S}_{t_k}| \vee 1}\right].
		\end{aligned}
	\end{equation*}
	As for the quantities $I^B$, we obtain the following result in the same way:
	\begin{equation*}
		I^B  =\mathbb E\left[\sum_{j:t_0^j\leq t_k, \theta^j =\mathrm A}\mathbb E_{\theta^j=\mathrm A}\left\{\left.\dfrac{\mathbb{I}\{\delta_{t_k}^j = \mathrm B\}}{|\mathcal{S}_{t_k}| \vee 1}\right|  \mathcal{G}_{t_k}^{1:(j-1)}\right\}\right] 
		\leq \mathbb E\left[\sum_{j:t_0^j\leq t_k, \theta^j =\mathrm A}\dfrac{\bar{\alpha}_{t_k}^{j,B}}{|\mathcal{S}_{t_k}| \vee 1}\right].
	\end{equation*}
	Combining the results above, we have that
	\begin{equation*}
		\mathrm{FSR}_{t_k}  \leq \mathbb E\left[\sum_{j:t_0^j\leq t_k, \theta^j = \mathrm B} \dfrac{\bar{\alpha}_{t_k}^{j,A}}{|\mathcal{S}_{t_k}| \vee 1}+\sum_{j:t_0^j\leq t_k, \theta^j = \mathrm A} \dfrac{\bar{\alpha}_{t_k}^{j,B}}{|\mathcal{S}_{t_k}| \vee 1}\right] 
		\leq \mathbb E\left[\sum_{j:t_0^j\leq t_k} \dfrac{\bar{\alpha}_{t_k}^{j,A} \vee \bar{\alpha}_{t_k}^{j,B}}{|\mathcal{S}_{t_k}| \vee 1}\right] \leq \alpha,
	\end{equation*}
	which concludes the proof of part (a) of Theorem \ref{thm:fsr}. 
	
	\medskip
	\noindent \textit{\textbf {Proof of part (b)}}. 
	We first outline the proof idea at a high level. We rewrite the numerator of the mFSR and apply the law of total probability to bound the conditional probability of false selection from above. The derivation of the upper bound leverages the definition of always-valid directional p-values and the monotonicity of test levels. 
	
	Without loss of generality, the always-valid directional p-values are still assumed to be non-increasing in $t$ for all $j\in\mathbb J$.
	For any decision time point $t_k\in \mathbb T$, mFSR$_{t_k}$ is of the form:
	\begin{equation}
		\label{eq:mfsrdivide}
		\mathrm{mFSR}_{t_k} = \dfrac{\mathbb E\left[\sum_{j\colon t_0^j\leq t_k,\, \theta^j = \mathrm B}\mathbb{I}\{\delta_{t_k}^j = \mathrm A\}\right]}{\mathbb E(|\mathcal{S}_{t_k}|\vee 1)} + \dfrac{\mathbb E\left[\sum_{j\colon t_0^j\leq t_k,\,\theta^j = \mathrm A}\mathbb{I}\{\delta_{t_k}^j = \mathrm B\}\right]}{\mathbb E(|\mathcal{S}_{t_k}|\vee 1)}. 
	\end{equation}
	
	Using the same notation as discussed in the proof of part (a), we turn to deal with the first term of \eqref{eq:mfsrdivide}. Note that
	$$
	\mathbb E\left[\sum_{j\colon t_0^j\leq t_k,\, \theta^j = \mathrm B}\mathbb{I}\{\delta_{t_k}^j = \mathrm A\}\right]=  \sum_{j\colon t_0^j\leq t_k,\, \theta^j = \mathrm B}\mathbb E_{\theta^j=\mathrm B}\left[\left.\mathrm{Pr}_{\theta^j=\mathrm B}(\delta_{t_k}^j = \mathrm A \right| \mathcal{G}_{t_k}^{1:(j-1)})\right].
	$$
	By the same technique in \eqref{eq:denomi2}, we obtain $ \mathrm{Pr}_{\theta^j=\mathrm B}\left(\left.\delta_{t_k}^j = \mathrm A\right|\mathcal{G}_{t_k}^{1:(j-1)}\right)\leq \bar{\alpha}_{t_k}^{j,A}$.
	Therefore, we have
	$        \mathbb E\left[\sum_{j\colon t_0^j\leq t_k,\, \theta^j =\mathrm B}\mathbb{I}\{\delta_{t_k}^j = \mathrm A\}\right] \leq \mathbb E\left[\sum_{j\colon t_0^j\leq t_k,\, \theta^j =\mathrm B}\bar{\alpha}_{t_k}^{j,A}\right] .
	$
	Similarly, we can show that
	$    \mathbb E\left[\sum_{j\colon t_0^j\leq t_k,\, \theta^j = \mathrm A}\mathbb{I}\{\delta_{t_k}^j = \mathrm B\}\right]
	\leq \mathbb E\left[ \sum_{j\colon t_0^j\leq t_k,\, \theta^j = \mathrm A}\bar{\alpha}_{t_k}^{j,B}\right].
	$ Therefore, 
	\begin{equation*}
		\begin{aligned}
			\qquad & \mathbb E\left[\sum_{j\colon t_0^j\leq t_k,\, \theta^j = \mathrm B}\mathbb{I}\{\delta_{t_k}^j = \mathrm A\}\right] + \mathbb E\left[\sum_{j\colon t_0^j\leq t_k,\,\theta^j = \mathrm A}\mathbb{I}\{\delta_{t_k}^j = \mathrm B\}\right] \\
			& = \mathbb E\left[\sum_{j\colon t_0^j\leq t_k,\, \theta^j = \mathrm B}\bar{\alpha}_{t_k}^{j,A} + \sum_{j\colon t_0^j\leq t_k,\, \theta^j = \mathrm A}\bar{\alpha}_{t_k}^{j,B} \right] \\
			& \leq  \mathbb E\left[ \dfrac{\sum_{j\colon t_0^j\leq t_k}(\bar{\alpha}_{t_k}^{j,A} \vee \bar{\alpha}_{t_k}^{j,B} )}{|\mathcal{S}_{t_k}|\vee 1}(|\mathcal{S}_{t_k}|\vee 1)\right]\leq \alpha \mathbb E[|\mathcal{S}_{t_k}|\vee 1],
		\end{aligned}	
	\end{equation*}
	proving the desired result $\mathrm{mFSR}_{t_k} \leq \alpha$.
	\qedsymbol

	\subsection{Proof of Lemma \ref{lem:deltatilde}}\label{proof-b.2}
	We focus on the proof of part (a), as part (b) can be proved in a similar manner.
	We prove the statement using the method of induction on the given index $j$. Let $\tilde{\alpha}_t^{i,A}$ and $\tilde{\alpha}_t^{i,B}$ denote the test levels for task $i$ at the decision time $t$ when $\tilde{\pmb \delta}_t^{-j}$ is used to the construction of test levels. According to \eqref{eq:leaveone}, we have $\delta^i_{t} = \tilde \delta^{i,-j}_{t}$ for all $i<j$ and $t \in\mathbb T$, and $\mathbb I\{\tilde \delta_{t}^{j,-j} = \mathrm A \text{ or } \mathrm B\} = 1$ for all $t\in\mathbb T$. Consider task $j+1$. Since $\alpha_t^{(j+1),A}$ and $\alpha_t^{(j+1),B}$ are non-decreasing functions of $(\mathbb I\{\delta_t^i = \mathrm A \text{ or } \mathrm B\})_{i=1}^{j}$, $t\in\mathbb T$, we obtain 
	\begin{equation*}
		\alpha_s^{(j+1),A} \leq \tilde{\alpha}_s^{(j+1),A},\quad\text{ and }\quad\alpha_s^{(j+1),B} \leq \tilde{\alpha}_s^{(j+1),B}, \quad\text{ for all } s=t_{s^{j+1}},\dots, t_k.
	\end{equation*}
	In particular, since $\mathbb I\{\delta_{t_k}^i = \mathrm A \text{ or } \mathrm B\} = \mathbb I\{\tilde \delta_{t_k}^{i,-j} = \mathrm A \text{ or } \mathrm B\}$ on the event $\{\delta^j_{t_k} = \mathrm A\}$ for all $i\leq j$, it follows that $\alpha_{t_k}^{(j+1), A} = \tilde \alpha_{t_k}^{(j+1), A}$ and $\alpha_{t_k}^{(j+1),B} = \tilde \alpha_{t_k}^{(j+1),B}$. Then we consider the relationship between $\delta_{t_k}^{j+1}$ and $\tilde{\delta}_{t_k}^{j+1,-j}$. 
	
	Consider the following cases:
	\begin{itemize}
		\item If $\delta_{t_k}^{j+1} = \mathrm D$ then the task $j+1$ has been dropped up to $t_k$. As the drop time has no bearing on concrete values of samples, task $j+1$ has been dropped up to $t_k$ when considering \eqref{eq:leaveone}. Hence, we have $\mathbb I\{\delta_{t_k}^{j+1} = \mathrm A \text{ or } \mathrm B\} = \mathbb I\{\tilde{\delta}_{t_k}^{j+1,-j} = \mathrm A \text{ or } \mathrm B\} = 0$.
		\item If $\delta_{t_k}^{j+1} = \mathrm C$, then none of always-valid directional p-values of task $j+1$ falls below the test levels before time $t_k$. According to monotonicity of test levels (Lemma \ref{lem:monotonethre}) and always-valid directional p-values, $p_{t_i}^{(j+1),A} > \alpha_{t_k}^{(j+1),A}$ and $p_{t_i}^{(j+1),B} > \alpha_{t_k}^{(j+1),B}$ for all $i=s^{j+1}, \dots, k$. The fact that $\alpha_{t_k}^{(j+1),A} = \tilde{\alpha}_{t_k}^{(j+1),A}$ and $\alpha_{t_k}^{(j+1),B} = \tilde{\alpha}_{t_k}^{(j+1),B}$ leads to $\mathbb I\{\delta_{t_k}^{j+1} = \mathrm A \text{ or } \mathrm B\} = \mathbb I\{\tilde{\delta}_{t_k}^{j+1,-j} = \mathrm A \text{ or } \mathrm B\} = 0$.
		\item If $\delta_{t_k}^{j+1} = \mathrm A \text{ or } \mathrm B$, then some always-valid directional p-value falls below corresponding test level at some time $t_i \leq t_k$, according to the decision rule. Since test levels with input \eqref{eq:input} are always not less than original ones by Lemma \ref{lem:monotonethre}, we directly have $\tilde{\delta}_{t_k}^{j+1,-j} = \mathrm A \text{ or } \mathrm B$ and thus $\mathbb I\{\delta_{t_k}^{j+1} = \mathrm A \text{ or } \mathrm B\} = \mathbb I\{\tilde{\delta}_{t_k}^{j+1,-j} = \mathrm A \text{ or } \mathrm B\}$. 
	\end{itemize}
	
	In conclusion, we have $\mathbb I\{\delta_{t_k}^{j+1} = \mathrm A \text{ or } \mathrm B\} = \mathbb I\{\tilde{\delta}_{t_k}^{j+1,-j} = \mathrm A \text{ or } \mathrm B\}$. Now assume that $\mathbb I\{\delta_{t_k}^{i} = \mathrm A \text{ or } \mathrm B\} = \mathbb I\{\tilde{\delta}_{t_k}^{i,-j} = \mathrm A \text{ or } \mathrm B\}$ for all $i = j+1, j+2, \dots, l-1$. We now show that $\mathbb I\{\delta_{t_k}^{l} = \mathrm A \text{ or } \mathrm B\} = \mathbb I\{\tilde{\delta}_{t_k}^{l,-j} = \mathrm A \text{ or } \mathrm B\}$. Given these conditions, the property of test levels guarantees that $\alpha_{t_k}^{l,A} = \tilde{\alpha}_{t_k}^{l,A}$ and $\alpha_{t_k}^{l,B} = \tilde{\alpha}_{t_k}^{l,B}$. Using the same arguments, we conclude that $\mathbb I\{\delta_{t_k}^{l} = \mathrm A \text{ or } \mathrm B\} = \mathbb I\{\tilde{\delta}_{t_k}^{l,-j} = \mathrm A \text{ or } \mathrm B\}$. Therefore, it holds that $\mathbb I\{\delta_{t_k}^{i} = \mathrm A \text{ or } \mathrm B\} = \mathbb I\{ \tilde\delta_{t_k}^{i,-j} = \mathrm A \text{ or } \mathrm B\}$ on the event $\{\delta_{t_k}^{j} = \mathrm A\}$ for each $i = j+1,\dots, n_k$. The proof is completed. \qedsymbol

	\subsection{Proof of Proposition \ref{prop:level-arm-spec}}\label{proof-prop-a.2}
	
	Similar to the proof of Proposition \ref{prop:sava-lors}, we first demonstrate that the test levels satisfy Conditions \ref{assump2} and \ref{assump5}, then show that they also satisfy Condition \ref{assump4}.
	
	\noindent\textbf{Part 1. Verification of Condition \ref{assump2}.}
	
	For each $j \in \mathbb{J}$, it is straightforward to verify that the test levels at any $t \in \mathbb{T}$ depend exclusively on the observed data streams from tasks that commenced prior to task $j$. Consequently, the test levels are $\mathcal{G}_t^{1:(j-1)}$-measurable.
	
	\noindent\textbf{Part 2. Verification of Condition \ref{assump5}.}
	
	According to \eqref{eq:testlevelposneg}, $\alpha_t^{j,A}$ is fully determined by the decisions $\mathbb{I}\{\delta_t^1 = \mathrm{A}\}, \dots, \mathbb{I}\{\delta_t^{j-1} = \mathrm{A}\}$, and a similar result holds for $\alpha_t^{j,B}$. We rewrite the test levels $\alpha_t^{j,A}$ and $\alpha_t^{j,B}$ as functions:
	\[
	f_t^{j,A}(\mathbb{I}\{\delta_t^1 = \mathrm{A}\}, \dots, \mathbb{I}\{\delta_t^{j-1} = \mathrm{A}\}) \quad \text{and} \quad f_t^{j,B}(\mathbb{I}\{\delta_t^1 = \mathrm{B}\}, \dots, \mathbb{I}\{\delta_t^{j-1} = \mathrm{B}\}).
	\]
	Following the same approach as in the proof of Proposition \ref{prop:sava-lors}, we establish that $f_t^{j,A}$ and $f_t^{j,B}$ are non-decreasing in each coordinate.
	
	\noindent\textbf{Part 3. Verification of Condition \ref{assump4}}
	
	We focus on the proof of $\widehat{\mathrm{FSR}}^{A}_t\leq \alpha^A$, and $\widehat{\mathrm{FSR}}^{B}_t\leq \alpha^B$ can be derived from a similar argument.
	Similar to the proof of Proposition \ref{prop:sava-lors}, it follows that the estimate for $\mathrm{FSR}^A$ reduces to 
	$$\widehat{\mathrm{FSR}}^A_t = \dfrac{\sum_{j:t_0^j\leq t}\alpha_t^{j,A}}{|\mathcal S_t^A| \vee 1}.
	$$
	Let $l^A_j$ denote the smallest index in a non-empty $\mathcal S^A_{t,j-}$, and $n_t = |\{j\in\mathbb J: t_0^j \leq t\}|$ the number of tasks having commenced prior to $t$. We consider the following cases: (i) $|\mathcal S_t^A|\leq 1$; (ii) $|\mathcal S_t^A| := s > 1$.
	For case (i), we have $\widehat{\mathrm{FSR}}_t \leq \alpha^A$ in the same way as discussed in the proof of Proposition \ref{prop:sava-lors}.
	
	For case (ii), we have
	\begin{equation*}
		\begin{aligned}
			\sum_{j:t_0^j\leq t}\alpha^{j,A}_t 
			&=\dfrac{\alpha^A}{k^A}\sum_{j:t_0^j\leq t}\mathbb I\{j\leq k^A\} + \dfrac{\alpha^A}{k^A}\sum_{i:i\in\mathcal S^A_{t,j-}, i> l^A_{n_t}}\sum_{j:t_0^j\leq t}\mathbb I\{ i<j\leq i+k^A\} \\
			& \leq \alpha^A + \alpha^A|\{i:i\in\mathcal S^A_{t}, i> l^A_{n_t}\}| = \alpha^A|\mathcal S^A_t|.
		\end{aligned}
	\end{equation*}
	Therefore, in the second case we obtain 
	\begin{equation*}
		\widehat{\mathrm{FSR}}^A_t = \dfrac{\sum_{j:t_0^j\leq t}\alpha^A_t}{|\mathcal S^A_t|}\leq \dfrac{\alpha^A|\mathcal S^A_t|}{|\mathcal S^A_t|} = \alpha^A.
	\end{equation*}
	The desired result follows by combining the results for the two cases. \qedsymbol

	\subsection{Proof of Theorem \ref{thm:distinct}}\label{proof-thm-a.2}
	\textit{\textbf {Proof of part (a)}}.
	The proof is similar to the proof of Theorem \ref{thm:fsr}, and we use the same notations and technique there. 
	The extended version of leave-one-out technique is again used. 
	Let $\pmb \delta_{t_k} = (\delta_{t_k}^1, \dots, \delta_{t_k}^{n_k})$ denote the vector of decisions for the $n_k$ tasks at time $t_k$. For any $j\leq n_k$, let
	$$\tilde{\pmb \delta}_{t_k}^{-j,A} = (\tilde{\delta}^{1,-j,A}_{t_k},\tilde{\delta}^{2,-j,A}_{t_k},\dots,\tilde{\delta}^{n_k,-j,A}_{t_k})$$
	represent the vector of decisions when the decision rule \eqref{eq:decisionrule} is applied to
	\begin{equation*}
		\mathbf p_{t_k}^{1,A},\dots, \mathbf p_{t_k}^{(j-1),A},\mathbf 0, \mathbf p_{t_k}^{(j+1),A},\dots,\mathbf p_{t_k}^{(n_k),A} \quad \text{and} \quad \mathbf p_{t_k}^{1,B}, \mathbf p_{t_k}^{2,B}, \dots, \mathbf p_{t_k}^{(n_k),B}.
	\end{equation*}
	We have $\mathbb I\{\delta^i_{t_k} = \mathrm A\}\leq \mathbb I\{\tilde{\delta}^{i,-j,A}_{t_k} = \mathrm A\}$ for all $i\leq n_k$, in the same reason discussed in the proof of Theorem \ref{thm:fsr}. Define the function $g^A:\{\mathrm{A,B,C,D}\}^n\rightarrow \mathbb{R}$ as $g(x_1,x_2,\dots,x_n)=(\sum_{i=1}^n \mathbb I\{x_i = \mathrm A\}) \vee 1$. As a trivial consequence of Lemma \ref{lem:deltatilde}, $\mathbb I\{\delta_{t_k}^{i} = \mathrm A\} = \mathbb I\{\tilde{\delta}_{t_k}^{i,-j,A} = \mathrm A\}$ on the event $\{\delta^j_{t_k} = \mathrm A\}$ for all  $i\leq n_k$, implying that
	\begin{equation*}
		\dfrac{\mathbb I\{\delta_{t_k}^j = \mathrm A\}}{g^A(\pmb \delta_{t_k})} = \dfrac{\mathbb I\{\delta_{t_k}^j = \mathrm A\} }{g^A(\tilde{\pmb \delta}_{t_k}^{-j,A})}.
	\end{equation*}
	Combining the results above and using the technique in \eqref{eq:denomi} and \eqref{eq:denomi2}, we obtain
	\begin{equation*}
		\begin{aligned}
			\mathrm{FSR}^A_{t_k}&= \mathbb E\left[\sum_{j:t_0^j\leq t_k, \theta^j = \mathrm B}\mathbb E_{\theta^j=\mathrm B}\left\{\left.\dfrac{\mathbb{I}\{\delta_{t_k}^j = \mathrm A\}}{|\mathcal{S}_{t_k}^{A}| \vee 1}\right|  \mathcal{G}_{t_k}^{1:(j-1)}\right\}\right] \\
			& \leq \mathbb E\left[\sum_{j:t_0^j\leq t_k, \theta^j = \mathrm B}\mathrm{Pr}_{\theta^j=\mathrm B}(\left.\delta_{t_k}^j = \mathrm A \right| \mathcal G_{t_k}^{1:(j-1)})\mathbb E_{\theta^j=\mathrm B}\left\{\left.\dfrac{1}{|\mathcal{S}_{t_k}^{A}| \vee 1}\right|  \mathcal{G}_{t_k}^{1:(j-1)}\right\}\right] \\
			& \le \mathbb E\left[\sum_{j:t_0^j\leq t_k, \theta^j = \mathrm B}\mathbb E_{\theta^j=\mathrm B}\left\{\left.\dfrac{\bar{\alpha}_{t_k}^{j,A}}{|\mathcal{S}_{t_k}^{A}| \vee 1}\right|  \mathcal{G}_{t_k}^{1:(j-1)}\right\}\right] \\
			& = \mathbb E\left[\sum_{j:t_0^j\leq t_k, \theta^j = \mathrm B}\dfrac{\bar{\alpha}_{t_k}^{j,A}}{|\mathcal{S}_{t_k}^{A}| \vee 1}\right] \leq \alpha^A.
		\end{aligned}
	\end{equation*}
	In the same way, we can also obtain that $\mathrm{FSR}^B_{t_k}\leq \alpha^B$, which concludes the proof of part (a) of Theorem \ref{thm:distinct}.
	
	\noindent\textit{\textbf {Proof of part (b)}}.
	In the same way as discussed in the proof of part (b) in Theorem \ref{thm:fsr}, we obtain:
	\begin{equation*}
		\begin{aligned}
			\quad &\quad \mathbb E\left[\sum_{j:t_0^j\leq t_k, \theta^j = \mathrm B}\mathbb I\{\delta^j_{t_k} = \mathrm A\}\right]\\
			& = \sum_{j:t_0^j\leq t_k,\theta^j = \mathrm B}\mathbb E_{\theta^j=\mathrm B}\left[\mathrm{Pr}_{\theta^j=\mathrm B}(\delta^j_{t_k} = \mathrm A \mid \mathcal G_{t_k}^{1:(j-1)})\right] \\
			& = \sum_{j:t_0^j\leq t_k,\theta^j = \mathrm B}\mathbb E_{\theta^j=\mathrm B} \left[\mathrm{Pr}_{\theta^j=\mathrm B}(T^{j,A}\leq t_k, T^{j,A}<\min\{T^{j,B},T^j_{\rm{drop}}\} \mid \mathcal G_{t_k}^{1:(j-1)})\right] \\
			&\leq \sum_{j:t_0^j\leq t_k,\theta^j = \mathrm B}\mathbb E \left[\bar{\alpha}_{t_k}^{j,A}\right]  = \mathbb E\left[\dfrac{\sum_{j:t_0^j\leq t_k} \bar{\alpha}_{t_k}^{j,A}}{|\mathcal S_{t_k}^{A}|\vee 1}(|\mathcal S_{t_k}^A|\vee 1)\right] \\
			&\leq \alpha^A\mathbb E[|\mathcal S_{t_k}^A| \vee 1],
		\end{aligned}
	\end{equation*}
	which leads to $\mathrm{mFSR}^A_{t_k}\leq \alpha^A$ for all $k$. The result that $\mathrm{mFSR}^B_{t_k}\leq \alpha^B$ can be obtained in the same way, which concludes the proof of part (b) of Theorem \ref{thm:distinct}. 
	\qedsymbol

	\section{Counterexamples}\label{sec:ctexam}

	\setcounter{algorithm}{0}
	\renewcommand\thealgorithm{E.\arabic{algorithm}}
	
	\setcounter{theorem}{0}
	\renewcommand\thetheorem{E.\arabic{theorem}}
	\setcounter{proposition}{0}
	\renewcommand\theproposition{E.\arabic{proposition}}
	\setcounter{corollary}{0}
	\renewcommand\thecorollary{E.\arabic{corollary}}
	\setcounter{condition}{0}
	\renewcommand\thecondition{E.\arabic{condition}}
	\setcounter{algorithm}{0}
	\renewcommand\thealgorithm{E.\arabic{algorithm}}
	\setcounter{equation}{0}
	\renewcommand\theequation{E.\arabic{equation}}
	\setcounter{lemma}{0}
	\renewcommand\thelemma{E.\arabic{lemma}}
	\setcounter{figure}{0}
	\renewcommand\thefigure{E.\arabic{figure}}

	In this section, we present two counterexamples to demonstrate that the conditions outlined in Theorem \ref{thm:fsr} are, to some extent, necessary.
	
	\subsection{Counterexample 1}\label{sec:counterexample1}
	
	The first counterexample demonstrates that removing the maximum values for the test levels in \eqref{eq:fsrhat} may lead to inflated mFSR and FSR levels.

	Since $p^{j,A}_t\equiv p^{j,A}$ is trivially an always-valid p-value for testing $H_0^j :\mu^j = \mu^j_A -\mu^j_B < 0$ if $p^{j,A}$ is a valid p-value for testing $H_0^j :\mu^j = \mu^j_A-\mu^j_B< 0$, we assume always-valid directional p-values do not update as data accumulate.
	This counterexample involves 500 tasks, where the means of the distributions of their data streams are generated from $ \mathrm{Ber}(0.5) - 0.5$. We assume that $\theta^j = \mathrm A$ if $\mu^j \geq 0$, and $\theta^j = \mathrm B$ otherwise. The data collection step is omitted, and the always-valid directional p-values are generated directly as follows.
	\begin{itemize}
		\item When $\theta^j = \mathrm A$, $p^{j,A} = 1 - \Phi(Y^j)$, where $Y^j\sim N(\mu^j, 0.25)$ and $\Phi$ is the distribution function of $N(0,1)$. Concurrently, $p^{j,B}$ is assigned a uniform distribution $U(0,1)$. 
		\item When $\theta^j = \mathrm B$, $p^{j,B}=\Phi(Y^j)$, where $Y^j\sim N(\mu^j, 0.25)$, while $p^{j,A}$ is assigned a uniform distribution $U(0,1)$. 
	\end{itemize} 
	
	We now aim to design an alpha-investing rule that satisfies Condition \ref{assump2} while violating Condition \ref{assump1}. However, we will ensure that a weaker version of Condition \ref{assump1} is fulfilled:
	\begin{equation}\label{eq:counter2fsrhat}
		\dfrac{\sum_{j:t_0^j\leq t_k}\alpha_{t_k}^{j,A}\vee \alpha_{t_k}^{j,B}}{|\mathcal S_{t_k}|\vee 1} \leq \alpha
	\end{equation}
	holds for all $t_k\in\mathbb T$. 
	The basic strategy is that test levels for task $j$ are determined by its arrival time $t_0^j$ and the decision times of selections. Initially, the test levels for tasks $j\in\{1,\dots, k\}$ are assigned values that change at different decision times, while ensuring that the sum of the test levels for these tasks always equals $\alpha$. Moving forward, each selection leads to an increase in the test level by a time-change value for the $k$ tasks that arrive after the selected task.   
	
	Let $T_{n}(\mathcal S_{t,j-})$ represent the time when task $I_{n}(\mathcal S_{t,j-})$ is stopped. Fix a tuning parameter $k\in\mathbb N$, and define the function $g_k(i) = 2^{-i}\mathbb I\{i<k\} + 2^{-(k-1)}\mathbb I\{i = k\}$. Let $s^j = \min\{i: t_i \geq t^j_0\}$ denote the index of first decision time met by task $j$ after its arrival. 
	Each selection $i\in\mathcal S_{t,j-}$ contributes to the test levels of task $j$ if $j\in\{i+1,\dots,i+k\}$. However, in this case, the contribution changes with decision time, so the test levels are no longer monotonic.
	Formally, the test levels for active tasks at decision time $t_i$ are given by:
	\begin{equation}\label{eq:countex2}
		\begin{aligned}
			\alpha_{t_i}^{j,A} = \alpha_{t_i}^{j,B} & = \alpha\cdot g_k([(t_i - t_{s^j})\bmod k] + 1)\mathbb I\{j\leq k\} \\
			& \quad + \sum_{n\geq 2}\alpha \cdot g_k([(t_i - T_{n}(\mathcal S_{t_i,j})) \bmod k] + 1)\mathbb I\{I_n(\mathcal S_{t_i,j}) < j \leq I_n(\mathcal S_{t_i,j}) + k\},
		\end{aligned}
	\end{equation}
	where $a \bmod b$ represents the remainder after dividing $a$ by $b$. According to \eqref{eq:countex2}, for each selection $l\in\mathcal S_{t,j}$, the total contribution to the test levels of task in $\{i+1, \dots, i+k\}$ is fixed and equals to $\alpha$. However, the allocation to each task in this set changes depending on the decision time. 
	This construction uses information only from tasks 1 through $j-1$, and thus  Condition \ref{assump2} is satisfied.
	However, the estimate in \eqref{eq:fsrhat} can be very large, indicating a violation of Condition \ref{assump1}. The SAVA algorithm using this construction of test levels is referred to as ``Method 1''. 
	
	For comparison, we design test levels for valid SAVA algorithm with a similar form as follows:
	\begin{equation}\label{eq:savaspecial}
		\begin{aligned}
			\alpha_{t_i}^{j,A} = \alpha_{t_i}^{j,B}
			& = \alpha\cdot g_k(j)\mathbb I\{j\leq k\} \\
			& + \sum_{n\geq 2}\alpha \cdot g_k(j - I_n(\mathcal S_{t_i,j})) \mathbb I\{I_n(\mathcal S_{t_i,j}) < j \leq I_n(\mathcal S_{t_i,j}) + k\}.
		\end{aligned}
	\end{equation}
	This construction is monotonic in $t$, as can be proven similarly to the proof in Section \ref{sec:monotoneproof}, ensuring that Condition \ref{assump1} holds. The remaining conditions are straightforward to verify. Therefore, the SAVA algorithm with these test levels controls FSR and mFSR for all $t\in\mathbb T$.
	
\begin{figure}[tb]
		\centering
		\includegraphics[width=0.90\linewidth]{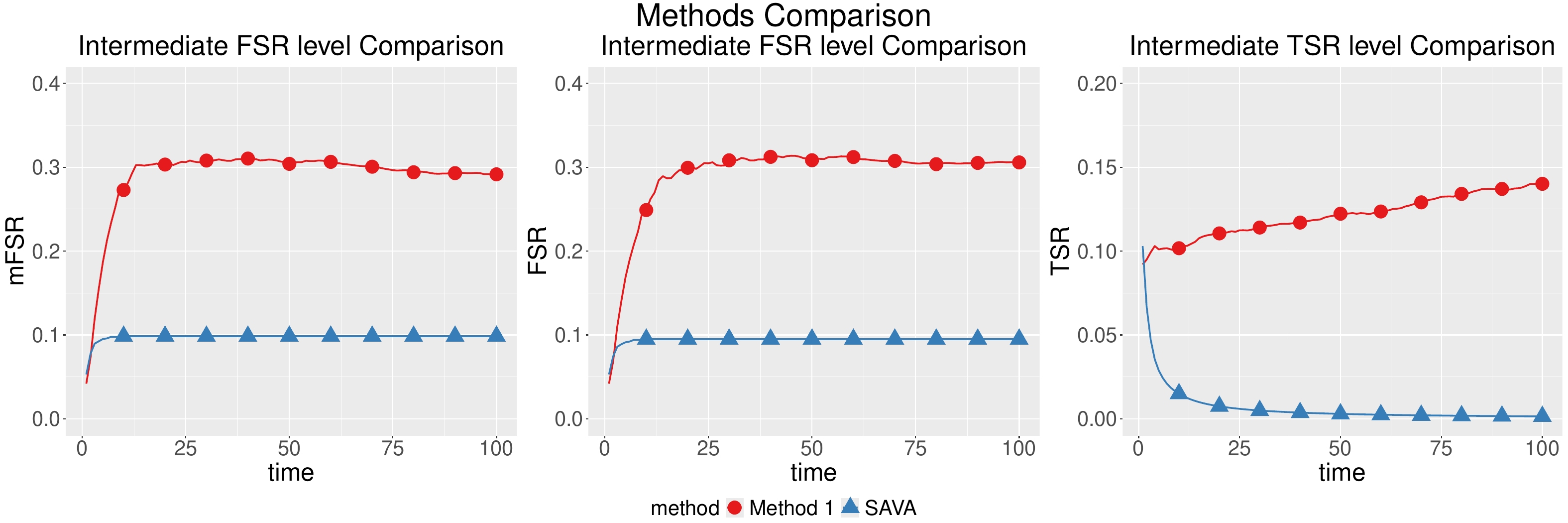}
		\caption{\small Performances of Method 1 and SAVA.}
		\label{fig:ctexam 1}
	\end{figure}

	The performance of these algorithms is presented through simulations with a setup similar to that discussed in Section \ref{subsec:online-rule}.
	The target (m)FSR level is set to 0.1, and set $T = 100$. The new task arrives with probability $p=1$. We repeated the experiments 1000 times and summarized the averaged results in Figure~\ref{fig:ctexam 1}. Algorithm \ref{alg:SAVA} with test levels defined in \eqref{eq:savaspecial} is denoted as ``SAVA''. It is evident that Method 1 fails to control both mFSR and FSR, underscoring the importance of the maximum constraints on test levels in \eqref{eq:fsrhat}. Since the test levels are not monotonic, the probability of falsely selecting arm A is $\mathrm{Pr}_{\theta^j = \mathrm B}(\delta_t^j = \mathrm A) = \mathrm{Pr}_{\theta^j = \mathrm B}(\text{There exists } s\leq t: p^{j,A}\leq \alpha_s^{j,A}, s\in\mathbb T)$. However, the test level $\alpha_t^{j,A}$ in \eqref{eq:countex2} can underestimate this probability, leading to inflated error rates.

	\subsection{Counterexample 2}\label{sec:counterexample2}
	
	The second counterexample indicates that the dependence of test levels on more data streams can impact mFSR and FSR performances. This is illustrated with an extreme case where test levels $(\alpha_{t_k}^{j,A},\alpha_{t_k}^{j,B})$ are allowed to depend on all information of data streams arrived before $t_k$.

	As discussed in Section \ref{sec:counterexample1},  always-valid directional p-values are not updated. 
	This counterexample involves 100 tasks, with the means of the distributions of their data streams given as follows:
	$$ \mu^j = \left\{\begin{array}{rl}
		2.5, & j=1,2,\dots,20, \\
		0.01, & j = 21,22, \dots,50, \\
		2.5, & j = 51,52,\dots,60,\\
		0.001, & j = 61,62,\dots, 70,\\
		2.5, & j=71,72,\dots, 80.\\
		0.001, & j = 81,82,\dots, 100.\\
	\end{array}\right.
	$$
	We assume that $\theta^j = \mathrm A$ if $\mu^j \geq 0$, and $\theta^j = \mathrm B$ otherwise. Since the design of test levels is central to the SAVA algorithm, the data collection step is omitted, and the always-valid directional p-values are generated directly as follows:
	\begin{itemize}
		\item When $\theta^j = \mathrm A$, we assign $p^{j,A}=1-\Phi(Y^j)$, with $Y^j\sim N(\mu^j,0.25)$, where $\Phi$ is the distribution function of $N(0,1)$. Concurrently, $p^{j,B}$ is assigned a uniform distribution $U(0,1)$. 
		\item When $\theta^j = \mathrm B$, $p^{j,B}=\Phi(Y^j)$, where again $Y^j\sim N(\mu^j,0.25)$, while $p^{j,A}$ is assigned a uniform distribution $U(0,1)$. 
	\end{itemize}
	
	We aim to construct test levels such that $(\alpha_{t_k}^{j,A},\alpha_{t_k}^{j,B})$ are $\sigma(\mathcal G_{t_k}^{1:(j-1)},\mathcal G_{t_{k-1}}^{j:n_{k-1}})$-measurable, where $n_{k-1} = |\{j\in\mathbb J: t_0^j \leq t_{k-1}\}|$ represents the total number of tasks arrived up to $t_{k-1}$. 
	The basic strategy for construction is as follows: after the first decision time at which task $j$ is evaluated, the test levels for task $j$ are set to the smaller p-value, $\min\{p^{j,A}, p^{j,B}\}$, provided that the alpha-wealth is sufficiently large.
	
	At decision time $t_1$, the alpha-wealth $W_{t_1}$ is initialized to $\alpha$. At any decision time, say $t_k$, we first use the function $r_{t_k}(\cdot)$ defined in Section \ref{subsec:pro-alg} to rank the active tasks. Then, we define $W_{t_k,r_{t_k}(1)} \coloneqq W_{t_k}$. For each task $r_{t_k}(i)$, if it was active at $t_{k-1}$, and its test levels $\alpha_{t_{k-1}}^{j,A} = \alpha_{t_{k-1}}^{j,B} = 0$, and if the wealth $W_{t_k,r_{t_k}(i)}$ is larger than $\min\{p_{t_{k-1}}^{r_{t_k}(i),A}, p_{t_{k-1}}^{r_{t_k}(i),B}\}$, then we assign $\alpha_{t_k}^{j,A} = \alpha_{t_k}^{j,B} = \min\{p_{t_{k-1}}^{j,A},p_{t_{k-1}}^{j,B}\}$, and let the wealth for next task $W_{t_k,r_{t_k}(i+1)} \coloneqq W_{t_k, r_{t_k}(i)} - \min\{p_{t_{k-1}}^{j,A},p_{t_{k-1}}^{j,B}\}$. Otherwise, we set test levels $\alpha_{t_k}^{j,A} = \alpha_{t_k}^{j,B} = 0$. 
	After assigning test levels for all active tasks, we use decision rule \eqref{eq:decisionrule} to make decisions for active tasks. The wealth for next decision time is assigned a value as $W_{t_{k+1}} = W_{t_k, r_{t_k}(|\mathcal A_{t_k}|)} + \alpha(\max\{1,|\mathcal S_{t_k}|\} - \max\{1,|\mathcal S_{t_{k-1}}|\})$. Finally, we output the decisions for the active tasks. The procedure is continuously executed through all decision times, and is summarized as Algorithm \ref{alg:counter1}. 
	\begin{algorithm}
		\caption{}\label{alg:counter1}
		\vskip6pt
		\begin{algorithmic}
			\State \textbf{Input}: a decision time grid $\{t_i\}$, initial time $\{t_0^j\}_{j\in\mathbb J}$, a target FSR level $\alpha$, tolerance durations $b_j = +\infty$ for all $j$, a variable $W = \alpha$.
			\State \textbf{For each } $t_i \in \mathbb T$, \textbf{do:}
			\State \quad \textbf{Step 1}: Update the active tasks $\mathcal A_{t_{i}} $ defined in \eqref{eq:active-set}. 
			\State \quad \textbf{Step 2}: For active tasks that arrived between $t_{i-1}$ and $t_i$, set test levels as zero.
			\State \quad \textbf{Step 3}: Calculate always-valid directional p-values $\{(p_{t_i}^{j,A}, p_{t_{i}}^{j,B})\}_{j\in \mathcal A_{t_{i}}}$ for active tasks as discussed above.
			\State \quad \textbf{Step 4}: \textbf{For} $j = r_{t_i}(1),\dots,r_{t_i}(|\mathcal A_{t_i}|)$, \textbf{do:}
			\State \qquad \textbf{Step 4.1}: If $j\in\mathcal A_{t_{i-1}}$, $\alpha_{t_{i-1}}^{j,A} = \alpha_{t_{i-1}}^{j,B} = 0$, and $W\geq \min\{p_{t_{i-1}}^{j,A},p_{t_{i-1}}^{j,B}\}$, then set 
			\[\alpha_{t_{i-1}}^{j,A} = \alpha_{t_{i-1}}^{j,B} = \min\{p_{t_{i-1}}^{j,A},p_{t_{i-1}}^{j,B}\},
			\]
			\qquad and update $W\leftarrow W - \min\{p_{t_{i-1}}^{j,A},p_{t_{i-1}}^{j,B}\}$.
			\State \qquad \textbf{Step 4.2}: Make decisions $\delta_{t_i}^j$ according to decision rule \eqref{eq:decisionrule}.
			\State \quad \textbf{Step 5}: For tasks $j$ such that $\delta_{t_{i-1}}^j \neq \text{C}$, update $\delta_{t_i}^j = \delta_{t_{i-1}}^j$, and stop sampling for these tasks.
			\State \quad \textbf{Step 6}: Update $W\leftarrow W + \alpha(\max\{1,|\mathcal S_{t_i}|\} - \max\{1,|\mathcal S_{t_{i-1}}|\})$.
			\State  \quad \textbf{Output}: decision-making states $\{\delta_{t_i}^j:j\in\mathcal A_{t_i}\}$. 
			\State \textbf{End for}
			
		\end{algorithmic}
	\end{algorithm}
	
	The test levels only rely on information prior to $t_{k}$ and are therefore $\mathcal G_{t_{k-1}}^{1:n_{k-1}}$-measurable, and naturally $\sigma(\mathcal G_{t_k}^{1:(j-1)},\mathcal G_{t_{k-1}}^{j:n_{k-1}})$-measurable. Equation \eqref{eq:fsrhat} is still satisfied during the allocation of alpha-wealth, so only Condition \ref{assump2} is violated.
	
\begin{figure}[tb]
		\centering
		\includegraphics[width=0.90\linewidth]{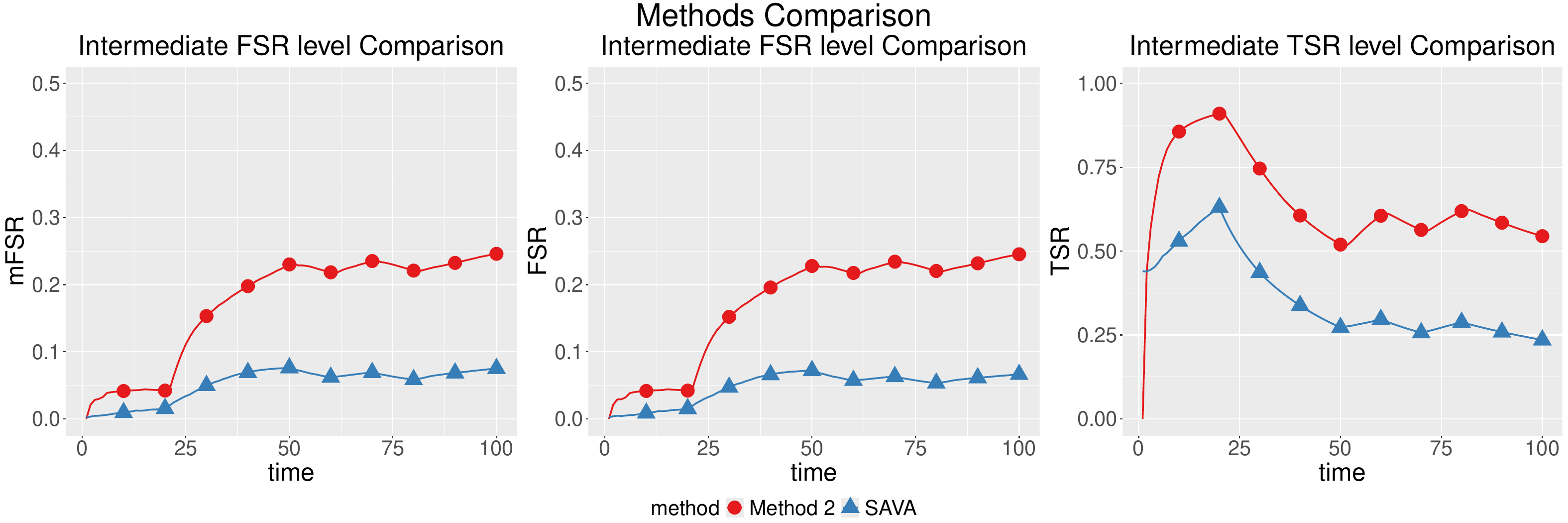}
		\caption{\small Performances of Method 2 and SAVA.}
		\label{fig:ctexam 2}
	\end{figure}

	The performance of these algorithms is presented through simulations with a setup similar to that discussed in Section \ref{subsec:online-rule}.
	The target (m)FSR level is set to 0.1, and set $T = 100$. The new task arrives with probability $p=1$.
	We repeat the experiment 1000 times and summarize the averaged result in Figure~\ref{fig:ctexam 2}, Algorithm \ref{alg:counter1} is denoted as Method 2 in the figure. Figure \ref{fig:ctexam 2} shows that SAVA effectively controls the FSR, primarily because its test levels fulfill Condition \ref{assump2}, thereby preventing inflation of the false selection rate.

	However, Algorithm \ref{alg:counter1} does not guarantee mFSR and FSR control at all decision times. This is primarily due to the strong dependence of the test levels on the always-valid directional p-values, leading to the false selection rate being much larger than the given level.
	Consider the following analysis for illustration: the probability of falsely selecting arm B in this setup can be calculated approximately as $\mathrm{Pr}_{\theta^j = \mathrm A}(\delta^j_{t} = \mathrm B)=\mathrm{Pr}_{\theta^j = \mathrm A}(p^{j,B} \leq \alpha_t^{j,B}, p^{j,A} > \alpha_t^{j,A}) = \mathrm{Pr}_{\theta^j = \mathrm A}(p^{j,B} \leq \min\{p^{j,A},p^{j,B}\}, \alpha_t^{j,A} = \alpha_t^{j,B} > 0)=\mathrm{Pr}_{\theta^j = \mathrm{A}}(p^{j,B}\leq p^{j,A},\alpha_t^{j,A} = \alpha_t^{j,B} = p^{j,B})$. When $p^{j,A}$ is powerless—particularly in extreme cases where the distribution of $p^{j,A}$ has a high probability mass near 1—the probability $\mathrm{Pr}_{\theta^j = \mathrm{A}}(p^{j,B} \leq p^{j,A}, \alpha_t^{j,A} = \alpha_t^{j,B} = p^{j,B})$ can exceed the test level $\alpha_t^{j,B} = p^{j,B}$. This results in a poor estimate of the false selection probability.
	Without Condition \ref{assump2}, the estimate in \eqref{eq:fsrhat} becomes invalid unless additional conditions on the always-valid directional p-values are imposed.

			\section{Comparison with Multi-armed Bandits}\label{sec:mab}
		
		We discuss important distinctions between SAVA and multi-armed bandits, a framework that has been extensively studied in sequential decision-making to efficiently identify promising arms from a set of potential candidates \citep{explorebandits, robustband16, MBP, maxobj24}. 
		
		First, SAVA and online FDR rules emphasize uncertainty quantification and error rate control in selective and sequential inference. In contrast, the multi-armed bandits framework addresses the problem from an operational perspective. For example, the algorithm presented in \cite{Degenne2019NonAsymptotic} focuses on identifying the top $M$ arms without examining the associated error rates or statistical risks in sequential decisions. 
		
		Second, the sampling and operational schemes in these two lines of work differ significantly. In the literature on multi-armed bandits, the primary objective is to minimize the regret. In contrast, our goal is to efficiently identify promising tasks while providing theoretical guarantees on FDR/FSR control. The operation of SAVA carefully calibrates the trade-off between exploration and exploitation.
		
Finally, the error metrics differ between these frameworks. Multi-armed bandit analyses typically consider (a) the probability of finding the best arm \citep{Chen2015}, (b) identifying all ``good'' arms \citep{Mason2020}, or (c) ensuring that all selected arms are ``good'' \citep{katz20}. In contrast, SAVA targets large-scale selection tasks with thousands of candidates, where it is often acceptable if most (e.g., 95\%) of the selections are promising. Framing the problem in terms of FDR/FSR control therefore provides a more practical and relevant objective.

\end{document}